\documentclass[onecolumn,11pt]{article}
\pdfoutput=1
\usepackage{jheppub}
\usepackage[T1]{fontenc}

\usepackage[font={small}]{caption}
\usepackage{subfigure}
\usepackage{graphicx}
\usepackage{dcolumn}

\usepackage{float}
\usepackage[normalem]{ulem}
\usepackage{mathrsfs}
\usepackage{multicol}
\usepackage{cancel}
\usepackage{mathtools}
\usepackage{hyperref}

\usepackage{subfigure}
\usepackage{xcolor}
\usepackage{simplewick}
\usepackage{bm}
\usepackage{amsmath}
\usepackage{amsfonts}
\usepackage{amssymb}
\usepackage{graphicx}
\usepackage{simplewick}
\usepackage{hyperref}
\usepackage{multirow}
\usepackage{enumitem}
\usepackage{physics}
\usepackage{booktabs}
\usepackage{xcolor}
\usepackage{dsfont}
\definecolor{ao}{rgb}{0, 0.7, 0}

\usepackage[nohyperlinks,nolist]{acronym}

\newcommand{\fs}{\mathrm{FS}}
\newcommand{\pf}{\mathrm{PF}}
\newcommand{\J}{\mathcal{J}}
\newcommand{\I}{\mathcal{I}}

\newcommand{\unitMat}{\mathds{1}}
\def\le{\left(}
\def\ri{\right)}
\newcommand{\tar}{\textrm{T}}

\newcommand{\inner}[2]{\langle #1,#2 \rangle}
\newcommand{\ham}{\hat{H}}

\newcommand{\trH}{{\inner{\ham}{\ham}}}

\newcommand{\fdi}{\mathcal{F}_{\text{bi-inv}}}

\newcommand{\fpi}{\mathcal{F}_{\mathcal{I}}}
\newcommand{\ffs}{\mathcal{F}_{\fs}}
\newcommand{\fpf}{\mathcal{F}_{\pf}}

\newcommand{\beq}{
\begin{equation}}
  \newcommand{\eeq}{
\end{equation}}
\newcommand{\bea}{
  \begin{equation}
  \begin{aligned}}
    \newcommand{\eea}{
    \end{aligned}
\end{equation}}

\usepackage{mathtools}

\newcommand{\argdot}{{\hspace{0.18em}\cdot\hspace{0.18em}}}

\DeclareMathOperator{\arctanh}{arctanh}

\usepackage{xcolor,pifont}
\newcommand*\colourcheck[1]{%
  \expandafter\newcommand\csname #1check\endcsname{\textcolor{#1}{\ding{52}}}%
}
\colourcheck{blue}
\colourcheck{green}
\colourcheck{red}

\newcommand*\colourmark[1]{%
  \expandafter\newcommand\csname #1mark\endcsname{\textcolor{#1}{\ding{56}}}%
}
\colourmark{blue}
\colourmark{green}
\colourmark{red}

\newcommand{\lb}{\bar{\lambda}}
\newcommand{\lbs}{\bar{\lambda}^*}
\newcommand{\lls}{\lambda^*}

\newcommand{\eg}{{e.g.,}\ }
\newcommand{\ie}{{i.e.,}\ }

\newtheorem{thm}{Theorem}[section]
\newtheorem{theo}[thm]{Theorem}

\newtheorem{deff}[thm]{Definition}

\title{CFT Complexity and Penalty Factors}
\author{Stefano Baiguera${}^{1,2}$, Nicolas Chagnet${}^3$, Shira Chapman${}^4$,
Osher Shoval${}^4$}
\affiliation{$^1$INFN Sezione di Perugia, Via A. Pascoli, 06123 Perugia, Italy \\
$^2$Dipartimento di Matematica e Fisica, Universit\`{a} Cattolica del Sacro Cuore, \\ Via della Garzetta 48, 25133 Brescia, Italy\\
$^3$ Institute Lorentz for Theoretical Physics, $\Delta$-ITP, Leiden University, \\
Niels Bohrweg 2, Leiden, the Netherlands \\
$^4$ Department of Physics, Ben-Gurion University of the Negev,
  \\ David Ben Gurion Boulevard 1, Beer Sheva 84105, Israel.}

\emailAdd{stefano.baiguera@pg.infn.it}
\emailAdd{chagnet@lorentz.leidenuniv.nl}
\emailAdd{schapman@bgu.ac.il}
\emailAdd{oshersho@post.bgu.ac.il}

\abstract{
Quantum complexity of conformal field theory (CFT) states has recently gained significant attention, both as a diagnostic tool in condensed matter systems and in connection with holographic observables probing black hole interiors. 
Previous studies have primarily focused on cases where all generators of the conformal group contribute equally to the cost of building a circuit. 
In this work, we present a general framework for studying the complexity of circuits in generic Lie groups, where penalty factors assign relative weights to different generators. Our approach constructs a metric on the coset space of quantum states, induced from a (pseudo-)Riemannian norm on the space of unitary circuits. The geodesics of this metric are interpreted as optimal circuits. The method builds on the formalism of (pseudo-)Riemannian submersions and connects naturally to other prescriptions in the literature, including cost function minimization along stabilizer directions and constructions based on coadjoint orbits.
As a concrete application, we compute state complexity for states in one- and two-dimensional CFTs. For specific choices of penalty factors, our prescription yields a positive-definite metric with a viable interpretation as complexity; in other cases, the resulting metric is indefinite. In the viable regime, we derive analytic results when a specific penalty factor is turned off, develop perturbative expansions for small values of the penalty factors, and provide numerical results in the general case.
We comment on the relation of our measure of complexity to holography. }

\begin{document}

\begin{acronym}
  \acro{fs}[FS]{Fubini-Study}
  \acro{mc}[MC]{Maurer-Cartan}
  \acro{cft}[CFT]{conformal field theory}
  \acro{ads}[AdS]{Anti-de Sitter}

  \acrodefplural{cft}{conformal field theories}
\end{acronym}

\maketitle

\section{Introduction}

\sloppy

The notion of quantum complexity has become ubiquitous in recent studies of quantum information
and quantum gravity, \eg  see the reviews \cite{Susskind:2018pmk,Chapman:2021jbh,Baiguera:2025dkc}.
Complexity  provides a useful tool to estimate the difficulty of decoding the radiation of an evaporating
black hole
\cite{Harlow:2013tf,Kim:2020cds,Hayden:2007cs,Yoshida:2017non}, it 
can be used to distinguish phases of matter \cite{Liu:2019aji,Yang:2023qxx},
and even helps us understand the
interior of black holes in holography  
\cite{Susskind:2014rva,Stanford:2014jda,Brown:2015bva,Brown:2015lvg,Lehner:2016vdi,Carmi:2016wjl,Chapman:2016hwi,Couch:2016exn,Carmi:2017jqz,Belin:2021bga,Belin:2022xmt}.\footnote{Other notions of quantum complexity exist in the literature, including Krylov and spread complexity, see \eg  \cite{Parker:2018yvk,Nandy:2024evd,Rabinovici:2025otw}, but we will not be dealing with those here.}

\paragraph{Complexity geometry.} 
The circuit complexity $\mathcal{C}(U)$ associated with a unitary task $U$ in a
discrete set-up is typically defined as the minimal number of gates $g_i$ (\ie
the elementary unitary operators) needed to build a circuit that implements the
unitary operator $U$.
One can extend the notion of complexity to the space of states, by studying the
optimal way to reach from a \emph{reference state} (usually not entangled) to a \emph{target
state}.
In this context, one usually refers to a finite-dimensional system composed of qubits where each unitary can be approximated arbitrarily well by using a universal set of gates consisting, \eg of Hadamard, phase shift, and CNOT gates.
In practical settings, the gates involved in constructing  quantum circuits are not necessarily equally difficult to implement. To encode this aspect in the definition of complexity, one equips the different gates with relative weights, which often go under the name of \emph{penalty factors}. 
While generic features of the time evolution of complexity can be identified by means of simple counting arguments
\cite{Susskind:2014jwa,Chapman:2021jbh,Susskind:2018pmk,Brown:2016wib}, this discrete problem
is typically very hard to solve precisely due to the lack of analytic tools, and one has
to resort to heavy numerics to make progress. 

The remarkable insight by Nielsen and collaborators was to relate the previous discrete setting to a continuous formulation, where complexity is defined as the minimal length of a geodesic in a manifold defined over the unitary space and equipped
with an appropriate notion of distance (referred to as \textit{cost
function}) \cite{Nielsen1,Nielsen2,Nielsen3}.
This framework presents several advantages.
First, there is a large equivalence class of metrics that provide lower- and upper-bounds on the discrete counting of gates \cite{Nielsen1,Nielsen2,Nielsen3}, and whose Nielsen complexities all agree at large distances up to polynomial terms in the number of qubits and in the length of the path \cite{Brown:2021uov,Brown:2022phc}. 
Second, the definition of Nielsen complexity as a minimal distance on the unitary manifold allows, in some cases, to apply powerful analytic tools from Riemannian geometry.
Since any above-mentioned equivalence class includes members with modest curvature~\cite{Brown:2021uov,Brown:2022phc}, the computation of Nielsen's complexity becomes easier for such representatives.
Third, the continuous nature of this definition is suitable for the
generalization to the case of quantum field theories (QFTs).
Finally, Nielsen's framework is directly related to \textit{Trotterization}, a quantum-computational method for building a discrete circuit that can simulate the time evolution of a Hamiltonian~\cite{Lloyd_96_Universal}.

In Nielsen's formulation, each generator of the unitary algebra is equipped with a penalty
factor -- a parameter that measures the relative cost of moving along the corresponding
direction in the tangent space of the group manifold. This parallels the relative weights assigned to the different gates in the discrete setting of circuit complexity.
The geometric properties of Nielsen's complexity are encoded by the cost
function, which we shall denote by $\mathcal{F}$.
There exist a plethora of candidate cost functions, but the preferred choices
in the literature have been the so-called $\mathcal{F}_1$ and $\mathcal{F}_2$ norms 
\cite{Nielsen1,Brown:2017jil,Balasubramanian:2018hsu,Brown:2019whu,Bernamonti:2019zyy,Balasubramanian:2019wgd,Auzzi:2020idm,Caginalp:2020tzw,Basteiro:2021ene,Brown:2021uov,Balasubramanian:2021mxo,Brown:2022phc,Baiguera:2023bhm}.
The reason is that the former has the interpretation of counting the number of
gates being implemented, while the latter defines a Riemannian metric, which
is amenable to an analytic treatment.

In particular, some guiding principles have been proposed to choose a
physically sensible cost function.
One possibility is to assign penalty factors in such a way as to accurately reproduce an experimental setting, by assigning larger penalty factors to generators that exponentiate to unitary gates that are harder to implement.
Another possibility is to define a cost function that reproduces certain
physical properties of a system.
For instance, one could demand that close geodesics deviate from each other, which is a necessary condition to achieve chaos \cite{Brown:2017jil}. 
For a given Hamiltonian, which induces motion on the unitary group manifold via the Schr\"{o}dinger equation, one can define a norm on the tangent space such that the length of a geodesic (corresponding to Nielsen's complexity) reproduces the switchback effect, \ie a delay of complexity occurring when a perturbation is inserted \cite{Stanford:2014jda}.
The two above properties -- geodesic deviation and the switchback effect -- typically require the existence of a right-invariant (but not bi-invariant) metric on the group manifolds, which allows for regions with negative sectional curvature \cite{MILNOR1976293}.

The geometry of Nielsen's complexity has been thoroughly investigated for finite-dimensional Lie groups, mostly in the case of $\mathrm{SU}(N)$.
For the purposes of this work, we will also be interested in using the right-invariant metric on the space of unitary matrices to induce a metric on the space of quantum states. Indeed, Ref.~\cite{Brown:2019whu} suggested that an induced metric on the space of quantum states can be obtained by performing a local minimization over the degrees of freedom (referred to as
\textit{stabilizer} directions) which do not affect the reference state.
In the Riemannian case, the projection from the unitary space to the space of states, performed using the previous minimization, was shown to be a Riemannian submersion in Ref.~\cite{Auzzi:2020idm}.
This geometric interpretation allows to systematically define a metric over the space of states and to exploit properties of the submersions, for example O'Neill's theorem \cite{ONeill1,ONeill2}, which relates the sectional curvatures on the space of unitaries and those on the space of states.

\paragraph{Complexity in QFT,  CFT and holography.}
There has been a lot of interest in defining notions of complexity in QFT and conformal field theory (CFT). 
First, these notions allow to optimize the implementation of a task in quantum many-body systems. Second, CFT features resonate with cMERA, the continuous version of \textit{multiscale entanglement renormalization ansatz} (MERA)~\cite{Vidal:2007hda,Vidal:2008zz,Haegeman:2011uy,Nozaki:2012zj}, a class of tensor networks which efficiently simulates the ground states of critical systems.\footnote{A tensor network is a graphical representation of a quantum state in terms of a multilinear map~\cite{Cirac:2020obd}.} Finally, the notion of complexity in CFT plays an important role in holography. 

The holographic principle posits a duality between gravity in a bulk spacetime and a quantum system on its boundary. A well-established example is the AdS/CFT correspondence, relating gravitational dynamics in AdS$_{d+1}$ to a $d$-dimensional conformal field theory. Within this framework, a connection between the quantum complexity of CFT states and geometric quantities in the bulk has been proposed. The original conjecture identified the volume of the Einstein-Rosen bridge (ERB) with the complexity of the dual thermofield double (TFD) state~\cite{Susskind:2014rva}, with later works proposing various refinements~\cite{Brown:2015bva,Brown:2015lvg,Couch:2016exn,Couch:2018phr,Belin:2021bga,Belin:2022xmt}.

On the field theory side, numerous approaches have been developed to define quantum complexity~\cite{Jefferson:2017sdb,Chapman:2017rqy,Hashimoto:2017fga,Khan:2018rzm,Hackl:2018ptj,Chapman:2018hou,Camargo:2018eof,Bernamonti:2019zyy,Guo:2018kzl,Bhattacharyya:2018bbv,Magan:2018nmu,Bueno:2019ajd,Caceres:2019pgf,Chapman:2019clq,Ge:2019mjt,Bernamonti:2020bcf,Bernamonti:2021jyu,Chowdhury:2023iwg,Caputa:2017urj,Caputa:2017yrh,Bhattacharyya:2018wym,Takayanagi:2018pml,Cotler:2018ufx,Camargo:2019isp,Boruch:2020wax,Boruch:2021hqs,Caputa:2018kdj,Flory:2020eot,Flory:2020dja,Erdmenger:2020sup,Erdmenger:2024xmj,Chagnet:2021uvi}. Early studies focused on Nielsen's geometric approach for trajectories between Gaussian states, highlighting its ultraviolet (UV) structure and comparing it to holographic expectations~\cite{Jefferson:2017sdb,Chapman:2017rqy,Hashimoto:2017fga,Khan:2018rzm,Hackl:2018ptj,Chapman:2018hou,Camargo:2018eof,Guo:2018kzl,Cotler:2018ufx,Bernamonti:2019zyy,Caceres:2019pgf,Ge:2019mjt}. Other proposals, such as path-integral complexity~\cite{Caputa:2017urj,Caputa:2017yrh} and Krylov complexity~\cite{Parker:2018yvk,Nandy:2024evd,Baiguera:2025dkc,Rabinovici:2025otw}, offer complementary perspectives. Here, we will adopt Nielsen’s framework.\footnote{We refer the reader to reference~\cite{Camargo:2019isp} for a discussion of the relation between path-integral and Nielsen complexities; and references~\cite{Lv:2023jbv,Aguilar-Gutierrez:2023nyk} for the comparison between Krylov and Nielsen's approaches.}

While the problem of computing Nielsen's complexity is usually hard (due to the high dimensionality of the unitary group manifold), it can become solvable when one exploits the symmetry group $G$ underlying a given theory.
In this context, the case of CFTs is particularly important.
The reason is twofold. First, this setting can shed light on the study of quantum systems near critical points.
Second, CFT states are dual to black holes via the holographic map. Therefore, we can hope to improve our understanding of the conjectured relation to holographic observables probing the interior of black holes.

Progress in studying complexity for two-dimensional CFT states began with~\cite{Caputa:2018kdj}, where the authors analyzed the optimal implementation of continuous circuits using generators of the infinite-dimensional Virasoro algebra. This setup was further generalized in~\cite{Erdmenger:2024xmj} by introducing deformations through the insertion of a primary operator. Related constructions were also used to develop holographic duals for circuits built from conformal transformations~\cite{Erdmenger:2020sup,Flory:2020eot,Flory:2020dja,Erdmenger:2021wzc,Erdmenger:2022lov,deBoer:2023lrd}. A higher-dimensional generalization was proposed in~\cite{Chagnet:2021uvi}, where the authors established connections between Nielsen's complexity with isotropic cost functions and both the Kähler geometry of coadjoint orbits and geodesic distances in AdS spacetime. Unlike in two dimensions, where the relevant symmetry algebra is infinite-dimensional, the conformal group $\mathrm{SO}(d,2)$ in higher dimensions is finite-dimensional, leading to significant simplifications.

\paragraph{This work.}
Despite these exciting developments, the studies of complexity in CFT performed so far have been limited in that only cases where the cost functions are isotropic, \ie
the generators have the same cost, have been considered.
The main purpose of the present paper is to overcome this limitation by setting up a systematic procedure for defining a metric with non-trivial penalty factors
over the space of states for a theory invariant under any (possibly
non-compact) finite-dimensional Lie group. This procedure can be concisely phrased by exploiting the mathematical framework of (pseudo--)Riemannian submersions, as we do below.
In this way, we provide a systematic map from the unitary manifold to the Hilbert space that not only applies to the case of compact Lie groups (previously covered by Ref.~\cite{Auzzi:2020idm}), but also to non-compact Lie groups, which play a fundamental role in characterizing the spacetime symmetries of a theory. 
Furthermore, we explain how in this case the minimization technique of \cite{Brown:2019whu} becomes an extremization of the metric over the stabilizer directions.
We comment on a potential relation with the method of coadjoint orbits, that was employed in Ref.~\cite{Chagnet:2021uvi} to study the Nielsen complexity of CFTs without penalty factors.

We apply the above procedure to study the influence of penalty factors on the complexity of \ac{cft} states in one and two dimensions.
We find analytic results for the complexity in simple cases where one of the penalty factors is turned off. In this setting, Nielsen's complexity is a simple rescaling of the result that one obtains in the isotropic case, where the geometry on the Hilbert space is characterized by the Fubini-Study (FS) metric. In the other cases, we provide expansions for
small values of the penalty factors, and numerical results otherwise.

We find that the complexity metric obtained through our procedure is positive-definite only within certain ranges of the penalty factors. This gives rise to a kind of landscape and swampland for CFT complexities, where only specific choices of penalties yield a well-defined complexity interpretation. Understanding these constraints directly from the CFT perspective would be an interesting direction for future work. Within the viable range of penalties, the cost associated with the dilatation operator has a relatively minor effect on the complexity, while increasing penalties in other directions leads to more significant changes.

A convenient collection of the main results is reported in
table~\ref{tab:results}. We view our work as an additional step towards defining complexity for \ac{cft} states.
As an aside, in appendix~\ref{app:holographic_interpretation} we find a holographic connection between the CFT complexity defined by pseudo-Riemannian submersions and geodesics in AdS spacetime, in the case of an isotropic cost function. 
We do this by relating the FS metric with the bulk symplectic form of a massive particle moving in an AdS geometry (this relation was first pointed out in~\cite{Chagnet:2021uvi}), and then we interpret the projection over the coset space in terms of the vanishing of an appropriate bulk symplectic potential. 
This is a first step towards a complete connection to holography, that we hope to achieve in the future.

\begin{table}[ht]
  \begin{center}
    \begin{tabular}  {|c|c|c|} \hline  \textbf{Nielsen complexity} &
      \textbf{One-dimensional CFTs}    & \textbf{Two-dimensional CFTs} \\ \hline
      \rule{0pt}{4.9ex}\textbf{Analytic results}    &
      Sections~\ref{ssec:interlude:simple_metric}  and~\ref{ssec:analytic_1d}
        &  Section~\ref{ssec:2d_analytic} \\
      \rule{0pt}{4.9ex}\textbf{Perturbative results}    &
      Section~\ref{ssec:perturbative_1d}  &
      Sections~\ref{ssec:2d_perturbative}
       \\
      \rule{0pt}{4.9ex} \textbf{Numerical results}  &
      Section~\ref{ssec:numerical_1d}  &  Section~\ref{ssec:2d_numerical} and appendix~\ref{app:sec:cost_nontrivial_bdy}  \\[0.2cm]
      \hline
    \end{tabular}
    \caption{Summary of the main results of Nielsen's complexity for CFT states in
    the presence of penalty factors.}
    \label{tab:results}
  \end{center}
\end{table}

\paragraph{Outline.}
The paper is organized as follows.
We begin in section~\ref{sec:complexity_geometry} by briefly reviewing the
definition of complexity geometry by Nielsen.
Section \ref{ssec:projection_coset}  contains the main core of the paper: it derives a general procedure to project a metric with non-trivial penalty factors from a (possibly non-compact) Lie group to the coset space.
In section~\ref{sec:1dCFT}, we apply this method to investigate the state complexity of one-dimensional CFTs. 
In section~\ref{sec:2dCFT}, we use our method to analyse the complexity of two-dimensional CFT states. 
Conclusive remarks and a discussion of possible future directions are reserved to section~\ref{sec:discussion}.
The appendices contain additional mathematical details on (pseudo-)Riemannian submersions (appendix~\ref{app:submersions}), on the fundamental representation of the conformal group (appendix~\ref{app:fund_CFT}), and on numerical computations of complexity in certain two-dimensional CFTs (appendix~\ref{app:sec:cost_nontrivial_bdy}).
Finally, we discuss the holographic interpretation of our results in
appendix~\ref{app:holographic_interpretation}.
\section{Complexity geometry}
\label{sec:complexity_geometry}

In this section, we present Nielsen's geometric approach to quantum complexity \cite{Nielsen1,Nielsen2,Nielsen3}. 
We begin in subsection~\ref{ssec:Nielsen_complexity}, where we define the notion of complexity geometry for both unitaries and quantum states.
An important choice in defining the complexity geometry is that of a cost function, encoding the relative difficulty of applying different generators (or gates). 
subsection~\ref{ssec:cost_function},  we discuss some possible choices of cost functions. Throughout the section, we keep in mind  circuits in SU($N$) as our primary example, as this is the most intuitive case. In later sections, we  explain how to recast the quantum state complexity in terms of a Riemannian submersion and use this to study the complexity for generic, not necessarily compact, Lie groups.

\subsection{Nielsen complexity}
\label{ssec:Nielsen_complexity}

Let us define Nielsen's complexity geometry.
The problem of interest is to build the optimal path that connects the identity with a target unitary $U_T$.
A generic trajectory is described by
\beq\label{eq:unitarydef}
U(t)= \overleftarrow{\mathcal{P}} \exp \le -i\int_0^{t} \dd t^\prime \,
\ham(t^\prime) \ri \, ,
\eeq
where $t\in[0,1]$ is a circuit parameter such that $U(0)=\unitMat$ and
$U(1)=U_T$, and  $\overleftarrow{\mathcal{P}}$ denotes path ordering such
that the circuit is constructed from right to left.
In the previous expression, $\ham(t)$ is the instantaneous Hamiltonian at a
point on the curve, obtained from the Schr\"{o}dinger equation as follows:
\begin{equation}\label{eq:shroeq}
  \ham(t) = i \dot U(t) U(t)^{-1} \, ,
\end{equation}
where $\dot X \equiv \frac{\dd X}{\dd t}$ denotes the derivative with respect
to the path parameter.
In Nielsen's approach, the Hamiltonian is constructed out of an orthonormal
basis of Hermitian generators $\lbrace \omega_I \rbrace$, whose exponentiation
provides a continuous version of the elementary gates from the discrete complexity setting.
For instance, for $\mathrm{SU}(2^n)$,  
a basis of such generators is provided by tensor products of Pauli matrices and two-dimensional identity matrices acting on $n$ qubits.
In general, the Hamiltonian reads\footnote{We will later use the summation convention where repeated indices are assumed to be summed, even when the sum is not explicitly written.}
\beq\label{eq:HamiComb}
\ham(t) = \sum_I Y^I(t) \omega_I \, ,
\eeq
where $Y^I(t)$ are real parameters called \textit{velocities} or
\textit{control functions}, since they describe the tangent vector to a curve in the group manifold.

The relative difficulty of implementing a specific trajectory through the space of unitaries is assessed by a \emph{cost function} $\mathcal{F}[U,\ham]$, which depends on the unitary and the  Hamiltonian along the curve.
This function mimics the fact that certain operations are harder to realize than others in an experimental set-up.
We define \textit{unitary complexity} as the minimal length, computed according to the cost function, of a trajectory connecting the identity with the target unitary:
\begin{equation}
  \mathcal{C}_{\mathcal{F}}[U_T]=\min_{\lbrace U \, : \, U(0)=\unitMat,
  U(1)=U_T  \rbrace}  \int_0^{1} \dd t \, \mathcal{F}[U, \ham]  \, .
  \label{eq:unitary_complexity}
\end{equation}
Different cost functions have been considered in the literature
\cite{Jefferson:2017sdb,Chapman:2017rqy,Brown:2017jil,Balasubramanian:2018hsu,Magan:2018nmu,Caputa:2018kdj,Brown:2019whu,Balasubramanian:2019wgd,Bernamonti:2019zyy,Bueno:2019ajd,Auzzi:2020idm,Caginalp:2020tzw,Chagnet:2021uvi,Basteiro:2021ene,Brown:2021uov,Balasubramanian:2021mxo,Brown:2022phc,Erdmenger:2022lov,Baiguera:2023bhm,Craps1,Craps:2022ese},
but in this work we will focus on cases that lead to a pseudo-Riemannian geometry,
where the tools of differential geometry can be used.
We will discuss some relevant choices for the cost functions in
section~\ref{ssec:cost_function}. In some cases it is convenient to extremize\label{eq:squaredL}
\begin{equation}
    \mathcal{L}\equiv \mathcal{F}^2
\end{equation}
rather than $\mathcal{F}$, where $\mathcal{L}$ plays the role of the Lagrangian.

Next, let us assume that the unitary \eqref{eq:unitarydef}
acts on a Hilbert space $\mathcal{H}$, where $\ket{\psi_R}$ is a
reference state and $\ket{\psi_T}$ a target state that we aim to build.
The \textit{state complexity} is defined as the minimum of the unitary
complexities computed over all the paths connecting the reference to the
target state: 
\beq
\mathcal{C}^{\mathrm{state}}_{\mathcal{F}} [\ket{\psi_T}, \ket{\psi_R}] =
\min_{ \lbrace U \, : \, \ket{\psi_T} = U \ket{\psi_R} \rbrace} \mathcal{C}_{\mathcal{F}}
[U] \, .
\label{eq:def_compl_state}
\eeq
When discussing non-compact Lie groups, the minimization in this equation has to be changed to an \emph{extremization}.\footnote{When talking about unitaries in non-compact Lie groups, we mean unitary representations of the non-compact \emph{complex} Lie algebra.} This is because when focusing on cost functions defined using the notion of inner-product of the Lie group, distances between unitaries will no-longer be positive definite.
We will further comment on this issue in section~\ref{ssec:projection_coset}.

Typically, there exist different unitaries that move us between the same two rays in the Hilbert space of quantum states.  These unitaries are
all related to each other by actions of the stabilizer of the reference state.\footnote{Recall that the space of quantum states is defined in terms of rays in the Hilbert state, i.e., vectors which differ by a phase are identified. For this reason when talking about the stabilizer of a reference state, we mean, unitaries satisfying $U|\psi_R\rangle=e^{i\phi}|\psi_R\rangle$ for some phase $\phi$.} For example, focusing on SU$(N)$, the maximal subgroup is $\mathrm{SU}(N-1) \times \mathrm{U}(1)$ of $\mathrm{SU}(N)$.\footnote{Given an action $G\times X \rightarrow X$ of a
  group $G$ on a set $X$, the stabilizer group of the element $x$ is defined as
$\mathrm{stab} \, x = \lbrace g \in G \, : \, g \cdot x =x \rbrace$.}
In particular, one can define a map from the unitary space to the quotient
space obtained via the projection
\beq
\pi : \mathrm{SU}(N) \rightarrow \mathbb{CP}^{N-1} \equiv
\frac{\mathrm{SU}(N)}{\mathrm{SU}(N-1) \times \mathrm{U}(1)} \, .
\label{eq:quotient_map}
\eeq

The minimization with respect to the stabilizer can be performed at each step
along the trajectory, yielding a norm on the space of states.
This is done as follows. First, we fix $|\psi(t)\rangle$ and  $|\dot
\psi(t)\rangle$.
Using the Schr\"odinger equation $|\dot \psi(t)\rangle=i \hat H (t) |
\psi(t)\rangle$, we then minimize over the degrees of freedom (control
functions) in $\ham(t)$ which are not fixed by $|\psi(t)\rangle$ and  $|\dot
\psi(t)\rangle$. This gives the induced norm
\begin{equation}
  \mathcal{F}^{\text{state}}[| \psi(t)\rangle,\dot \psi(t)\rangle]  =
  \min_{\mathrm{ stab} \, \ket{\psi(t)}} \, \mathcal{F}[U, \ham] \, .
  \label{eq:cost_function_minimized}
\end{equation}
Finally, the state complexity can be re-expressed as
\begin{equation}
  \label{eq:statemin}
  \mathcal{C}^{\mathrm{state}}_{\mathcal{F}} [\ket{\psi_T},\ket{\psi_R}]
  =\min_{{
      \begin{array}{>{\scriptstyle}c}
        |\psi(t)\rangle \\
        |\psi(0)\rangle=\ket{\psi_R}  \\
        |\psi(1)\rangle=\ket{\psi_T}
    \end{array}}
  }  \int_0^1 \dd t \, \mathcal{F}^{\mathrm{state}} [\ket{\psi(t)},|\dot{ \psi}
  (t)\rangle] \, .
\end{equation}
If we consider a cost function on the space of states defined via the
minimization \eqref{eq:cost_function_minimized} of a Riemannian norm, we can study the geodesics of the resulting metric by using the tools of
calculus, including the Euler-Lagrange equations.

\subsection{Choice of the cost function}
\label{ssec:cost_function}

The geometric features of Nielsen complexity are encoded by the cost function $\mathcal{F}[U,\ham]$ entering the definition \eqref{eq:unitary_complexity}.
Our next goal is to select a class of cost functions for the evaluation of complexity. 
Let us begin with some possible cost functions in the definition of unitary complexity \eqref{eq:unitary_complexity}. The simplest possibility is provided by the bi-invariant $L_2$ norm
\begin{align}
\label{eq:cost_Cartan_Killing}
  \fdi^2 \equiv \pm \langle \ham, \ham \rangle   \, ,
\end{align}
where $\langle \argdot, \argdot \rangle$ denotes the Cartan-Killing form on the group, and the Hamiltonian $\hat H$ can be read of equation \eqref{eq:shroeq}.  
In practice, many of our algebraic manipulations will not depend on the exact representation of the algebra and we will therefore be able to use the fundamental representation and the inner product $\langle \argdot, \argdot \rangle$  given by the trace form. For the case of the unitary group $\mathrm{SU}(N)$, the Cartan-Killing form  will be positive-definite.
The choice of sign in equation \eqref{eq:cost_Cartan_Killing}  depends on whether we study a compact or non-compact group and on the signature of the relevant directions in the group. In practice, we select the sign such that the metric on the space of quantum states is positive definite. For instance, when dealing with SU($N$), the sign will be positive. 

We can re-express equation \eqref{eq:cost_Cartan_Killing} in terms of the control functions in equation \eqref{eq:HamiComb}. This is done as follows. Assume that the generators are orthogonal and normalized such that $\langle \omega_I, \omega_J \rangle = \eta_{IJ}$. The matrix $\eta_{IJ}$ is a diagonal matrix with entries $\pm 1$, reflecting the signature of the different group directions. For the case of SU($N$), we have $\eta_{IJ}=\delta_{IJ}$, the identity matrix.   Then from equation \eqref{eq:HamiComb}, we have for orthogonal generators\footnote{It is straightforward to generalize these expressions for the case where the generators are not orthogonal but instead obey another inner product $\langle \omega_I,\omega_J\rangle = A_{IJ}$. This is done by a simple change of basis.\label{foot:changebase}}
\begin{equation}
    Y^I = \langle \ham, \omega_I \rangle/\langle \omega_I, \omega_I \rangle
\end{equation}
where in the last equation the index $I$ is not summed. Using this expression for the control functions, the bi-invariant cost function can be expressed as
\begin{align}
  \fdi^2 = \pm Y^I \eta_{IJ} Y^J =\pm
  \inner{\ham}{\omega_I} \eta_{IJ} \inner{\ham}{\omega_J} \, ,
  \label{eq:cost_Cartan_Killing2}
\end{align}
where for SU($N$), we select the positive sign.

The cost function \eqref{eq:cost_Cartan_Killing} is bi-invariant and treats every generator equally.
However, as stated above, one is often interested in breaking down this isotropy.
The relative costs of the different generators are referred to as \emph{penalty factors}, and can be introduced in this framework
by replacing $\eta_{IJ} \to \I_{IJ}$, where $\I$ is a non-trivial symmetric and positive definite penalty matrix. For the case of a non-compact group, the penalty matrix will no-longer we positive definite, but will have the same signature as $\eta_{IJ}$.
This defines the cost function
\begin{align}
  \fpi^2 \equiv \pm Y^I \I_{IJ} Y^J = \pm\sum_{I,J} \frac{\inner{\ham}{\omega_I}}{\langle\omega_I,\omega_I\rangle}
  \I_{IJ}\frac{\inner{\ham}{\omega_J}}{\langle\omega_J,\omega_J\rangle} \, ,
  \label{eq:cost_Riemannian_penalties}
\end{align}
where the last equality is again only true for orthogonal generators.
We will eventually select the sign in the cases we study such that the metric obtained on the space of quantum states is positive definite.
In the presence of non-trivial penalty factors, the bi-invariance of the cost function is broken, but we still require that the cost function is
right-invariant, \ie independent of the location on the manifold
\cite{MILNOR1976293}.
In what follows, it is useful to point out that the cost function \eqref{eq:cost_Riemannian_penalties} corresponds to the
Riemannian metric
\beq
\dd s^2 = \pm \mathcal{I}_{IJ} \rho^I \rho^J  \, , \qquad
\rho^I \equiv Y^I \dd t ,
\label{eq:Riemannian_metric_penalties}
\eeq
where $\rho^I$ is a right-invariant form on the group manifold.
By construction the space is still homogeneous, but the presence of penalty
factors allows for some of the sectional curvatures to be negative. This is necessary (but not sufficient) in order for the complexity geometry to reproduce features typical of chaotic systems, where nearby geodesics deviate away from each other, and complexity manifests the switchback effect
\cite{Brown:2016wib,Brown:2017jil}.

Next, one can also define cost functions on the space of quantum states. This is done as follows. Consider a path $\ket{\psi(t)} = U(t) \ket{\psi_R}$. The simplest possibility is provided by the \ac{fs} norm defined by 
\beq
\ffs^2 = \ev{\dot U^\dagger \dot U}{\psi_R} - \left|\ev{U^\dagger \dot
U}{\psi_R} \right|^2 =
\ev{\ham^2}{\psi(t)} -  \ev{\ham}{\psi(t)}^2 \, ,
\label{eq:FS_cost_functional}
\eeq
where $\ket{\psi}$ denotes the generic state along the trajectory, and we used
the Schr\"{o}dinger equation \eqref{eq:shroeq} in the last equality.
The previous expression can be formally written in terms of the density matrix
$\rho_{\psi} = |\psi(t)\rangle \langle \psi(t) |$ for the (pure) state along the trajectory and using the Hilbert space trace operator as follows:
\beq
\label{eq:fs-in-inner-form}
\ffs^2 = \Tr_{\mathcal H} \bigl[ \rho_{\psi} \ham_0^2 \bigr] \, , \qquad
\ham_0 \equiv \ham - \Tr_{\mathcal H} \bigl[\rho_{\psi} \ham \bigr] \unitMat \, .
\eeq
\emph{En passant,} we notice that the cost function $\fdi$ in eq.~\eqref{eq:cost_Cartan_Killing}
can be rewritten in the form \eqref{eq:fs-in-inner-form} with the
replacements $\ham_0 \rightarrow \ham$, and $\rho_{\psi} \rightarrow
\rho_{\unitMat}$,  where $\rho_{\unitMat}$ is the maximally-mixed state.
The \ac{fs} cost function naturally ignores contributions from the stabilizer by its very definition, as it is a metric properly defined on the projective Hilbert space.

More generally, starting from a cost function involving penalties on the space of unitary matrices \eqref{eq:cost_Riemannian_penalties}, we have to follow the prescription outlined above equation \eqref{eq:cost_function_minimized} to obtain the associated metric on the space of quantum states. When applied to the bi-invariant metric on the space of unitaries, one obtains precisely the Fubini-Study metric. The explicit calculation for generic penalties involves minimizing over many degrees of freedom and can get quite cumbersome. In the next section, we expain how to do it systematically by recasting the metric on the space of quantum states in terms of a Riemanninan submersion.

\section{State complexity and Riemannian submersions}
\label{ssec:projection_coset}

In this section, we 
outline a systematic procedure for  
projecting the quantum complexity geometry from the space of unitary operators to the
space of states.
For the case of the unitary group SU($N$), the projection $\pi$ in
\eqref{eq:quotient_map} to the space of states was explicitly determined in
section~V of reference \cite{Auzzi:2020idm}, where it was shown that 
it is equivalent to the minimization \eqref{eq:cost_function_minimized}
    over the stabilizer of the cost function
    \eqref{eq:cost_Riemannian_penalties} (as proposed in \cite{Brown:2019whu}). The authors of \cite{Auzzi:2020idm} further showed that the map is a Riemannian submersion, see appendix~\ref{app:submersions} for a review.

Here, we review and extend these results to the case where the theory is invariant under a generic (possibly
non-compact) Lie group.
In particular, we connect the systematic procedure to the notion of
\emph{pseudo-Riemannian submersion}. 
We then make contact with the minimization method of \cite{Brown:2019whu} and comment on an a desired interpretation in terms of coadjoint orbits \cite{Chagnet:2021uvi}.

\subsection{General procedure}
\label{ssec:general_projection}
 
Consider a generic Lie group $G$ with Lie algebra $\mathfrak{g}$ spanned by generators $\omega_I$, and a reference state $\ket{\psi_R}$.  We  assume that the generators $\omega_I$ are orthonormal in the sense $\langle \omega_I ,\omega_J\rangle =\eta_{IJ}$, but the generalization is straightforward, see footnote \ref{foot:changebase}. 
We denote the maximal subgroup that leaves $\ket{\psi_R}$ invariant (the stabilizer) by $H$ (with subalgebra $\mathfrak{h}$).
We split the index labelling the generators $\omega_I \in \mathfrak{g}$ as
$I=(a,i)$, where $a \in \lbrace 1, \dots, \mathrm{dim} \, (H) \rbrace $ refers to the generators of the maximal subalgebra $\mathfrak{h} $, while $i \in \lbrace \mathrm{dim} \, (H)+1, \dots, \mathrm{dim} \, (G) \rbrace$ complements them to an orthogonal basis $\mathfrak{b}$ of $G$, labeling the coset directions.
Finally, we introduce real coordinates on the Lie manifold and we split them as $x^I = (\gamma^a, \theta^{i})$, where $\theta^{i}$ parametrize the coset space and $\gamma^a$ the other directions.
These conventions are conveniently summarized in
table~\ref{tab:separation_coset_stabilizer}. 

\begin{table}[ht]
  \begin{center}
    \begin{tabular}  {|c|c|c|} \hline
      \rule{0pt}{2.9ex}
      \textbf{Object on the full space} & \textbf{Maximal subgroup (stabilizer)}  &
      \textbf{Coset space}   \\ \hline
      \rule{0pt}{2.9ex}
      Generators $\omega_I$ & $\omega_a$ & $\omega_i$ \\
      \rule{0pt}{2.9ex}
      Coordinates $x^I$ & $\gamma^a$ & $\theta^i$  \\[0.2cm]
      \hline
    \end{tabular}
    \caption{Notation adopted to denote generators and coordinates on the coset
    space and on the maximal subgroup. }
    \label{tab:separation_coset_stabilizer}
  \end{center}
\end{table}

Consider now a unitary circuit 
\begin{equation}\label{eq:mycircuitpq}
    g(t) = p(t) q(t)\in G,
\end{equation}
where $q(t) \in H$ and $p(t)$
is a representative element in $G/H$, which brings the reference state to other states along the circuit.
In terms of the decomposition with $p$ and $q$, the Hamiltonian $\ham = i \dot g g^{-1}$ reads
\beq
\ham = i \le \dot{p} q + p \dot{q} \ri  q^{-1} p^{-1}
= i \le \dot p p^{-1} +  p  \dot q q^{-1} p^{-1} \ri \, .
\label{eq:decomposition_ham1}
\eeq
Next, we evaluate the velocities as follows
\begin{subequations}
  \beq
  Y^I = \frac{\langle \hat{H}, \omega_I \rangle
  }{\langle \omega_I, \omega_I \rangle}
  =\frac{ i  \langle \mathrm{Ad}_p \le p^{-1} \dot{p} + \dot{q} q^{-1} \ri, \omega_I
  \rangle
  }{\langle \omega_I, \omega_I \rangle}
  = \frac{i  \langle p^{-1} \dot{p} + \dot{q} q^{-1}, \mathrm{Ad}_{p^{-1}}
  (\omega_I)  \rangle 
  }{\langle \omega_I, \omega_I \rangle}\, ,
  \label{eq:velocities_submersion}
  \eeq
  \beq
  \qquad \mathrm{Ad}_{p} (\omega_I) \equiv  p \, \omega_I p^{-1}  
\equiv(\mathrm{Ad}_{p})_I^{\,\,\, M}  \omega_M\, ,\label{eq:defadj_submersion}
  \eeq
\end{subequations}
where we introduced the definition of adjoint action, and we used the fact that the operators satisfy the property $\langle X, \mathrm{Ad}_{p}(Y) \rangle = \langle \mathrm{Ad}_{p^{-1}}(X), Y\rangle$ 
under the inner product $\langle \argdot,\argdot \rangle$.\footnote{
We refer to appendix~\ref{app:inner_product_Lie} for more details about the definition and the properties of the inner product.
  The property $\langle X, \mathrm{Ad}_{p}(Y) \rangle = \langle \mathrm{Ad}_{p^{-1}}(X), Y\rangle$  is stated in eq.~\eqref{eq:self_adjoint_app}.}
This allows to re-express the cost function
\eqref{eq:cost_Riemannian_penalties}
as\footnote{We will eventually select the sign in the cases we study such that the metric obtained on the coset space is positive definite. In this way, the associated distance can be interpreted as a reasonable notion of complexity.}
\beq\label{eq:costv1}
\fpi^2 = \pm \mathcal{I}_{IJ} Y^I Y^J  =  \pm \tilde{\mathcal{I}}_{IJ} \le u^I + v^I \ri
\le u^J + v^J \ri \, ,
\eeq
where we used the following convenient redefinitions: 
\beq
\begin{aligned}
  & \tilde{\mathcal{I}}_{IJ} \equiv \mathcal{I}_{MN}
  (\mathrm{Ad}_{p^{-1}})_M^{\,\,\,\,\, I}   (\mathrm{Ad}_{p^{-1}})_N^{\,\,\,\,\, J} \, ,
  & \\
  & u^I \equiv \frac{-i \langle p^{-1} \dot{p}, \omega_{I}  \rangle
  }{\langle \omega_I, \omega_I \rangle}\, , \qquad
  v^I \equiv \frac{-i \langle \dot{q} q^{-1}, \omega_I \rangle
  }{\langle \omega_I, \omega_I \rangle}\, , &
\end{aligned}
\label{eq:convenient_defs_cost}
\eeq
and $\tilde{\mathcal{I}}_{IJ}$ is again a symmetric matrix.
The dependence of these quantities on the coordinates of the group manifold is
summarized in table~\ref{tab:dependence_cost}.
Notice that this procedure allowed us to completely separate the dependence of the cost function on the velocities associated with coordinates on the coset space $\dot \theta^i$, and those associated with directions
along the stabilizer subgroup $\dot \gamma^a$.

\begin{table}[ht]
  \begin{center}
    \begin{tabular}  {|c|c|c|c|c|c|} \hline
      \rule{0pt}{2.9ex}
      \textbf{Object/Coordinates} & $\theta^{i}$ & $\dot{\theta}^{i}$ &
      $\gamma^{a}$ & $\dot{\gamma}^{a}$  \\ \hline
      \rule{0pt}{2.9ex}
      $\tilde{\mathcal{I}}_{IJ}$ & \greencheck  & \redmark & \redmark &
      \redmark \\
      \rule{0pt}{2.9ex}
      $u^I$ & \greencheck & \greencheck & \redmark & \redmark \\
      \rule{0pt}{2.9ex}
      $v^I$ & \redmark & \redmark & \greencheck & \greencheck  \\[0.2cm]
      \hline
    \end{tabular}
    \caption{Dependence of the quantities defined in
      eq.~\eqref{eq:convenient_defs_cost} on the coordinates of the group
     manifold. }
    \label{tab:dependence_cost}
  \end{center}
\end{table}

To obtain a norm on the space of states one has to extremize the cost function with respect to the stabilizer directions. We now explain how this can be done explicitly, under certain assumptions on the structure of the algebra. 
Denote by $\mathfrak{b}$ the set of generators associated with the coset space $G/H$, \ie the orthogonal complement of $\mathfrak{h}$ in $\mathfrak{g}$ with respect to the inner product.
The key point which allows us to make progress is that $\dot{q} q^{-1}$ has non-vanishing components just along the stabilizer directions, as long as the algebra satisfies\footnote{The notation $[\mathfrak{h},  \mathfrak{b}] \subset \mathfrak{b}$ means that if we pick generators $\omega_1 \in \mathfrak{h}$ and $\omega_2 \in \mathfrak{b}$, then their Lie bracket satisfies $[\omega_1, \omega_2] \in \mathfrak{b}$.}
\beq
[\mathfrak{h}, \mathfrak{h}] \subset \mathfrak{h} \, , \qquad
[\mathfrak{h}, \mathfrak{b}] \subset \mathfrak{b} \, , \qquad
[\mathfrak{b}, \mathfrak{b}] \subset \mathfrak{h} \, .
\label{eq:commutator_reductive}
\eeq
 The structure of commutators can be brought to the form \eqref{eq:commutator_reductive} whenever the coset $G/H$ is reductive and symmetric \cite{Castellani:1991et}. This typically happens for simple Lie groups $G$ and their maximal compact subgroup $H$. The quotient spaces $\frac{\mathrm{SU}(N)}{\mathrm{SU}(N-1) \times \mathrm{U}(1)}$ and $\frac{\mathrm{SO}(p,q)}{\mathrm{SO(p) \times \mathrm{SO}(q)}}$, considered in the present work, are reductive and symmetric for any integer $(p,q)$, see table~12.1 of reference~\cite{gilmore2012lie}.
Splitting the indices of the Lie algebra as outlined at the beginning of this subsection,
this implies $v^i=0$, \ie the vector $v^I$ has vanishing components along the coset directions.

After some manipulations, we find 
\beq
\fpi^2 =
\begin{bmatrix}
  u^i & u^a + v^a
\end{bmatrix}
\begin{bmatrix}
  \tilde{\mathcal{I}}_{ij} & \tilde{\mathcal{I}}_{ib} \\
  \tilde{\mathcal{I}}_{aj} & \tilde{\mathcal{I}}_{ab}
\end{bmatrix}
\begin{bmatrix}
  u^j \\
  u^b + v^b
\end{bmatrix}
= \le \tilde{\mathcal{I}}_{ij} - \tilde{\mathcal{I}}_{ic}
(\tilde{\mathcal{I}}^{-1})^{ca} \tilde{\mathcal{I}}_{aj}  \ri u^i u^j
+ \tilde{\mathcal{I}}_{ab} f^a f^b \, ,
\label{eq:projection_cost_compact}
\eeq
where
\beq
f^a \equiv u^a + v^a + (\tilde{\mathcal{I}}^{-1})^{ad} \tilde{\mathcal{I}}_{dj}
u^j \, .
\eeq
One can check that the map
\beq
\begin{aligned}
  \pi  : \quad  & G \rightarrow G/H  \\
  & (\theta^i, \gamma^a) \underset{f_a=0}{\longmapsto} \theta^i
\end{aligned}
\label{eq:submersion_map}
\eeq
defines a smooth surjective submersion.
This condition is then sufficient to prove that $\pi$ is a (pseudo-)Riemannian
submersion that defines a unique cost function over the coset
space:\footnote{More precisely, we apply the
  theorem~\ref{thm:induced_metric_submersion2}. 
  The reader can find more details
in appendix~\ref{app:submersions}.}
\beq
(\fpi^{\mathrm{state}})^2 = \le \tilde{\mathcal{I}}_{ij} -
\tilde{\mathcal{I}}_{ic} (\tilde{\mathcal{I}}^{-1})^{ca}
\tilde{\mathcal{I}}_{aj}  \ri u^i u^j  \, .
\label{eq:state_functional_compact}
\eeq
We stress that this statement is valid for any Lie group, either compact or
non-compact.
In particular, the above procedure can be applied to project a cost function
from the conformal group $\mathrm{SO}(d,2)$ to the coset space
$\frac{\mathrm{SO}(d,2)}{\mathrm{SO}(2) \times \mathrm{SO}(d)}$ in any number
of the spacetime dimensions $d$.

\subsection{Abelian example and relation to conserved charges} 

Let us illustrate the above procedure for the specific case where the stabilizer is abelian.
We will be able to demonstrate in this case that obtaining the metric on the coset space via a Riemannian submersion is equivalent to equating certain conserved charges to zero.

Consider the unitary circuit \eqref{eq:mycircuitpq} with $q = \prod_a e^{i \gamma^a \omega_a}$. Using equation \eqref{eq:convenient_defs_cost}, we observe that the control functions $v^I$ are given by: $v^i=0$ and $v^a=\dot \gamma^a$, while the control functions $u^I$ do not depend on $\gamma^a$ nor on $\dot\gamma^a$. Therefore, the cost function \eqref{eq:costv1} does not depend on $\gamma^a$. As mentioned around 
\eqref{eq:unitary_complexity}, we can associate a Lagrangian 
\begin{equation}\label{eq:CommentLagrangianSign}
    \mathcal{L}_\mathcal{I}= 
\mathcal{F}^2_\mathcal{I}
\end{equation}
with 
 the complexity extremization problem for this cost function.
  The independence of the Hamiltonian on the coordinate $\gamma^a$ implies that there is a conserved charge associated with the cost \eqref{eq:projection_cost_compact}:
\begin{equation}\label{eq:conserved_charge}
      K_a = \frac{1}{2} \frac{\partial \mathcal{L}_\mathcal{I}}{\partial \dot \gamma^a} = \frac{1}{2} \frac{\partial \mathcal{L}_\mathcal{I}}{\partial v^a}= \tilde {\mathcal{I}}_{ab}f^a,
 \end{equation}
 where we included a factor $1/2$ for convenience. 
Therefore, setting $K_a=0$ (which is equivalent to $f^a=0$) defines the unique metric over the coset space which coincides with the one obtained from the pseudo-Riemannian submersion.

\subsection{Relation to the minimization method}
\label{ssec:relation_minimization}

For compact Lie groups, the map $f^a=0$ amounts to the minimization
\eqref{eq:cost_function_minimized}, since the last term is positive-definite
and contains all the dependence on the coordinates $\gamma^a$ along the
stabilizer \cite{Auzzi:2020idm}.
In the case of a single qubit with symmetry group $\mathrm{SU}(2),$ the metric
\eqref{eq:state_functional_compact} coincides with the result obtained in
\cite{Brown:2019whu}. 
When the cost function on the unitary space is bi-invariant, see eq.~\eqref{eq:cost_Cartan_Killing}, the penalty matrix
degenerates to $\mathcal{I}_{IJ} \rightarrow \delta_{IJ}$, and the cost function on the coset space reduces to the \ac{fs} metric on the projective space $\mathbb{CP}^{N-1}$.

In the case of a non-compact Lie group, provided that $\I$ is invertible, the quadratic bilinear form
defining the cost function \eqref{eq:projection_cost_compact} over $G$ is still non-degenerate and symmetric. However, since it is now indefinite, when brought to a diagonal form, the different contributions to eq.~\eqref{eq:projection_cost_compact} could have mixed signs.
We note that, the cost function splits into two parts: one fully defined on the coset $G/H$, and the second one which has dependence on the stabilizer $H$.
For this reason, extremizing eq.~\eqref{eq:projection_cost_compact}
over the stabilizer degrees of freedom amounts to setting $f_a=0$.
As we have just seen, the projection of the metric from the Lie group
to the coset space in terms of a Riemannian submersion is determined by setting $f^a=0$, but now this prescription coincides with an extremization rather than a minimization of the metric
\eqref{eq:projection_cost_compact}.

\subsection{Comment on the  relation with coadjoint orbits}

Reference~\cite{Chagnet:2021uvi} observed a direct connection between the \ac{fs} metric for states connected by circuits in the global conformal group in arbitrary dimensions and the K\"{a}hler metric induced by a
coadjoint action. 
In our language, the \ac{fs} metric corresponds to the projected cost function \eqref{eq:state_functional_compact} obtained via
a (pseudo)-Riemannian submersion with trivial penalties $\mathcal{I}_{IJ} = \eta_{IJ}$. It is  plausible that a similar connection can be made with the case including penalties. We comment on our attempts in this direction below.

Let us start with a few key notions. Consider the dual space $\mathfrak{g}^*$
consisting of linear
maps on $\mathfrak{g}$:
\begin{equation}
     \lambda = \langle X, \argdot \rangle \in \mathfrak{g}^* \, , \quad X,Y \in
  \mathfrak{g} \, , \quad g \in G.
\end{equation}
The coadjoint action
on the dual space is defined as
\beq
\begin{gathered}
  \mathrm{Ad}^*_g(\lambda)(Y) \equiv \langle  \mathrm{Ad}^*_g(X),Y \rangle =
  \langle X, \mathrm{Ad}^{-1}_g(Y) \rangle = \langle X, \mathrm{Ad}_{g^{-1}}(Y)
  \rangle = \lambda(\mathrm{Ad}_{g^{-1}}(Y)) \, .
\end{gathered}
\eeq
The coadjoint orbit of a  dual algebra element $\lambda
\in \mathfrak{g}^*$ is given by 
\beq
\mathcal{O}_{\lambda} = \lbrace \mathrm{Ad}^*_{g}(\lambda) \, : \, g \in G
\rbrace \, .
\eeq
The coadjoint orbit can be identified with the coset space $G/H_{\lambda}$, where $H_{\lambda}$ is the
stabilizer of the element $\lambda$. 
One can then define a pre-symplectic form
$\mathcal{A}_\lambda = \lambda(\Theta)$ and a the Kirillov-Kostant symplectic form $\omega_\lambda \equiv \lambda(d\Theta) $ where $\Theta \equiv g^{-1} \dot g$. 
The advantage of this formulation is that the coadjoint orbit is naturally
associated with a symplectic manifold (on which an action can be defined) and a K\"{a}hler metric on the coset space \cite{ALEKSEEV1989719,Alekseev:2018pbv}.
Indeed, this metric precisely coincides with  the \ac{fs} metric for the case of
the global conformal group in arbitrary dimension. For more details, see sections 4 and 5 of \cite{Chagnet:2021uvi}.
To obtain these matches,  the element $\lambda$ in the above definitions is taken to be
\begin{equation}
    \lambda(\mathcal{O}) = i \Tr [|\psi_R \rangle \langle \psi_R| R(\mathcal{O})] \, ,
\end{equation}
where the trace is taken in the infinite-dimensional
Hilbert space and $R(\mathcal{O})$ denotes the representation of the operator $\mathcal{O}$ on the Hilbert space. We drop the $R$ below to simplify the notation.  
The coadjoint action of the circuit unitary $g(t)$ on $\lambda$ reads:
\begin{equation}
\mathrm{Ad}^*_{g(t)}(\lambda) (\mathcal{O}) = 
    i \Tr [|\psi_R \rangle \langle \psi_R| g^{-1}\mathcal{O}g] = i \langle \psi_R| g^{-1}\mathcal{O}g|\psi_R \rangle \, .
\end{equation}
Now, consider the time derivative of this coadjoint action:
\begin{equation}
\begin{split}
     \frac{d}{dt}[\mathrm{Ad}^*_{g(t)}(\lambda) (\mathcal{O})] &= -i \langle \psi_R| g^{-1}\dot g g^{-1} \mathcal{O}g|\psi_R \rangle + i \langle \psi_R| g^{-1} \mathcal{O}\dot{g}|\psi_R \rangle
     \\
     &= -\langle \psi_R| g^{-1} [\hat H, \mathcal{O}]g|\psi_R \rangle  \, .
\end{split}
\end{equation}
Selecting $\mathcal{O} =\omega_I$, we can now express this as
\begin{equation}
   \frac{d}{dt}[\mathrm{Ad}^*_{g(t)}(\lambda) (\mathcal{O})]  =
   Y^{J} f_{IJ}^{K}\langle \psi_R| g^{-1} \mathcal{\omega}_K g|\psi_R \rangle 
   =
Y^{J} f_{IJ}{}^{K}(\mathrm{Ad}_{p^{-1}})_{K}{}^M \langle \psi_R|\omega_M|\psi_R \rangle
\end{equation}
where $f_{IJ}{}^K$ are the group's structure constants, and the third equality is obtained by using the definitions \eqref{eq:mycircuitpq}, \eqref{eq:defadj_submersion}. In the cases we focus on in this paper, the only non-trivial contribution to the expectation value comes from the stabilizer generators. 
Depending on the group structure, this equation could potentially be inverted to extract the $Y^I$-s in terms of the co-adjoint action, and the procedure of section \ref{ssec:general_projection} can then be re-interpreted in these terms. It is not immediately obvious that the metric obtained on the space of states in this way has a simple interpretation in terms of the symplectic manifold associated with the coadjoint orbit of $\lambda$. This is an interesting question which we leave for future work.

\section{One-dimensional CFT}
\label{sec:1dCFT}

Our next goal is to apply the machinery developed in
section~\ref{ssec:projection_coset} to one-dimensional \acp{cft}.
For simplicity, we take our quantum circuits to lie along the global part of the one-dimensional conformal group, thus taking the generators to be translation, special conformal transformation and dilatation. 
This construction is performed in section~\ref{ssec:CFT1_projection_coset}, where we compute the metric on the conformal group with arbitrary penalty factors, and project it over the coset space of a primary reference state.
This procedure generalizes the analysis performed in \cite{Chagnet:2021uvi}, where the cost function was isotropic.
In the next two subsections, we compute the state complexity of arbitrary states obtained by acting with the above circuits on the primary reference state. 
We start with the simple example where the dilatation generator $D$ is free of cost in 
section~\ref{ssec:interlude:simple_metric}. 
This is motivated by the fact that $D$ plays the role of the system's Hamiltonian for CFTs in radial quantization, and therefore has a privileged role in terms of the  natural evolution of the system. 
We then come back to the general cost function on the coset manifold and
compute the state complexity for arbitrary states in section~\ref{ssec:state_CFT1}. The results of this section also prepare the ground for our studies of complexity in two-dimensional CFTs in the next section.

\subsection{Complexity geometry for CFT$_1$}
\label{ssec:CFT1_projection_coset}

We focus on one dimensional ($d=1$) CFTs and on unitary circuits in a representation of the global conformal group $G=\mathrm{SO(2,1)}$ in Lorentzian signature.\footnote{While one-dimensional \acp{cft} (also called conformal quantum mechanics, CQM) do not admit a
local energy-momentum tensor, one can still consider a theory invariant under $\mathrm{SO}(1,2) \simeq \mathrm{SL}(2,\mathbb{R})$ with fields satisfying the
same structure of correlation functions that a standard \ac{cft} would have. CQM was originally formulated in \cite{deAlfaro:1976vlx}.
A discussion on the problems to introduce energy excitations in CQM is given in \cite{Maldacena:1998uz,Jensen:2011su,Jensen:2016pah}.  
The study of one-dimensional CFTs sets the stage for the study of CFT$_2$ complexity later on. This is because CFT$_2$ can be seen as a direct product of two copies of a CFT$_1$.}
The global generators of the conformal group are those that induce conformal transformations that are globally well-defined on the Riemann sphere. 
Such transformations can be spanned by the generators $\lbrace D, P, K \rbrace$ consisting of the dilatation, momentum and special conformal transformation generators, respectively.
Alternatively, one can use the Hermitian basis $\lbrace L_0, L_{\pm} \rbrace$ defined by
\beq
L_0 = D~, \qquad  L_+ = \frac{1}{2} \le P+K \ri~, \qquad L_- = \frac{i}{2} (P - K)~,
\label{eq:hermitian_SL2R}
\eeq
spanning the SL$(2,\mathbb{R})$ group, which is isomorphic to  $\mathrm{SO}(1,2)$. 
The generators satisfy the following conjugation rules:
\beq
D^{\dagger} = D \, , \qquad
P^{\dagger} = K \, , \qquad
L_{\pm}^{\dagger} = L_{\pm} \, .
\label{eq:conjugation_so12gen}
\eeq
Since the manipulations in this paper are mostly algebraic, we can, in practice, perform our calculations using matrices in the fundamental representation. Conventions for the algebra and an explicit representation are collected in
appendix~\ref{app:ssec:algebra_so12}, see, \eg equation \eqref{eq:generators_so12} for the explicit form of the matrices.\footnote{In this section we will not use an explicit notation $R(O)$ for the fundamental representation of operators, but it will be understood that we are always using it in our explicit calculations.} However, note that some care has to be taken when computing the conjugate generators in the fundamental representation, as this is not a simple complex conjugation operation on the matrices, see eq.~\eqref{eq:HermConj}.
Here, we point out that we are building a representation of the Lorentzian conformal group by means of the \emph{Euclidean} generators $\{D,P,K\}$, as discussed in reference~\cite{Minwalla:1997ka} and Appendix~A of \cite{Chagnet:2021uvi}. 
This choice is crucial because it allows us to build unitary representations of the conformal algebra.\footnote{This is analogous to the use of ladder operators $J_{\pm}$ to build representations of angular momentum in quantum mechanics.}  
In the following, we will also need the inner product of the different algebra generators, see appendix \ref{app:inner_product_Lie}: 
\beq
\inner{P}{K} = \inner{K}{P} = -2 \, , \qquad
\inner{D}{D} = \inner{L_0}{L_0}= 1 \, , \qquad
\inner{L_\pm}{L_\pm} = -1 \, .
\label{eq:normalization_generators_so12Main}
\eeq

Exponentiating the Euclidean generators $\{D,P,K\}$, one can build a unitary circuit as follows 
\beq
\label{eq:circuits-ansatz}
U(t) = e^{i \lambda(t) P} e^{\gamma_I(t) D} e^{i \rho(t) K} e^{i \gamma(t)
D}~,
\eeq
where $\lambda, \rho \in \mathbb{C}$ and $\gamma_I, \gamma \in \mathbb{R}$ are 
parameters that change along the circuit.
Imposing the unitarity condition $U^\dagger = U^{-1}$, we find the constraints
$\rho=\lambda^*$ and $\gamma_I = \log\left(1-|\lambda|^2 \right)$, together
with the inequality $|\lambda|^2 <1$. 

A natural reference state $\ket{\psi_R}$ is provided by a primary state $\ket{\Delta} = \mathcal O_\Delta \ket{0}$, where $\mathcal O_\Delta$ is a scalar operator with conformal dimension $\Delta$ acting on the vacuum
\cite{Caputa:2018kdj,Chagnet:2021uvi}. A primary state satisfies $K|\Delta\rangle=0$ and $D|\Delta\rangle=\Delta |\Delta\rangle$.
The stabilizer subgroup associated with a scalar primary state is given by $H = \mathrm{SO}(2)$, generated by the dilatation operator $D$, while the coset directions are associated with the generators $P$ and $K$. 
Applying the operator \eqref{eq:circuits-ansatz} to this reference state generates a generalized coherent state \cite{perelomov1977generalized}
\beq
\ket{\lambda (t)} = (1-|\lambda(t)|^2)^{\Delta} \, e^{i \gamma(t) \Delta} e^{i
\lambda(t) P} \ket{\Delta} \, ,
\label{eq:state_CFT1}
\eeq
where $\gamma(t)$ is an overall phase coming from the action of the stabilizer subgroup on the reference state.
From now on, we will refer to the direction parametrized by $D$ (equivalently,
$L_0$) as the stabilizer, and the directions along the generators $P, K$
(equivalently, $L_{\pm}$) as the coset ones.
The coherent states \eqref{eq:state_CFT1} are labeled by points on the unit
disk
\begin{equation}\label{eq:statedisk}
    \mathcal D = \lbrace \lambda \in \mathds{C} \, : \, |\lambda|^2 < 1 \rbrace .
\end{equation}
Sometimes, it will be convenient to parametrize the unit disk in angular coordinates
\begin{equation}\label{eq:angulardecomp}
    \lambda = r e^{i \theta},
\end{equation}
with $r<1$ the radial coordinate, and $\theta$ an angle.  

In the following, our goal is to compute the metric over the coset space $\mathrm{SO}(1,2)/\mathrm{SO}(2)$. We begin with the case of homogeneous penalty factors (\ie all penalties equal to
one for the Hermitian generators) to connect with the previous literature, and then we move to the case of general penalties.

\subsubsection{Homogeneous penalty factors}

There is a direct route to find the metric on the space of states when the
penalty factors are homogeneous: one can apply the state-dependent \ac{fs} cost
functional \eqref{eq:FS_cost_functional} to the state \eqref{eq:state_CFT1},
and then compute the infinitesimal line element using $\dd s_{\fs} =
\mathcal{F}_{\fs} \, \dd t$. This gives
\cite{Chapman:2017rqy,Chagnet:2021uvi}
\beq
\label{eq:fs-metric-1d}
\dd s^2_\fs = 4 \, \dfrac{\dd \lambda \dd \lambda^\ast}{(1 - |\lambda|^2)^2}~.
\eeq
This metric describes the Poincar\'{e} unit disk $\mathcal P \cong \mathrm{SO}(1,2)/\mathrm{SO}(2)$.\footnote{The symbol $\cong$ means that there exists an isomorphism between the Poincar\'{e} disk and this quotient space of orthogonal groups.}
The geodesics of this metric are found by extremizing the Lagrangian $\mathcal{L} =
\ffs^2$, and solving the Euler-Lagrange equations with boundary conditions:
\begin{equation}
    \lambda(0) = 0, \qquad \lambda(1) =\lambda_\tar= r_\tar e^{i \theta_\tar},
\end{equation}
where $(r_\tar,\theta_\tar)$ are the polar parameters that identify an arbitrary target state $\lambda_\tar$.\footnote{In
  practice, one would like to find stationary solutions associated with $\mathcal{L} =
  \mathcal F$ for a given cost function. However it is generally easier to do
  this for the squared Lagrangian $ \mathcal{F}^2$, which then yields affinely
parametrized solutions.}
This procedure yields solutions of the form
\beq
\label{eq:fsSol}
\lambda_0(t) = e^{i \theta_\tar} \tanh \left[ \arctanh(r_\tar) \, t \right]~.
\eeq
We used the subscript $``0"$ to indicate that this is the optimal trajectory. Let us denote the 
reference state as $\ket{\lambda (0)}= \ket{\Delta}$ and the target state as $\ket{\lambda (1)}= \ket{\lambda_{\rm T }}$.
The complexity associated with this target state can be evaluated by integrating the \ac{fs} cost function along the optimal trajectory \eqref{eq:fsSol} as follows:
\begin{align}
  \label{eq:complexity-fs}
  \mathcal C^{\mathrm{state}}_{\fs} [\ket{\lambda_{\rm T}}, \ket{\Delta}] = \int_0^1 \ffs
  \, \dd t = 4 \int_0^1 \sqrt{\dfrac{|\lambda_0^\prime(t)|^2}{(1 -
  |\lambda_0(t)|^2)^2}} \, \dd t = 2 \, \arctanh(r_\tar)~.
\end{align}
This result corresponds to integrating the proper length of the \ac{fs} metric along
a geodesic with fixed endpoints.
Since the solutions \eqref{eq:fsSol} are affinely parametrized, the on-shell
value of $\ffs = 2 \, \arctanh(r_\tar)$ is independent of the circuit time
parameter, we simply obtain $\mathcal C^{\mathrm{state}}_{\fs}
[\ket{\lambda_T}, \ket{\Delta}] = \ffs \vert_{\text{on-shell}}$. Note that the result for the complexity diverges near the boundary of the disk \eqref{eq:statedisk}, as $r_T\rightarrow1 $.

As discussed in section~\ref{ssec:general_projection}, it is possible to recover the previous result by performing a projection over the coset space of the metric
associated with the bi-invariant cost function $\fdi$ on the group manifold.
This is done as follows. First, plug the ansatz \eqref{eq:circuits-ansatz} for the unitary into the cost
function \eqref{eq:cost_Cartan_Killing}. The corresponding infinitesimal line element reads
\begin{align}
  \label{eq:susskindBeforeMin}
  \dd s^2 = -\trH \dd t^2  = \dd s_{\rm FS}^2 - \underbrace{\Bigl[ \dd \gamma - i    \dfrac{\lambda^\ast \dd \lambda-\lambda \dd \lambda^\ast}{1-|\lambda^2|}
  \Bigr]^2 }_{K_D^2 \dd t^2} ~,
\end{align}
where $t$ is the circuit parameter, and the overall minus sign in the middle equality was selected to obtain a positive-definite metric on the space of states, see the explanation below eq.~\eqref{eq:cost_Cartan_Killing}. In the above equation, we identified the conserved quantity $K_D$, associated with the direction of the stabilizer
group $\mathrm{SO}(2)$.
In other words, denoting the Lagrangian with $ \mathcal{L}_{\text{bi-inv}} = -\trH$, we defined $K_D = \frac{1}{2} \frac{\partial \mathcal{L}_{\text{bi-inv}}}{\partial \dot{\gamma}}$, cf. eq.~\eqref{eq:conserved_charge}. 
The line element can then be projected on the coset space by setting $K_D=0$, or equivalently by minimizing with respect to $d\gamma$,
recovering the \ac{fs} metric \eqref{eq:fs-metric-1d}.

\subsubsection{General penalties - Hermitian basis}
Let us now add non-trivial penalty factors, following the systematic procedure outlined in section~\ref{ssec:projection_coset}, starting with the cost function \eqref{eq:cost_Riemannian_penalties}. We
consider the Hermitian basis of orthogonal generators $\omega_I= \lbrace L_0, L_+, L_-
\rbrace$, and focus on Hamiltonians of the form $\ham = h_0 L_0 + h_+ L_+ + h_- L_-$. This defines a Lagrangian 
\beq
\mathcal{L}_{\mathcal{I}} = \mathcal{F}_{\mathcal{I}}^2= - \langle \ham, \omega_I \rangle \mathcal{I}_{IJ} \langle
\ham, \omega_I \rangle = - \mathcal{I}_0 h_0^2 +  \mathcal{I}_- h_-^2 +  \mathcal{I}_+ h_+^2 \, ,
\label{eq:cost_CFT1_so12}
\eeq
where we used the penalty matrix 
$ \mathcal{I}_{IJ} = \mathrm{diag} \, (\mathcal{I}_0, -\mathcal{I}_+, -\mathcal{I}_-) $, and the relative signs inside $L_{\mathcal{I}}$ arise from the normalization of the generators
reported in eq.~\eqref{eq:normalization_generators_so12Main}.
Notice that the relative signs are consistent with the signature expected for the metric on this non-compact Lie group manifold, as discussed above eq.~\eqref{eq:cost_Riemannian_penalties}.  
The case with homogeneous penalty factors in eq.~\eqref{eq:susskindBeforeMin}
is trivially recovered by setting $\mathcal{I}_0=\mathcal{I}_{\pm}=1$. The coefficients $h_0,h_{\pm}$, describing the linear decomposition of the
Hamiltonian $\ham$ on the Hermitian basis of generators,  read\footnote{The identities \eqref{eq:hamiltonian-basis} can be obtained by using $h_I = \frac{\langle\hat H,\omega_I \rangle}{\langle\omega_I,\omega_I \rangle}$, which hold because the generators are orthogonal in this basis.}
\begin{align}
  \label{eq:hamiltonian-basis}
  \begin{gathered}
    \quad h_+ = - \dfrac{\dot \lambda + \dot \lambda^* - i (\lambda
    - \lambda^*) \dot \gamma}{1- |\lambda|^2}~, \quad h_- =
    \dfrac{i(\dot \lambda - \dot \lambda^*) + (\lambda + \lambda^*) \dot
    \gamma}{1- |\lambda|^2}~,\\
    h_0 = -\dfrac{\dot \gamma (1 + |\lambda|^2) + i(\lambda^* \dot \lambda -
    \lambda \dot \lambda^*) }{1-|\lambda|^2}~.
  \end{gathered}
\end{align}
We use these formulas later on in subsection \ref{subsec:GeneratorsTrajectory} to develop some intuition as to which generators are active when moving in different directions in the coset space.

Substituting the relations \eqref{eq:hamiltonian-basis} into the Lagrangian  \eqref{eq:cost_CFT1_so12} and using the angular decomposition \eqref{eq:angulardecomp} in the space of states, we obtain
\begin{equation}\label{eq:metricpmbasisfinal}
    \dd s^2  = -\frac{\I_0 \bigl(1+r^2\bigr)^2 - 2 r^2
      \alpha(\theta)}{\bigl(1-r^2\bigr)^2} \dd A^{\prime 2} +  \frac{ 2 \I_0 \alpha(\theta)
      r^2}{\I_0 \bigl(1+r^2\bigr)^2 - 2 r^2 \alpha(\theta)} \dd B^2 + \frac{8
        \I_-
      \I_+}{\alpha(\theta) \bigl(1-r^2\bigr)^2} \dd r^2~,
\end{equation}
with
  \begin{subequations}
    \begin{align}
 \dd A^\prime & \equiv \dd \gamma - 2 r^2 \frac{\I_0 (
      1+r^2) - \alpha(\theta)}{\I_0 (1+r^2)^2 - 2 \alpha(\theta) r^2}\dd \theta + 2 r
      \frac{(\I_- -
      \I_+) \sin(2 \theta)}{\I_0 (1+r^2)^2 - 2 \alpha(\theta) r^2} \dd r~,\\
      \dd B & \equiv \dd \theta + \frac{1+r^2}{1-r^2}\frac{(\I_- - \I_+) \sin(2
      \theta)}{r \alpha(\theta)} \dd r~,
      \\
      \alpha(\theta) &\equiv \I_+ + \I_- + (\I_- - \I_+) \cos(2 \theta).\label{eq:alphatheta}
    \end{align}
  \end{subequations}
Note that the function  $\alpha(\theta) \geq 0$ as long as the penalty factors have the expected signature, i.e., $\I_\pm \geq 0$.  
In this form, it is easy to see that the $\dd A'$ and $ \dd B$ directions generally  contribute with opposite signs. 

Computing the conserved charge $K_0$ associated with the cyclic coordinate $\gamma$, see eq.~\eqref{eq:conserved_charge}, yields 
\begin{equation}\label{eq:phase-coordinate-pf2}
    K_0  = -\frac{\I_0 \bigl(1+r^2\bigr)^2 - 2 r^2
      \alpha(\theta)}{\bigl(1-r^2\bigr)^2} \dot A' \, ,
\end{equation}
and setting it to zero via $ \dd A'=0$ (alternatively solving for the cyclic coordinate $d\gamma$ in terms of the other variations $d\theta$, $dr$), results in the following metric on the space of states
\begin{equation}\label{eq:metricpmbasisfinal2}
    \dd s^2  =   \frac{ 2 \I_0 \alpha(\theta)
      r^2}{\I_0 \bigl(1+r^2\bigr)^2 - 2 r^2 \alpha(\theta)} \dd B^2 + \frac{8
        \I_-
      \I_+}{\alpha(\theta) \bigl(1-r^2\bigr)^2} \dd r^2~,
\end{equation}
where $\alpha(\theta)$ was defined in eq.~\eqref{eq:alphatheta}.

One may wonder which requirements make the cost function~\eqref{eq:metricpmbasisfinal2} well-suited to characterize the complexity geometry of physical systems.
One approach, that we will adopt in the bulk of this work (including section~\ref{sec:2dCFT} for the case of two-dimensional CFTs), is to impose that the resulting metric is positive definite across all the space of states.
In the present case, this constraint can be satisfied everywhere on the unit disk as long as
\begin{equation}\label{eq:pfIpm0condition}
    \mathcal{I}_0> \max(\mathcal{I}_+,\mathcal{I}_-).
\end{equation} 
Another possibility is that the penalty factors are kept arbitrary, but the space of states admits a non-trivial boundary delimiting the region where the metric is positive-definite.
In other words, not all the states are  accessible, and there is a physical obstruction to overcoming the boundary signaled by a singularity where the metric changes signature.
We restrict our analysis below to choices of penalties satisfying \eqref{eq:pfIpm0condition} and leave the discussion of this other possibility to  appendix~\ref{app:sec:cost_nontrivial_bdy}.

Let us add a final comment. Substituting the condition $ \dd A'=0$ into the equations for the Hamiltonian coefficients \eqref{eq:hamiltonian-basis}, we can fix the motion in the
stabilizer direction and  express the Hamiltonian just in terms of
$(\lambda, \lambda^*)$ and their derivatives. The result \eqref{eq:hamiltonian-basis} allows to infer some features of the geodesics over the coset space.
If we take a path such that $\lambda \in \mathbb R$, we find that $h_0 = h_- = 0$, implying that the velocity vector is entirely oriented along the $L_+$ generator.
On the other hand, if we take a purely imaginary path $\lambda \in i \mathbb{R}$, then $h_0 = h_+ = 0$, and in this case the velocity is entirely oriented along the $L_-$ generator. We extend this interpretation to more general trajectories in subsection \ref{subsec:GeneratorsTrajectory}.

\subsubsection{General penalties - physical basis}
It is also instructive to show how the metric
looks like in the basis $\lbrace D,P, K \rbrace$. This is a natural basis of generators implementing  conformal transformations of the one-dimensional line, \ie  dilatation, translation and special conformal transformation. 
We will obtain the metric in this physical basis by perfoming the transformation \eqref{eq:hermitian_SL2R}, together with the following map
\begin{subequations}
  \label{eq:map-I-J}
  \begin{align}
    & \J_1 = \I_0~, \qquad \J_2 = \frac{\I_- - \I_+}{4}~, \qquad \J_3 =
    \frac{\I_- + \I_+}{2}~, \\
    \text{ or equivalently, }
    \qquad &  \I_0 = \J_1~, \qquad \I_- = \J_3 + 2 \J_2~, \qquad \I_+ = \J_3 -
    2 \J_2~m.
  \end{align}
\end{subequations}
Note that the conditions $\mathcal{I}_{\pm},\mathcal{I}_0\geq0$ imply 
\begin{equation}\label{eq:firstconstraintJ}
\mathcal{J}_3>|2\mathcal{J}_2|.
\end{equation}
We can now re-write the cost function \eqref{eq:cost_CFT1_so12} over the
$\mathrm{SO}(1,2)$ group manifold as
\begin{align}
  \label{eq:susskind-penalty-factors}
  \mathcal{L}_{\mathcal{J}} = - \inner{\ham}{\tilde\omega_I}  \J_{IJ} \inner{\ham}{\tilde \omega_J}
  \quad \text{ with } \quad \J =
  \begin{bmatrix}
    \J_1 & 0 & 0\\
    0 & \J_2 & -\J_3/2\\
    0 & -\J_3/2 & \J_2
  \end{bmatrix}~,
\end{align}
where now $\tilde \omega_I = \lbrace D, P, K \rbrace$.
Since this is simply a change of variables, we have $\mathcal{L}_{\mathcal{J}}=\mathcal{L}_{\mathcal{I}}$. 
Moreover, we observe that a diagonal penalty matrix in the basis $\lbrace L_0, L_+, L_- \rbrace$ implies that the stabilizer and the coset part of the
geometry are factorized, which can be seen in the rotated basis $\tilde \omega_I$ too, since the coefficients along the $(K,D)$ and $(P,D)$
directions vanish.
The homogeneous cost function~\eqref{eq:susskindBeforeMin} is recovered by
setting 
\begin{equation}\label{eq:FubiniJtrivial}
    \{\J_1, \J_2, \J_3\}=\{\J_1^\fs, \J_2^\fs, \J_3^\fs\} \equiv \{1,0,1\}.
\end{equation}
In the above expression, we observe that a non-trivial $\J_2$ turns on a new direction in the tangent space, which
was not present in the case of a homogeneous cost function.
Therefore, we expect that it drastically changes the optimal trajectories.
In terms of the Hermitian set of generators, a non-trivial penalty
$\mathcal{J}_2$ describes an anisotropy in the $(L_+, L_-)$ plane.

We are now ready to find the metric on the coset space.
First, we compute the cost function \eqref{eq:susskind-penalty-factors} over
the $\mathrm{SO}(1,2)$ Lie group using the Hamiltonian associated with the
unitary operator \eqref{eq:circuits-ansatz}. This yields
\begin{align}
  \label{eq:pf-0-metric}
  \mathcal{L}_{\mathcal{J}} = \mathcal{L}_{\pf} - \frac{\left(1- |\lambda|^2\right)^2}{B} (\hat{K}_D)^2 \, , \qquad
   \hat{K}_D =
   -\frac{B}{\left(1- |\lambda|^2\right)^2} \left[\dot \gamma +
  \frac{i}{B} \left( D \dot \lambda - D^* \dot \lambda^*\right)\right]~
\end{align}
where we have defined the Lagrangian
\begin{equation}
    \mathcal{L}_{\pf} =  4 \, \dfrac{A \dot \lambda^2+ A^\ast \dot \lambda^{\ast 2} + C \dot
  \lambda \dot \lambda^\ast}{\left( 1 - |\lambda|^2 \right)^2 B} ~,
\end{equation}
and the functions $A,B,C,D$ as follows:
\begin{subequations}\label{eq:defB_so12}
  \begin{align}
    A(\lambda) & = \bigl( 4 \J_2^2 + (\J_1 - \J_3) \J_3 \bigr) (\lls)^2 - \J_1
    \J_2 \bigl(1 + (\lls)^4\bigr)~,\\
    B(\lambda) & =  \J_1 \left( 1 + |\lambda|^2\right)^2 - 4 \Bigl[\J_3
    |\lambda|^2 + \J_2 \left( \lambda^2 + \lambda^{\ast 2}\right) \Bigr]~,\\
    C(\lambda) & = 2 \left( 4 \J_2^2 - \J_3^2 \right) |\lambda|^2 + \J_1 \J_3
    \left[ 1 + |\lambda|^4 - 2 \left( \lambda^2 + \lambda^{\ast 2}\right)
    \J_2/\J_3 \right]~,\\
    D(\lambda)& = \lambda^* (\J_1 -2 \J_3 + \J_1 |\lambda|^2) - 4 \J_2
\lambda~.
  \end{align}
\end{subequations}
The quantity $\hat{K}_D$ in eq.~\eqref{eq:pf-0-metric} is simply the conserved momentum conjugate to the cyclic
coordinate $\gamma$, which identifies the direction along the generator $D$ in
the stabilizer group. 
The projection to the coset space is done by  setting $\hat{K}_D=0$, which amounts to substituting
\begin{align}
  \label{eq:phase-coordinate-pf}
  \dot \gamma =  - \frac{i}{B}\left( D \dot \lambda - D^* \dot \lambda^*\right)
  \, ,
\end{align}
in eq.~\eqref{eq:pf-0-metric}. This finally gives the cost function over the quotient space as 
\begin{align}
  \label{eq:metricPF}
  \mathcal{L}_{\pf} = \fpf^2 =  4 \, \dfrac{A \dot \lambda^2+ A^\ast \dot \lambda^{\ast 2} + C \dot
  \lambda \dot \lambda^\ast}{\left( 1 - |\lambda|^2 \right)^2 B} ~.
\end{align}
The metric \eqref{eq:metricPF} admits two reflection symmetries $\mathrm{Re}
\lambda \to - \mathrm{Re} \lambda$ and $\mathrm{Im} \lambda \to - \mathrm{Im}
\lambda$ (alternatively, they can be written as $\lambda \to - \lambda^*$ and
$\lambda \to \lambda^*$, respectively). 
This means that if we measure the complexity of every point in the upper-right quadrant of the
Poincar\'{e} disk, then without loss of generality, we can use the above-mentioned symmetry to infer what happens in the other quadrants.\footnote{This does not mean that trajectories are constrained to a specific quadrant.} 

The metric above will be positive-definite as long as we impose \eqref{eq:pfIpm0condition}, which in terms of the penalties $\mathcal{J}_1,\mathcal{J}_2,\mathcal{J}_3$ reads
\begin{equation}\label{eq:penaltycondJ}
   \mathcal{J}_1> \max(\mathcal{J}_3\pm 2\mathcal{J}_2).
\end{equation}

\subsubsection{Hamiltonian generators along coset space trajectories}\label{subsec:GeneratorsTrajectory}

\begin{figure}[ht]
  \centering
  \includegraphics[width=0.9\textwidth]{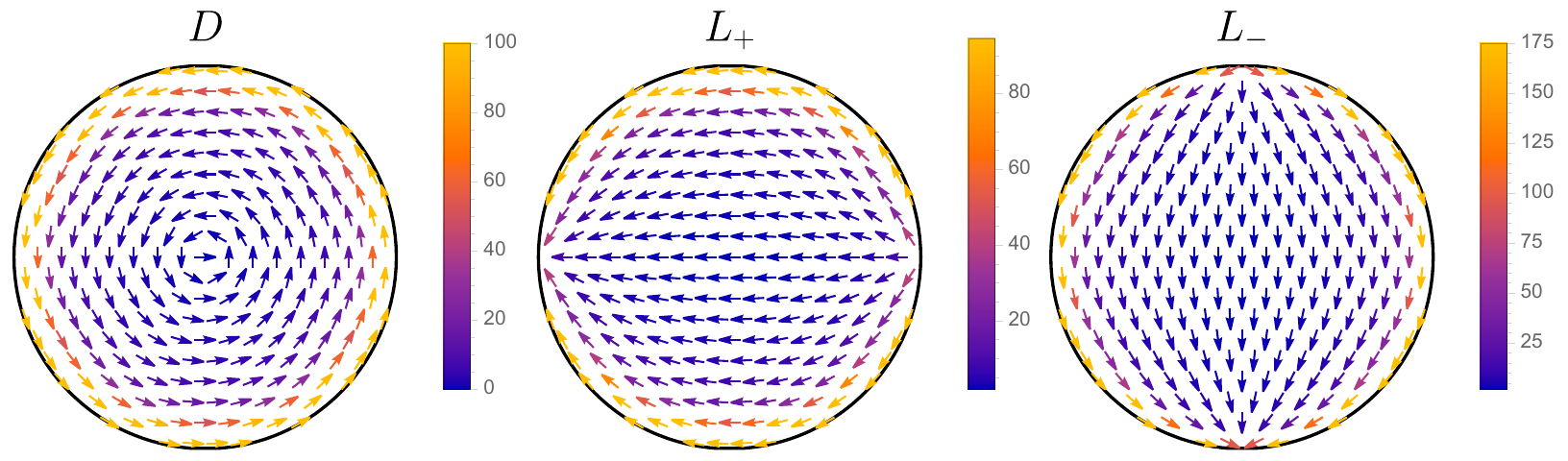} 
  \caption{Hamiltonian generator activation when moving along different coset space trajectories, when taking into account the submersion map constraint \eqref{eq:submersion_map} with penalties $\lbrace \mathcal{J}_1, \mathcal{J}_2, \mathcal{J}_3 \rbrace =  \lbrace 1,0,1 \rbrace$. Starting from a representatitve state \eqref{eq:state_CFT1} in the coset space and fixing a shift in the state $\lambda \rightarrow \lambda+d\lambda$, we can see which generators are active by taking the projection of the desired direction of motion and the vector field in the figure. 
The action of $h_+$ ($h_-$) near the origin is akin to translating the state on the real
    (imaginary) axis, while the action of  $h_0$ rotates the phase of the state
    at
  fixed radius. The colors indicate the magnitude of the vector fields according to the heat chart on the right of each figure.}
  \label{fig:direction-effect-lambda}
\end{figure}

\begin{figure}[ht]
  \centering
  \includegraphics[width=0.9\textwidth]{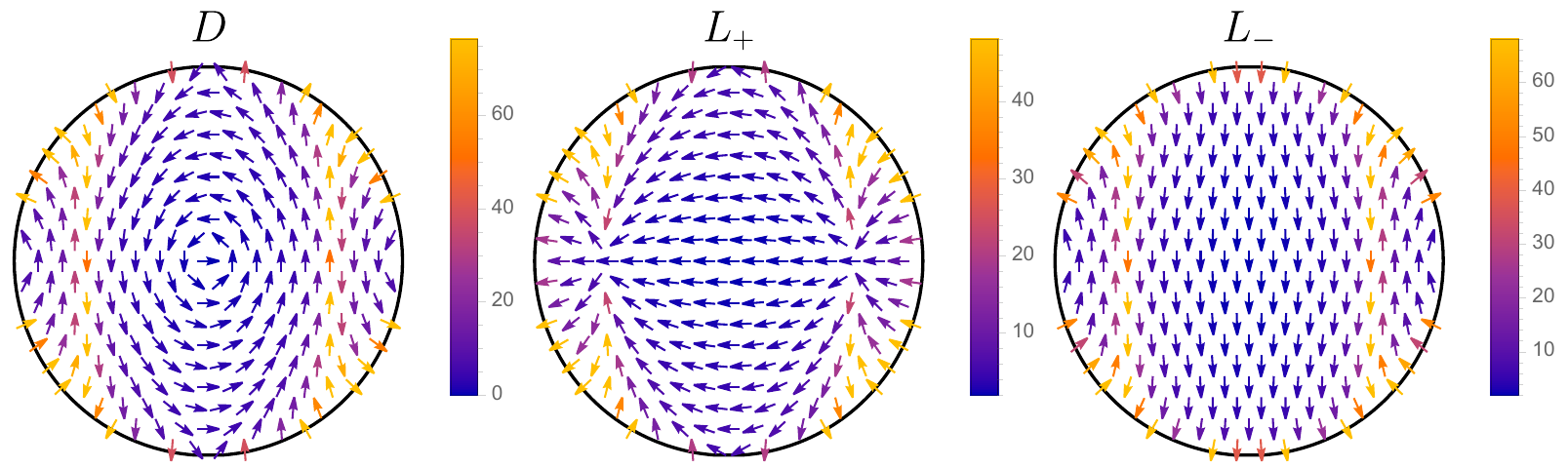}
  \caption{Hamiltonian generator activation when moving along different coset space trajectories, when taking into account the submersion map constraint \eqref{eq:submersion_map} with penalties $\lbrace \mathcal{J}_1, \mathcal{J}_2, \mathcal{J}_3 \rbrace =  \lbrace 1,0.1,1 \rbrace$. Starting from a representative state \eqref{eq:state_CFT1} in the coset space and fixing a shift in the state $\lambda \rightarrow \lambda+d\lambda$, we can see which generators are active by taking the projection of the desired direction of motion and the vector field in the figure. 
    The action of $h_+$ ($h_-$) near the origin is akin to translating the state on the real
    (imaginary) axis, while the action of  $h_0$ rotates the phase of the state. The colors indicate the magnitude of the vector fields according to the heat chart on the right of each figure.}
  \label{fig:direction-effect-lambdaN}
\end{figure}

We can now go back and develop some intuition as to which generators are active when moving along different directions in the coset space. We start by substituting our solution for the cyclic coordinate $\dot \gamma$ from equation \eqref{eq:phase-coordinate-pf} into the different Hamiltonian coefficients $h_0$, $h_{\pm}$, in equation \eqref{eq:hamiltonian-basis}, specifying the relative magnitudes of the $L_0$, $L_+$ and $L_-$ generators along a given trajectory. The resulting expressions will of course depend on the penalties and will take the following schematic form
\begin{equation}
    h_I = \vec{\mathcal{V}}_I(\vec{\lambda}, \mathcal{J}_i) \cdot \dot{\vec{\lambda}},
\end{equation}
where for convenience we defined a vector with the real and imaginary components  $\vec{\lambda} = (\text{Re}\,\lambda,\text{Im}\, \lambda)$. 
In figures \ref{fig:direction-effect-lambda} and \ref{fig:direction-effect-lambdaN}, we plot the vectors $\vec{\mathcal{V}}_I$ at different points along the the coset space for trivial and non-trivial penalties, respectively. The coset space is represented by a disk in a two-dimensional space, with the real (imaginary) part of the state parameter $\lambda$ referring to the horizontal (vertical) axis.
The figures encode the level of activation of the generators along the various directions of motion in the space of states. Note that when imposing the submersion map constraint \eqref{eq:submersion_map} on the motion in the stabilizer direction, only a two-dimensional section of the possible unitary motions can be realized. 
We can read the plots as follows. Fix a starting point on the space of states and a vector that induces a motion towards a nearby state. Project this vector on the vector fields in the different figures to read the coefficients of the different generators $\lbrace L_0, L_+ L_- \rbrace$ active in the Hamiltonian which induces motion in the desired direction.

From the figure, we observe that the $L_0$ generator near the origin induces angular rotations in the polar parametrization \eqref{eq:angulardecomp}. Similarly, as long as we are not too close to the boundary of the disk representing the space of states \eqref{eq:statedisk}, the $L_+$ generator mostly modifies the real part of $\lambda$, while the $L_-$ generator mostly affects its imaginary part. We therefore expect that when increasing the penalty $\mathcal{I}_-$, the minimal length solutions will disfavor the direction generated by $L_-$, and will prefer  motions along the real axis as much as possible.
Contrarily, increasing the $\mathcal{I}_+$ penalty will give rise to trajectories with preference to move along the imaginary axis.

\subsection{State complexity with free dilatations}
\label{ssec:interlude:simple_metric}

Before diving into the study of the geodesics of the metric
\eqref{eq:metricPF} in general, let us focus on the specific case $\mathcal{I}_0=0$. This is equivalent to require that the dilatation generator $D$ acts for free at any step along the
circuit.  The intuition behind this choice stems from the privileged status of $D$ as the Hamiltonian of the underlying \ac{cft}, as determined by applying the state-operator correspondence.
In other words, since the dilatation operator merely evolves the system, we may expect that it should not have any cost as a gate. Setting $\mathcal{I}_0=0$ in eq.~\eqref{eq:metricpmbasisfinal}, we obtain
\begin{align}
  \label{eq:metric_simple_projection_so12}
  \begin{gathered}
    \dd s^2 = \frac{2}{\bigl( 1 - r^2 \bigr)^2} \Bigl[\frac{4 \I_- \I_+ \dd r^2
    }{\alpha(\theta)} + r^2 \alpha(\theta) \dd A^2 \Bigr]~,\\
    \alpha(\theta) = \I_+ + \I_- + (\I_- - \I_+) \cos(2 \theta)>0 ~,\quad
    \dd A = \dd \gamma - \dd \theta - \frac{(\I_- - \I_+) \sin(2 \theta)}{r
    \alpha(\theta)} \dd r~,
  \end{gathered}
\end{align}
where we rewrote the definition of $\alpha(\theta)$ for convenience.
We therefore see that the effect of the limit $\I_0 \to 0$ in eq.~\eqref{eq:metricpmbasisfinal} is to eliminate the second term, proportional to $\dd B^2$, and to uniquely fix the sign of the first term,   leading
to a positive-definite metric. 
Note that even without applying our prescription to go to the coset space (which amounts to setting $\dd A=0$), each term in the sum has positive coefficients and the cost function is positive-definite. We could therefore directly study geodesics in this metric. 
Since the metric is positive-definite, any minimal path will have $\dd A =0$, a requirement which fully constrains the coordinate $\gamma$.

In this way, we are able to recover our coset procedure from section \ref{ssec:projection_coset}, which instead removed, by its definition, the direction $\dd A$. Note that in the special case $\mathcal{I}_0\rightarrow 0$, we do not need to impose the condition \eqref{eq:pfIpm0condition} to obtain positive definiteness, since the relevant term $\dd B^2$ is further multiplied by $\mathcal{I}_0$. 
The advantage of the coset reduction method of section \ref{ssec:projection_coset} lies in the possibility to use it for arbitrary values of $\I_0$, as we did in eq.~\eqref{eq:pf-0-metric}.
However, as we see, for the 
specific case $\mathcal{I}_0=0$, there exists an alternative way to project the cost function \eqref{eq:pf-0-metric} over the coset space which also yields a positive definite metric. 

In any event, with either way of reducing the metric 
\eqref{eq:metric_simple_projection_so12} to the coset space, the geodesics should minimize the following line element:
\beq
\dd s^2_{\mathrm{min}} = \frac{8 \I_- \I_+ \dd r^2 }{\alpha(\theta) ( 1 - r^2 )^2}
\, .
\label{eq:minimal_metric_simple}
\eeq
We observe that the expression is independent of $\dd \theta$, since phase
shifts in the coherent state representative \eqref{eq:state_CFT1} correspond
to the action of the dilatation operator $D$, which has vanishing cost.
This is also clear from the schematic depiction in
figure~\ref{fig:direction-effect-lambda}.
Solving the Euler-Lagrange equations associated with the metric
\eqref{eq:minimal_metric_simple} gives the same radial path as the \ac{fs} case,
that is $r(t) = \tanh(\arctanh (r_\tar)  t)$, where $r_\tar$ is the radial
coordinate characterizing the target state. 

The geodesics are piece-wise: the first part is a straight line at
fixed $\theta_0$, connecting the origin with the circle of radius $r_\tar$; the
second part is an arc of the circumference, corresponding to a free rotation
until the final angular coordinate of the target state $\theta_\tar$.
Since the radial cost of the geodesic depends on the function $\alpha(\theta)$,
the angle $\theta_0$ characterizing the orientation of the straight line is
chosen in such a way to follow the least penalized direction.
This direction corresponds to the real axis ($\theta=0$) when $\I_- > \I_+$,
and to the imaginary axis ($\theta=\pi/2$) when $\I_+ > \I_-$.
Plugging these results in the cost function, we obtain the state complexity
\beq
\mathcal C^{\mathrm{state}}_{\eqref{eq:metric_simple_projection_so12}}
[\ket{\lambda}, \ket{\Delta}] = 2 \, \arctanh (r_\tar) \sqrt{\min(\I_-, \I_+)} \,
.
\eeq
Comparing with eq.~\eqref{eq:complexity-fs}, we notice that this result coincides
with the \ac{fs} cost up to a rescaling by the penalty factor stemming from the motion along the radial direction.

\subsection{State complexity with general penalties}
\label{ssec:state_CFT1}

Next, let us study the geodesics associated with the cost
function~\eqref{eq:metricPF} inside the unit circle \eqref{eq:statedisk}. 
In general, this problem reduces to solving the following ordinary differential
equation (ODE) system
\begin{align}
  \label{eq:ode-system}
  \begin{gathered}
    \partial_t \left(\dfrac{\partial \mathcal{L}_\pf}{\partial \dot \lambda}\right) - \dfrac{\partial
    \mathcal{L}_\pf}{\partial \lambda} = 0~,\quad \partial_t \left(\dfrac{\partial
    \mathcal{L}_\pf}{\partial
    \dot \lambda^*}\right) - \dfrac{\partial \mathcal{L}_\pf}{\partial \lambda^*} = 0~,\\
    \lambda(0) = 0~, \quad \lambda(1) = r_\tar e^{i \theta_\tar}~,
  \end{gathered}
\end{align}
where $ \mathcal{L}_\pf$ was introduced in eq.~\eqref{eq:pf-0-metric}. Note that these equations extremize the Lagrangian $\mathcal{L}_\pf = \fpf^2$, instead of extremizing the cost function $\fpf$ itself, which results in affinely-parametrized geodesics. Those  will suffice for our purposes.   
In this form, the equations governing the optimal quantum circuit which connects two states can be thought of as  describing the motion of a classical two-dimensional trajectory in an effective  potential, reflecting the curved
geometry governed by the penalty factors.

We will evaluate the complexity $\mathcal C_{\pf}$ by substituting the geodesic solutions into eq.~\eqref{eq:statemin}. 
We will often compare the complexity $\mathcal C_{\pf}$ with non-trivial penalty factors to the \ac{fs} case, keeping the same boundary conditions.
For this comparison, the relevant quantity that we will compute is the ratio
\beq
\hat{\mathcal C} \equiv \frac{\mathcal C_\pf}{\mathcal C_\fs} \, .
\label{eq:def_Chat}
\eeq
In the remainder of this subsection, we analyze the geodesics of the cost
functional \eqref{eq:metricPF}, starting from specific analytic results, then moving to an expansion around the case of trivial penalties, and finally performing a numerical study of the Euler-Lagrange equations
\eqref{eq:ode-system}.

\subsubsection{Analytical solution}
\label{ssec:analytic_1d}

We begin with the special case $\mathcal{J}_2=0$ (equivalently, $\mathcal{I}_+=\mathcal{I}_-$), which leaves the cost
functional isotropic along the coset directions.
We show that this setting allows us to find an exact solution for the optimal
trajectories connecting the reference and the target states.
The Lagrangian reads
\begin{align}
  \label{eq:lag-pf-no-j2}
  \mathcal{L}_\pf = 4 \J_3 \left[ \frac{\dot r^2}{(1-r^2)^2}+\frac{\dot \theta ^2
  r^2}{(1+r^2)^2 -\frac{4\mathcal{J}_3}{\J_1} r^2} \right] \, ,
\end{align}
where we used cylindrical coordinates $\lambda(t) = r(t) e^{i \theta(t)}$.
The angular direction $\theta$ is cyclic, therefore there exists a corresponding conserved charge defined by\footnote{Note that this is an additional conserved charge, not the one used for the coset reduction.}
\begin{align}
  K_{\theta} = \frac{1}{2}\dfrac{\partial \mathcal{L}_\pf}{\partial \dot\theta} = \dfrac{4 \J_3 r^2}{(1
  + r^2)^2  - \frac{4 \J_3}{\J_1} r^2} \, \dot \theta~.
\end{align}
At $t=0$, imposing the boundary condition $\lambda(0) = \lls(0) = 0$ gives
$r(0)=0$, which further implies $K_\theta = 0$.
Since the charge is conserved during the evolution, $K_{\theta}$ must remain vanishing at all times, but since $r$ is no longer vanishing anywhere outside the origin, this implies $\dot \theta=0$,  \ie $\theta$ is constant along the circuit.
Plugging this result back into the Lagrangian \eqref{eq:lag-pf-no-j2}, we get
the following simplification
\begin{align}
  \mathcal{L}_\pf = 4 \J_3 \dfrac{\dot r(t)^2}{\left(1-r(t)\right)^2} = \J_3 \ffs^2
  \equiv \J_3 \mathcal{L}_{\fs}~.
\end{align}
Since this expression is proportional to the \ac{fs} Lagrangian, we immediately
conclude that the geodesics are given by the same stationary
solutions~\eqref{eq:fsSol}, the only difference being that the complexity is
rescaled by an overall factor:
\begin{align}
  \label{eq:complexityConservedQtty}
  \mathcal{C}_\pf = \sqrt{\J_3} \mathcal{C}_\fs \, .
\end{align}
Therefore, we deduce that as long as $\J_2 = 0$, the state complexity is independent of the penalty $\J_1$, and only receives rescaling corrections from the penalty $\J_3$. 
Intuitively, this can be understood from the form of $h_0$ in
eq.~\eqref{eq:hamiltonian-basis}, evaluated using the constraint \eqref{eq:phase-coordinate-pf}, which in polar coordinates takes the form
\beq
h_0 \sim \mathcal{A}(r,\theta, \mathcal{J}_i) \J_2 \dot r + \mathcal{B}(r,\theta,\mathcal{J}_i)\dot \theta \, ,
\eeq
where the functions $\mathcal{A}$, $\mathcal{B}$ behave regularly in the limit $\J_2\rightarrow 0$.  
When $\J_2 = 0$, we notice that $h_0 \propto K_{\theta}=0$ along the optimal trajectory, implying that the
corresponding $\mathcal{J}_1$ penalty does not affect the optimal trajectory's cost.

\subsubsection{Perturbation theory}
\label{ssec:perturbative_1d}

When $\J_2\ne 0$, we did not find a simple analytic expression for the geodesics of the cost function~\eqref{eq:metricPF}, 
but we were able to study the problem perturbatively.
This is done as follows. Expand the penalty factors and the trajectory around the solution obtained
in the \ac{fs} case
\begin{align}
  \J_i = \J_i^\fs + \varepsilon \beta_i ~, \qquad
  X^\mu(t) =
  X^\mu_0(t) + \varepsilon X^\mu_1(t) + \varepsilon^2 X^\mu_2(t) +
  \mathcal{O}(\varepsilon^3)~,
  \label{eq:pert_expansion_CFT1}
\end{align}
where $\J_i^\fs=\{1,0,1\}$ were defined in eq.~\eqref{eq:FubiniJtrivial}, and $X_0^{\mu} = (\lambda_0,\lambda_0^*)$ is the extremal solution obtained in eq.~\eqref{eq:fsSol}. The parameter $\varepsilon$ is assumed to be small. 
Plugging this ansatz inside the differential equations \eqref{eq:ode-system},
we can iteratively solve for the optimal path order by order in $\varepsilon$.
The explicit form for the optimal circuit is rather cumbersome and we obtained it using Wolfram Mathematica. This leads to a complexity ratio \eqref{eq:def_Chat} which reads
\begin{align}
  \label{eq:complexity-perturbation-theory}
  \hat{\mathcal C} & = 1 + \varepsilon \left[ \beta_3 - 2 \beta_2 \cos(2
  \theta_\tar) \right]\nonumber\\
  &- \varepsilon^2 \left[ \dfrac{\left(\beta_3 - 2 \beta_2 \cos(2
    \theta_\tar)\right)^2}{8} - \dfrac{2 \beta_2^2  \sin(2 \theta_\tar)^2}{r_\tar}\left(
      (1
  + r_\tar^2) \,\arctanh(r_\tar) - r_\tar \right) \right] + \mathcal{O}(\varepsilon^3)~.
\end{align}
We observe that, up to and including order $\varepsilon^2$, the complexity does not depend on the penalty 
$\beta_1$ along the direction of the dilation operator.
Instead, only the penalties for the generators $P, K$ play a non-trivial role.

As a consistency check, we observe that the limit $\mathcal{J}_2 = 0$ (implying
$\beta_2=0$) recovers the series expansion of the square root inside the
analytic result \eqref{eq:complexityConservedQtty} obtained in the previous
subsection, \ie
\begin{align}
  \hat{\mathcal C} = 1 + \varepsilon \beta_3 - \varepsilon^2 \beta_3/8
  +\mathcal O(\varepsilon^3) \approx \sqrt{1+\varepsilon \beta_3} =
  \sqrt{\J_3}~.
\end{align}
In the case $\beta_2 \ne 0$, we can try to recast the state complexity 
\eqref{eq:complexity-perturbation-theory} to a similar form, by re-expressing the series expansion as
\begin{align}
  \label{eq:perturbative-ratio-small-r1}
  \hat{\mathcal C}^{(2)} \approx \sqrt{\J_3 - 2 \J_2 \cos(2 \theta_\tar)} + 2
  \varepsilon^2 \beta_2^2  \sin(2 \theta_\tar)^2 \frac{(1 + r_\tar^2) \,\arctanh (r_\tar)
  - r_\tar}{r_\tar} + \mathcal{O}(\varepsilon^3) \, ,
\end{align}
where in the first term we assumed $\J_3 \approx 1 + \varepsilon\beta_3$ and $\J_2 \approx
\varepsilon\beta_2$.
When $r_\tar$ is small enough, the second term is of $\mathcal O(r_\tar^2)$,
therefore it is subleading compared to the first one.
We will further analyze this regime numerically below.

\subsubsection{Numerical results}
\label{ssec:numerical_1d}

Equipped with the experience gained from the perturbative analysis, we numerically investigate the solutions of the non-perturbative coupled ODE system \eqref{eq:ode-system}.
Our numerical setup is a shooting algorithm: we solve the system as an initial boundary value problem at $t = 0$, and then we vary over the initial velocity
so that the boundary condition at $t = 1$ is obeyed.
These numerical solutions will allow us to study the evolution of the
complexity over a wide range of values for each choice of the penalty factors, as well as for various choices of target states.
As mentioned below eq.~\eqref{eq:defB_so12}, the symmetry of the cost function \textemdash{} and therefore of the Lagrangian \textemdash{} implies that we can restrict the range of the angular coordinate parametrizing the
target state to $\theta_\tar \in [0, \pi/2]$. Note, however, that we do not assume that the trajectories are constrained within a single quadrant, but only their endpoints. We restrict our analysis to choices of penalty factors satisfying the constraints \eqref{eq:penaltycondJ} and \eqref{eq:firstconstraintJ}, for which the complexity metric over the space of states is positive-definite and thus admits a well-defined complexity interpretation.
Furthermore, the numerical investigations reported in this subsection are restricted to a region of the space of states with target radial coordinate  restricted by $|r_{\rm T}| \leq 0.5$. While it is not impossible to evaluate complexity closer to the boundary of the space of states where $|r|\lesssim 1$ for specific states, the numerical calculations takes a much longer time to run and the qualitative behavior remains unchanged. For this reason, we do not report these results here.

\begin{figure}[ht]
  \centering
  \includegraphics[width=\textwidth]{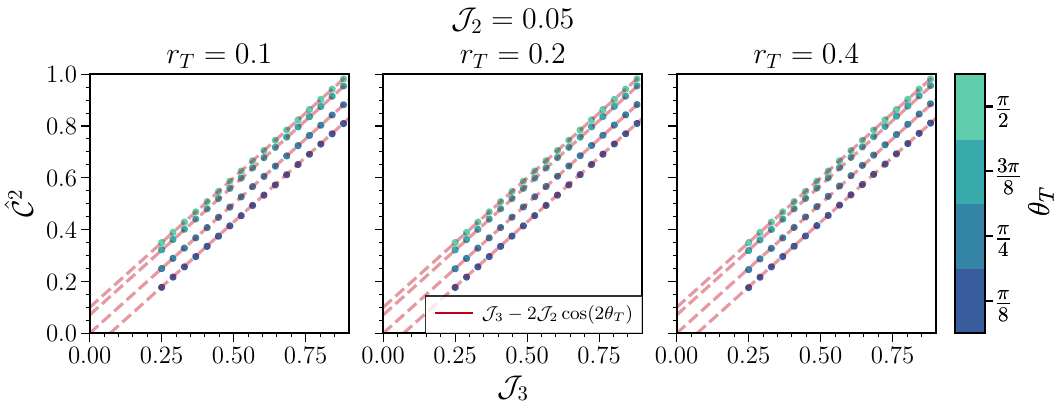}
  \caption{Squared ratio of the complexities \eqref{eq:def_Chat} as a function of
    $\J_3$, for fixed $\J_2 = 0.05$ and $\J_1 = 1$, varying $\theta_\tar \in
    [0,\pi/2]$. Each plot refers to a different value of $r_\tar$, but we observe that the data are nearly independent of this parameter. 
    In the plot, we compare the numerical result (dotted points) with the leading term in the perturbative expansion~\eqref{eq:perturbative-ratio-small-r1} (dashed red line). The difference between the corresponding data points in the three plots (differing by radius) is of the order of $\Delta \hat{\mathcal{C}}\lesssim 10^{-3}$, as explained in the main text. The penalty factors in the figures lie within the range that ensures a positive-definite complexity metric, see the condition~\eqref{eq:penaltycondJ}.}
  \label{fig:plotVaryJ3FixedJ2}
\end{figure}

We begin by fixing $\J_2$ and varying $\J_3$ to confirm the perturbative result obtained in eq.~\eqref{eq:perturbative-ratio-small-r1}.
Indeed, we observe in figure~\ref{fig:plotVaryJ3FixedJ2} that for small values of $\J_2$ and $r_\tar$, such that we are within the range of validity of perturbation theory, the complexity ratio is well approximated by
$\hat{\mathcal{C}}\approx \sqrt{\J_3 - 2 \J_2 \cos(2\theta_\tar)}$. The figures for different values of $r_\tar$ look nearly identical, but they differ at the order of $\Delta \hat{\mathcal{C}}\lesssim 2\mathcal{J}_2^2 r_T^2 \lesssim 10^{-3}$, as expected from expanding the last term in eq.~\eqref{eq:perturbative-ratio-small-r1} for small $r_\tar$.

\begin{figure}[ht]
  \centering
  \includegraphics[width=\textwidth]{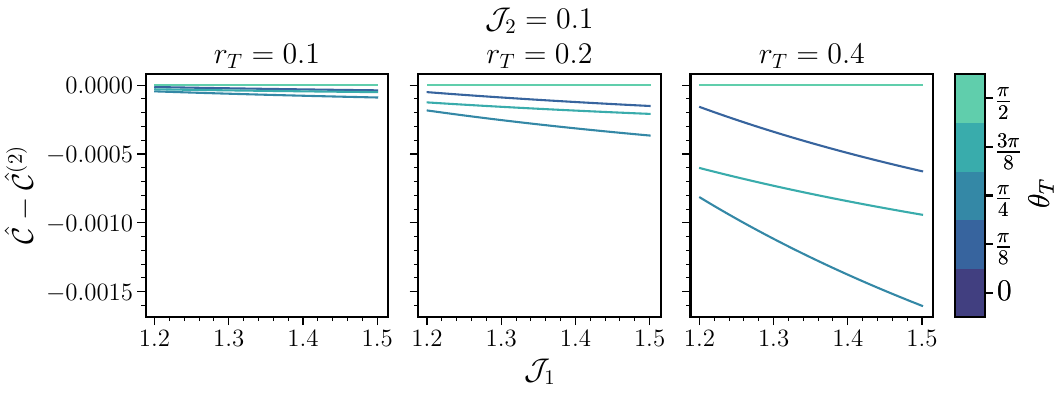}
  \caption{Difference between the ratio of complexities \eqref{eq:def_Chat} and its second-order expansion \eqref{eq:perturbative-ratio-small-r1} as a function of $\J_1$,
    for fixed $\J_2 = 0.1$ and $\J_3 = 1$, varying $\theta_\tar \in [0,\pi/2]$.
    Each plot refers to a different value of $r_\tar$.
    The dependence on $\J_1$, which comes from higher orders in the perturbative expansion, is enhanced for larger values of $r_\tar$. The penalty factors in the figures lie within the range that ensures a positive definite complexity metric, see the condition~\eqref{eq:penaltycondJ}.}
  \label{fig:plotVaryJ1FixedJ2}
\end{figure}

Furthermore, earlier we have seen that up to (and including) the second order in perturbation theory, the complexity does not depend on the penalty factor $\mathcal{J}_1$. Figure~\ref{fig:plotVaryJ1FixedJ2} studies the dependence of the complexity ratio $\hat{\mathcal{C}}$ on $\mathcal{J}_1$, by subtracting from it the second-order result $\hat{\mathcal{C}}^{(2)}$ in eq.~\eqref{eq:perturbative-ratio-small-r1}. From the figure we observe that  at finite $\J_2$, there is a dependence on the penalty factor $\J_1$, of the order $10^{-3}$. Such a contribution  comes from higher orders in the perturbative expansion of the complexity ratio, and could, \eg be the consequence of terms of the form $\Delta \hat{\mathcal{C}} \sim \mathcal{J}_1\mathcal{J}_2^2 r_\tar^2\sim 10^{-3}$. In any event, the influence of $\J_1$ on the complexity is very small compared to that of the other penalties.


\begin{figure}[ht]
  \centering
  \includegraphics[width=\textwidth]{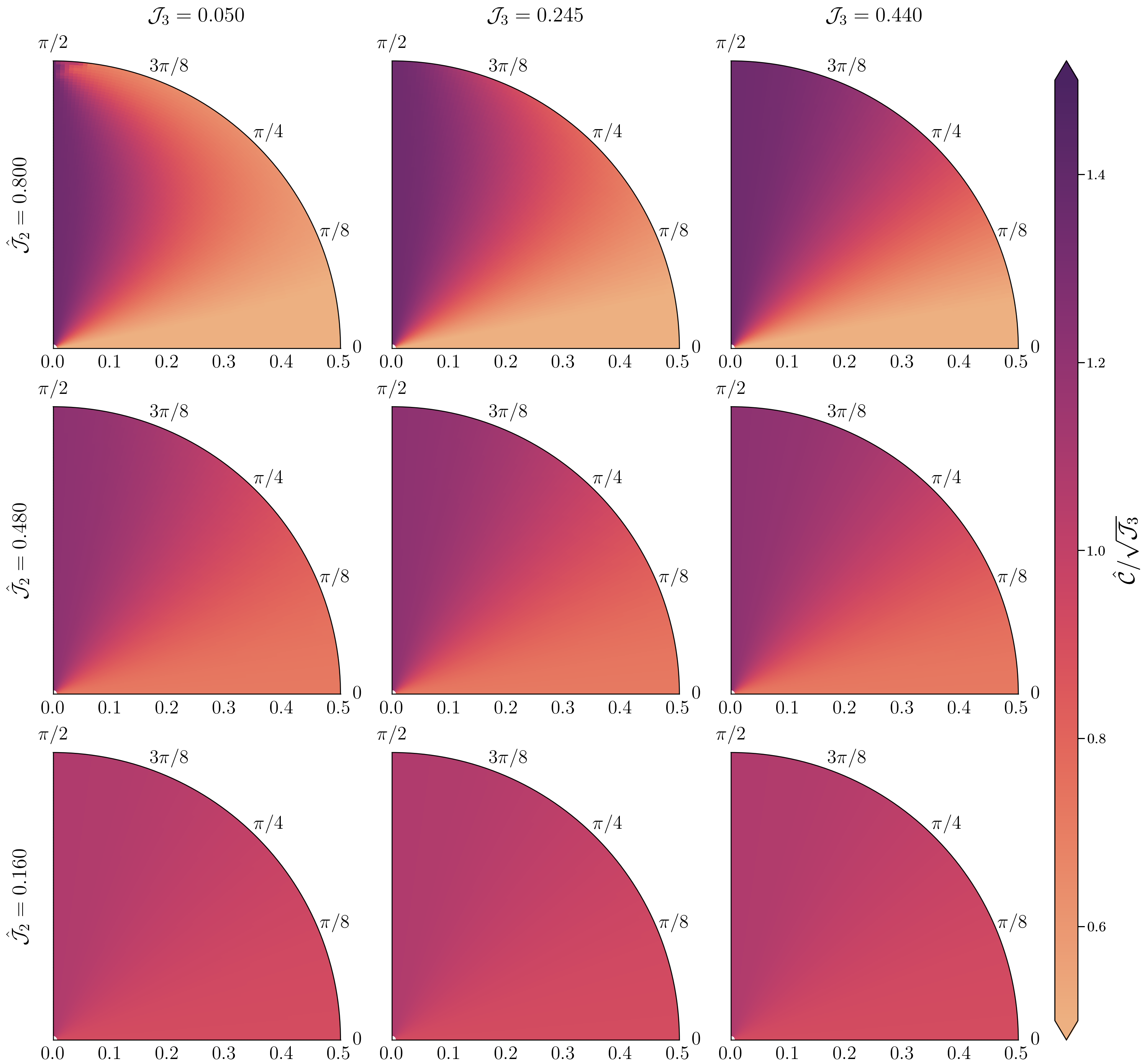}
  \caption{Normalized complexity ratio
    $\frac{\mathcal{C_{\pf}}}{\sqrt{\J_3}\mathcal{C_{\fs}}}$ as a function of
    $\J_3$ and $\hat{\mathcal{J}}_2 =\frac{2 \J_2}{\J_3}$, for fixed $\J_1 = 1$
    and for various values of $0 < r_\tar < 1/2$ and $0 < \theta_\tar < \pi/2$.}
\label{fig:overviewPlot}
\end{figure}

For this reason, we now focus on the case of $\J_1 = 1$, and instead vary both $\J_2$ and $\J_3$. 
This is the setting considered in figure~\ref{fig:overviewPlot}, where we plot
$ \hat{\mathcal C}/\sqrt{\J_3}$ while varying $0 < \J_3 < 0.44$ at fixed $\J_2$
(columns), and $0 \leq \hat{\J}_2 \equiv \frac{2 \J_2}{\J_3} < 0.8$ at fixed $\J_3$ (rows). 
The complexity of the target state is indicated by the different colors on the coset space represented as a quarter of a disk in a two-dimensional space, with the real (imaginary) part of the state parameter $\lambda$ parametrizing the horizontal (vertical) axis. For simplicity, we focused on $\J_2>0$, which implies that $\mathcal{I}_->\mathcal{I}_+$, i.e., the imaginary axis along the state space disk \eqref{eq:statedisk} is more penalized than the real axis. The case $\J_2<0$ is related to the latter via an exchange of the real and imaginary directions. 
Once again, we only selected penalty factors inside the range that ensures a positive definite state complexity metric, as per the constraint~\eqref{eq:penaltycondJ}.
From the perturbative result
\eqref{eq:perturbative-ratio-small-r1}, we observe that the value $\hat{\J}_2 = 1$ marks a locus of zero complexity along the real direction $\theta_\tar = 0$. The
geometry associated with the cost function \eqref{eq:metricPF} has a null direction at this value of $\hat{\J}_2 = 1$, since $\mathcal{I}_+=\J_3-2\J_2=0$, indicating a singularity in the parameter space of penalty factors. This is confirmed by evaluating the Ricci scalar of the metric on the coset space, \ie $R \propto
(\J_3 - 2 \J_2)^{-1}$.

Next, we observe in figure~\ref{fig:overviewPlot} that increasing $\J_3$ at fixed $\hat{\J}_2$
increases the cost of moving  along either the real or the  imaginary axes. We can see this in the figure by noting that the colors, which encode the complexity, are not changing near the axes, but since the plot is normalized as $\hat{\mathcal{C}}/\sqrt{\J_3}$, this means an increase of complexity in all directions. 
Recall from our intuitive analysis below
eq.~\eqref{eq:hamiltonian-basis} that the penalty coefficient
$\I_-$ ($\I_+$) corresponds to the generator $L_-$ ($L_+$) associated with a larger
imaginary (real) trend of the trajectories.
Rewriting the map \eqref{eq:map-I-J} as $\I_- = \J_3
(1 + \hat{\J}_2)$ and $\I_+ = \J_3 (1 - \hat{\J}_2)$, we observe that increasing $\J_3$ implies that the costs of both directions homogeneously increase, i.e., trajectories along both the real
and imaginary axes become more expensive, as we indeed observe in the figure. 

If we fix $\J_3$ instead, we notice that increasing $\hat{\J}_2$ generally leads to a higher relative anisotropy between the real and imaginary directions.
This increase manifests as a difference in colors between the real and imaginary axes. 
Our choice of $\J_2 > 0$ means $\I_- > \I_+$ via eq.~\eqref{eq:map-I-J}, which in turn
implies that the imaginary direction is more penalized than the real direction. We indeed see this in the figure, since the real axis appears in brighter colors.  
This shows that increasing $\hat \J_2$ implies that trajectories along the real
axis are favored, while the imaginary direction is more penalized.  We further observe that for small values of $\J_3$ and large values of $\hat \J_2>0$, i.e., in the upper left corner plot, shortcuts become available to reach states with a large imaginary component in an indirect way via the real axis. This detour becomes favorable due to the large anisotropy between the real and imaginary axes in this case and the smaller value of the real axis motion cost $\mathcal{I}_+$.

In summary, the analysis of the geodesics of the cost function \eqref{eq:metricPF} on the coset space revealed that penalizing the direction
corresponding to the dilatation operator does not significantly affects the
state complexity. Acting with the dilatation generator on a state simply changes it by a phase, see eq.~\eqref{eq:state_CFT1}, and so naively, we would not expect this operation to have any cost. However, our prescription for reducing to the coset space fixes the control function for the generator $D$ in terms of the other control functions for $P$ and $K$, and does not allow to fix it independently. Therefore, any operation of $D$ is accompanied with some action of $K$ and $P$. Nevertheless, the smaller effect of the associated penalty on complexity might be in line with it having a special role as the stabilizer direction, and more generally with its special role in CFT as the system's Hamiltonian. Instead, increasing the penalty factors along the $K, P$ directions generally
leads to a larger complexity change for coherent states \eqref{eq:state_CFT1}. 
More specifically, this phenomenon slightly differs depending on which penalty
factor is raised.
If we increase the anisotropy between the $L_{\pm}$ generators (as measured by
the penalty factor $\J_2$), the imaginary (or real) directions will develop costs which differ from each other.
Instead, when the penalty factors along the $L_{\pm}$ are increased
homogeneously (\ie when $\J_3$ is larger), the main trend is that complexity
becomes larger along both axes.

\section{Two-dimensional CFT}
\label{sec:2dCFT}

In this section, we study the Nielsen complexity associated with the global symmetry group of a two-dimensional \ac{cft}. To this end, we explicitly use the fact that the conformal group in $d =2$ neatly factorizes in two copies as $\mathrm{SO}(2,2) =
\mathrm{SO}(1,2)_L \times \mathrm{SO}(1,2)_R$, which we  refer to as left (holomorphic) and right (anti-holomorphic) copies, with Hermitian bases of generators  $ \lbrace L_0, L_{\pm} \rbrace \cup \lbrace\bar{L}_0, \bar{L}_{\pm} \rbrace$, respectively.  
The unitary operators and coherent states are defined  in terms of a direct product of the two factors, where each of the factors constructs the state in the same way as in eqs.~\eqref{eq:circuits-ansatz} and \eqref{eq:state_CFT1}, in terms of a pair of coordinates $(\lambda, \lb)$, associated with the two copies.\footnote{In the following, we denote all the quantities (such as coordinates and generators) of the right copy with a bar, but we stress that this is just a choice of notation: no complex conjugation is involved. Instead, the expression for each copy will involve the state parameter and its complex conjugate.} 
The boundary conditions on the reference and target states split as follows
\beq
\lambda(0) = \lb(0) = 0~, \quad \lambda(1) = r_\tar e^{i \theta_\tar}~, \quad \lb(1)
= \bar r_\tar e^{i \bar \theta_\tar}~.
\label{eq:bdy_cond_CFT2}
\eeq

In this setting, the simplest example of Nielsen complexity on
$\mathrm{SO}(2,2)$ corresponds to a cost function of the form
\beq
\mathcal{F}^2_{\mathrm{CFT}_2} =  \mathcal{F}^2_{\mathrm{CFT}_1} +
\bar{\mathcal{F}}^2_{\mathrm{CFT}_1} \, ,
\label{eq:naive_CFT2}
\eeq
where $\mathcal{F}_{\mathrm{CFT}_1}$ is a cost function defined on the Lie
group $\mathrm{SO}(1,2)$, for instance of the form \eqref{eq:pf-0-metric}.
In this case, the trivial factorization of the symmetry group also applies to the minimal paths in the geometry, such that the total complexity is simply the quadratic sum of the complexities associated with each copy, \ie 
\begin{equation}\label{eq:sumquadcft2}
\mathcal
C_{\mathrm{CFT}_2} = \sqrt{\bigl(\mathcal C_{\mathrm{CFT}_1}\bigr)^2 +
\bigl(\overline{\mathcal C}_{\mathrm{CFT}_1}\bigr)^2}.
\end{equation}
In summary, the presence of penalty factors in the two copies of
$\mathrm{SO}(1,2)$ separately keeps us inside a class of solutions where
trajectories are simply factorized.

In order to get non-trivial new geodesics, we need to consider mixed penalty factors between the left and right-handed copies.
The simplest way to achieve this goal is to complement the simple expression~\eqref{eq:naive_CFT2} with a coupling governed by a penalty factor $\mathcal{J}_0$ as follows,
\beq
\mathcal{F}^2_{\mathcal{J}_0, \mathrm{CFT}_2} =  \mathcal{F}^2_{\mathrm{CFT}_1} +
\bar{\mathcal{F}}^2_{\mathrm{CFT}_1} 
+ \J_0 \left( \mathcal Q \bar{\mathcal Q} + \mathrm{h.c.} \right)
\, ,
\label{eq:Lagrangian_CFT2}
\eeq
where $\mathcal Q \equiv \sum_a M_a \inner{H}{\omega_a}$ is a linear
combination (with arbitrary coefficients $M_a$) of the Hamiltonian components for a single CFT$_1$, and $\mathcal{F}^2_{\mathrm{CFT}_1}$ the cost function for a single copy, see \eqref{eq:pf-0-metric}; similar definitions apply to the anti-holomorphic copy.
This generalization allows for a rather large space of possibilities, depending on which operation costs are coupled between the two copies, and which penalty factors are turned on within each of the copies. 
To get a metric on the coset space (which we can then use to evaluate the complexity of states rather than unitaries) we apply the projection procedure outlined in section~\ref{ssec:general_projection} by setting the appropriate conserved quantities $K_n= \bar{K}_n=0$, as we did in the case of a CFT${}_1$. Equivalently, we extremize the expression for unitary complexity in terms of the stabilizer parameters of both copies.

In what follows, we consider a relatively simple example, where we only couple the dilatation operations between the two copies. 
First, we focus in section~\ref{ssec:2d_analytic} on the case when the anisotropy parameters of both copies are set to zero. That means that the costs associated with $L_+$ and $L_-$ are the same, and similarly, the costs associated with $\bar L_+$ and $\bar L_-$ are the same. In this case, we obtain analytic results based on conserved charges, similarly to the procedure outlined in section \ref{ssec:analytic_1d}.
In this regime, we show that the complexity is not sensitive to the coupling between the two theories. 
Next, we perform in section~\ref{ssec:2d_perturbative} a perturbative investigation around the case of trivial penalties, similarly to what was done in section~\ref{ssec:perturbative_1d}. We find that the Nielsen state complexity does not receive contributions from the coupling between the two copies up to (and including) second order in the perturbative expansion. 
In section~\ref{ssec:2d_numerical}, we explore the complexity numerically. We observe a small non-trivial dependence on the coupling between the two copies which likely comes from the third order in the perturbative expansion. We further study numerically cases at which the penalty factors are far from the range of the perturbative analysis. We observe that the coupling between the two copies can either increase or decrease the complexity, depending on the values of the different penalties. 
Finally, we comment on other possible couplings between the generators of the two copies of the $\mathrm{SO}(1,2)$ group in section~\ref{ssec:comments_other_coupling}.
While these latter cases do not admit a positive-definite metric over all the Poincar\'e disk, we interpret these configurations as describing physical systems where part of the state space is not accessible. 
We analyze their Nielsen complexity in appendix~\ref{app:sec:cost_nontrivial_bdy}.

\subsection{Coupling \texorpdfstring{$(D, \bar D)$}{(D,\bar D)}: coset space metric and analytic results}
\label{ssec:2d_analytic}

We study the cost function~\eqref{eq:Lagrangian_CFT2} in the case where we couple the dilatation operators of the two copies of the one-dimensional CFTs. 
In other words, we select $\mathcal{Q}= \langle H, D \rangle$ and $\bar{\mathcal{Q}} = \langle H, \bar{D} \rangle$ and define
\beq
\mathcal{L}_{\rm D \bar{D}} =  \mathcal{L}_{\mathrm{CFT}_1} +
\bar{\mathcal{L}}_{\mathrm{CFT}_1} 
+ \J_0  \langle  H, D \rangle  \langle H, \bar{D} \rangle
\, , \qquad
\mathcal{L}_{\rm CFT_1}  = \text{eq.~\eqref{eq:pf-0-metric}} \, .
\label{eq:Lagrangian_CFT2_DD}
\eeq
We take the Lagrangian in eq.~\eqref{eq:pf-0-metric} as the cost function $ \mathcal{F}^2_{\mathrm{CFT}_1}$ (and $ \bar{\mathcal{F}}^2_{\mathrm{CFT}_1}$) in each copy, equipped with generic \emph{real} penalty factors $(\mathcal{J}_1, \mathcal{J}_2, \mathcal{J}_3)$ and $(\bar{\mathcal{J}}_1, \bar{\mathcal{J}_2}, \bar{\mathcal{J}_3})$, respectively.

As argued in section~\ref{ssec:relation_minimization},the metric can be projected from the group manifold to the coset space by performing a pseudo-Riemannian submersion map. In practice this is done by one of the two equivalent methods:
\begin{enumerate}
    \item Compute the conserved charges associated with the cyclic coordinates $(\gamma, \bar{\gamma})$, one for each copy of CFT$_1$, by using the definition in eq.~\eqref{eq:conserved_charge}. Set both conserved charges to zero to obtain $\dot{\gamma}$ and  $\dot{\bar\gamma}$ in terms of the other coordinates and their derivative. Substitute back in the metric to obtain a metric in terms of the coset coordinates $\lambda$, $\bar \lambda$ and their derivatives.
    \item 
    \label{step2}
    Minimize the cost function with respect to $\dot{\gamma}$ and  $\dot{\bar\gamma}$ by completing the square. The order in which the minimization is performed is irrelevant, as we checked explicitly.
\end{enumerate}

After applying one of the above-mentioned steps, the cost function over the space of states is given by
\beq
\begin{aligned}
\mathcal{L}_{\rm PF ,D \bar{D}} & = 
 \frac{\left[ 4 \bar{B} A + \mathcal{V}^* \mathcal{J}_0^2  \le 1+|\bar{\lambda}|^2   \ri^2 \right]  
 \dot{\lambda}^2 
 +\left[ 4 \bar{B} A^* + \mathcal{V} \mathcal{J}_0^2  \le 1+|\bar{\lambda}|^2   \ri^2 \right]
 (\dot{\lambda}^*)^2
 +\mathcal{E} \dot{\lambda} \dot{\lambda}^*
 }{\le 1-  |\lambda|^2 \ri^2 \mathcal{T} } \, 
    \\
 &  + \frac{\left[ 4 B \bar{A} + \bar{\mathcal{V}}^* \mathcal{J}_0^2  \le 1+|\lambda|^2  \ri^2  \right] 
 \dot{\bar{\lambda}}^2
 +\left[ 4 B \bar{A}^* +  \bar{\mathcal{V}} \mathcal{J}_0^2  \le 1+|\lambda|^2  \ri^2  \right]
  (\dot{\bar{\lambda}}^*)^2
  +\bar{\mathcal{E}} \dot{\bar{\lambda}} \dot{\bar{\lambda}}^* 
 }{\le 1-|\bar{\lambda}|^2  \ri^2 \mathcal{T} } \, \\
  & - \frac{4 \mathcal{J}_0}{\le1- |\lambda|^2  \ri  \le 1-|\bar{\lambda}|^2  \ri \mathcal{T} } 
  \left(\mathcal{U}^* \dot{\lambda} -\mathcal{U} \dot{\lambda}^*\right) 
  \left( \bar{\mathcal{U}}^* \, \, \dot{\bar{\lambda}} - \bar{\mathcal{U}}   \, \dot{\bar{\lambda}}^* \right) 
 \, ,
\end{aligned}
\label{eq:cost_PF_CFT2_general}
\eeq     
where we defined the following  quantities (some of the definitions overlap with previous definitions in \eqref{eq:defB_so12}, but we reiterate them here to explicitly introduce the coupling dependence):
\begin{subequations}
\beq
\mathcal{V}(\lambda,\mathcal{J}_i) \equiv \mathcal{J}_2 \le 1+\lambda^4  \ri - \mathcal{J}_3 \lambda^2, \quad
\bar{\mathcal{V}} 
\equiv\mathcal{V}(\bar\lambda,\bar{\mathcal{J}}_i),
\eeq
\beq
A(\lambda,\mathcal{J}_i)\equiv 
\le 4 \mathcal{J}_2^2 - \mathcal{J}_3^2 \ri \lambda^*{}^2-\mathcal{J}_1 \mathcal{V}^* ,\quad
\bar A\equiv A(\bar \lambda,\bar{\J}_i),
\eeq
\beq
B(\lambda,\mathcal{J}_i) \equiv  \mathcal{J}_1 \le 1+|\lambda|^2 \ri^2 -4 \left[ \mathcal{J}_3 |\lambda|^2 + \mathcal{J}_2  \le \lambda^2 + \lambda^*{}^2 \ri  \right]   \, ,\quad
\bar{B} \equiv 
B(\bar\lambda,\bar{\mathcal{J}}_i)
\eeq
\beq
\mathcal{W}(\lambda,\J_i) \equiv \mathcal{J}_3 \le |\lambda|^4 +1 \ri - 2 \mathcal{J}_2 \le \lambda^2 + (\lambda^*)^2 \ri, 
\quad \bar{\mathcal{W}} \equiv \mathcal{W}(\bar\lambda,\bar{\J}_i)
\eeq
\beq
C(\lambda,\J_i)  \equiv  2 \left( 4 \J_2^2 - \J_3^2 \right) |\lambda|^2 + \J_1 
    \mathcal{W}(\lambda,\J_i), \quad \bar C=C(\bar\lambda,\bar{\J}_i) ~,
\eeq
\beq
\mathcal{E} \equiv 4 \bar{B} C- \mathcal{J}_0^2 \mathcal{W}  \le |\bar{\lambda}|^2 +1 \ri^2   \, ,\quad
\bar{\mathcal{E}} \equiv
4 B  \bar C - \mathcal{J}_0^2 \bar{\mathcal{W}}  \le |\lambda|^2 +1 \ri^2 
\eeq
\beq
\mathcal{U}(\lambda,\J_i) \equiv 2 \mathcal{J}_2 \le \lambda^3 - \lambda^* \ri  
+ \mathcal{J}_3 \lambda \le |\lambda|^2 -1 \ri \, , 
\quad
\bar{\mathcal{U}} = \mathcal{U}(\bar \lambda,\bar{\J}_i),
\eeq
\beq
\begin{aligned}
\mathcal{T} & \equiv B\bar{B} -(\mathcal{J}_0/2)^2 \le 1+ |\lambda|^2  \ri^2  \le 1+ |\bar{\lambda}|^2  \ri^2   
\end{aligned}
\label{eq:definition_T}
\eeq
\end{subequations}
Note that upon setting $\mathcal{J}_0=0$, we recover the sum of the CFT$_1$ state metrics for each copy. Hence in the limit $\mathcal{J}_0=0$, the two copies completely factorize and the trajectories can be studied in each copy separately, leading to a complexity of the form eq.~\eqref{eq:sumquadcft2}.
In the following, we will analyze the geodesics of the line element induced from the norm~\eqref{eq:cost_PF_CFT2_general} with the boundary conditions~\eqref{eq:bdy_cond_CFT2}.

\subsubsection{Positivity of the cost function}

As discussed below eq.~\eqref{eq:metricpmbasisfinal2}, one way to associate a meaningful notion of complexity to a cost function on the space of states is to require that the metric is positive-definite through all the state manifold.
This condition is \textit{not} satisfied by the cost  function~\eqref{eq:cost_PF_CFT2_general} with arbitrary penalty factors. Instead, it provides a constraint on the penalty factors. While it is difficult to  identify all the constraints on the penalty factors ensuring a positive definite metric in full generality, in this section, we outline some necessary set of conditions, and give several specific examples of choices of penalties for which the metric is positive definite on the full space of states. An alternative approach is to consider a space of states within a non-trivial boundary, where the metric is only positive definite inside the boundary and diverges on the boundary itself. In this case, we interpret the interior of the boundary as delimiting states which we are able to reach. We study some examples of this latter approach in appendix~\ref{app:sec:cost_nontrivial_bdy}.

It will be convenient in what follows to parametrize the trajectories in polar coordinates,
\beq
\lambda (t) = r(t) e^{i \theta(t)} \, , \qquad
\bar{\lambda} (t) = \bar{r} (t) e^{i \bar{\theta} (t)} \, , \qquad
t \in [0,1] \, .
\label{eq:parametrization_CFT2_trajectories}
\eeq
After performing this change of variables to the cost function~\eqref{eq:cost_PF_CFT2_general}, we obtain a four-dimensional metric $g_{ij}$ depending on the real coordinates $(r, \theta, \bar{r}, \bar{\theta})$.
According to Sylvester's criterion, a necessary and sufficient condition to have a positive-definite Hermitian matrix is that the determinant of all the leading principal minors is positive.
In our case, we need to require that the latter condition is satisfied inside all the Poincar\'{e} disk, namely when $r,\bar{r} \leq 1$ and for any $\theta, \bar{\theta}$. 
While it is difficult to precisely identify all the constraints from the principal minors, we found that a necessary set of conditions on the penalty factors is
\beq
\mathcal{J}_1, \bar{\mathcal{J}}_1 > 0  \, , \qquad
\mathcal{J}_3 > |2 \mathcal{J}_2| \, , \qquad
\bar{\mathcal{J}}_3 >  |2 \bar{\mathcal{J}}_2| \, .
\label{eq:constraints1_pen_CFT2}
\eeq
Furthermore, the following condition should hold 
\beq
\begin{rcases}
        \mathcal{J}_0^2 - 4 \mathcal{J}_1 (\bar{\mathcal{J}}_1 \pm  2 \bar{\mathcal{J}}_2 - \bar{\mathcal{J}}_3  ) \\
      \mathcal{J}_0^2 - 4 \bar{\mathcal{J}}_1(
      \mathcal{J}_1 \pm 2 \mathcal{J}_2   - \mathcal{J}_3)  \\
      \J_0^2 - 4 (\J_1 - \J_3) (\bar{\J}_1 \pm 2 \bar{\J}_2 - \bar{\J}_3) \\
       \J_0^2 - 4 (\bar{\J}_1 - \bar{\J}_3) (\J_1 \pm 2 \J_2 - \J_3) \\
      \J_0^2 - 4 (\J_1 \pm 2 \J_2 - \J_3) (\bar{\J}_1 \pm 2 \bar{\J}_2 - \bar{\J}_3)\\
      \J_0^2 - 4 (\J_1 - \J_3) (\bar{\J}_1 - \bar{\J}_3)\\
     \mathcal{J}_0^2 - 4 \mathcal{J}_1 \bar{\mathcal{J}}_1
\end{rcases}
\quad  \text{have the same sign} \, .
\label{eq:constraints12_pen_CFT2}
\eeq
In particular this tells us that 
\beq
\frac{\mathcal{J}_0^2 - 4 \mathcal{J}_1 \bar{\mathcal{J}}_1}{\mathcal{J}_0^2 -4 \le \mathcal{J}_1 - \mathcal{J}_3 \ri  \le \bar{\mathcal{J}}_1 - \bar{\mathcal{J}}_3 \ri} > 0 \, .
\label{eq:constraints2_pen_CFT2}
\eeq
From now on, we will impose the constraints~\eqref{eq:constraints1_pen_CFT2}--\eqref{eq:constraints2_pen_CFT2} on the penalty factors.
A consequence of~\eqref{eq:constraints2_pen_CFT2} is that either
\begin{itemize}
    \item the penalty factor that couples the holomorphic and anti-holomorphic sectors is large $(\mathcal{J}_0 > 2)$, in which case it is possible to have $\mathcal{J}_1 = \bar{\mathcal{J}}_1 = \mathcal{J}_3 = \bar{\mathcal{J}}_3=1$, 
    \item or there needs to exist a gap between the penalties $\mathcal{J}_1, \mathcal{J}_3$ (and their anti-holomorphic copies), in which case the coupling $\mathcal{J}_0$ between the two CFT$_1$ copies can be small.
\end{itemize}

In other words, these observations identify two distinct regimes for the penalty factors.
For instance, we checked that the cost function~\eqref{eq:cost_PF_CFT2_general} is positive-definite over all the space of states if any of the following two sets of penalty factors is chosen:
\begin{align}
\text{Case 1:} \qquad  & \mathcal{J}_0 = 0.1 \, , \quad
\mathcal{J}_1 = \bar{\mathcal{J}}_1 = 1.5 \, , \quad  
\mathcal{J}_2 = \bar{\mathcal{J}}_2 = 0.1 \, , \quad
\mathcal{J}_3 = \bar{\mathcal{J}}_3 = 0.5 \, ,
\label{eq:case1_penalties} \\
\text{Case 2:}  \qquad & \mathcal{J}_0 = 4.0 \, , \quad
\mathcal{J}_1 = \bar{\mathcal{J}}_1 = 1.0 \, , \quad
\mathcal{J}_2 = \bar{\mathcal{J}}_2 = 0.1 \, , \quad
\mathcal{J}_3 = \bar{\mathcal{J}}_3 = 1.0 \, .
\label{eq:case2_penalties}
\end{align}
We will numerically investigate these cases in section~\ref{ssec:2d_numerical}.
It would also be interesting to explore cases where the two copies are assigned different penalty factors (i.e., $\J_1 \neq \bar{\J}_1$, or $\J_3 \neq \bar{\J}_3$); we leave this for the future.

\subsubsection{Analytic solution: isotropic cost along the coset directions}
\label{ssec:isotropic_2dCFT}

We begin by considering the special case of the cost function~\eqref{eq:cost_PF_CFT2_general} where the penalty factors $(\mathcal{J}_1, \mathcal{J}_3)$ and $(\bar{\mathcal{J}}_1, \bar{\mathcal{J}}_3)$ are turned on, but we set $\mathcal{J}_2 = \bar{\mathcal{J}}_2=0$.
This is the two-dimensional generalization of the setting where the cost along the coset directions is isotropic, that we studied in section~\ref{ssec:analytic_1d} for one-dimensional CFTs.
Using the change of coordinates~\eqref{eq:parametrization_CFT2_trajectories} and plugging the above isotropy conditions on the penalties inside eq.~\eqref{eq:cost_PF2_CFT2}, we obtain the following cost function over the space of states:
\beq
\begin{aligned}
\mathcal{L}_{\rm PF ,D \bar{D},iso} = 
\frac{4\mathcal{J}_3 \dot{r}^2}{\le 1-r^2 \ri^2} + \frac{4\bar{\mathcal{J}}_3 \dot{\bar{r}}^2}{\le 1-\bar{r}^2 
\ri^2} 
-  \frac{1}{\mathcal{T}} \left[ \mathcal{J}_3^2 \bar{\mathcal{S}} r^2  \, \dot{\theta}^2
+ \bar{\mathcal{J}}_3 \mathcal{S}  \bar{r}^2 \, \dot{\bar{\theta}}^2 
- 16 \mathcal{J}_0 \mathcal{J}_3 \bar{\mathcal{J}}_3  r^2 \bar{r}^2 \, \dot{\theta}  \dot{\bar{\theta}} 
\right] \, ,
\end{aligned}
\label{eq:cost_PF2_CFT2}
\eeq    
where
\begin{subequations}
\beq
\mathcal{S} \equiv \mathcal{J}_0^2 \le 1+r^2  \ri^2
-4 \bar{\mathcal{J}}_1 \left[ \mathcal{J}_1 \le 1+r^2  \ri^2 -4 \mathcal{J}_3 r^2   \right] \, ,
\eeq
\beq
\bar{\mathcal{S}} \equiv \mathcal{J}_0^2 \le 1+\bar{r}^2  \ri^2
-4 \mathcal{J}_1 \left[ \bar{\mathcal{J}}_1 \le 1+\bar{r}^2  \ri^2 -4 \bar{\mathcal{J}}_3 \bar{r}^2   \right] \, ,
\eeq
\beq
 \mathcal{T}  \equiv 
 \left[ \mathcal{J}_1 \le 1+r^2  \ri^2 -4 \mathcal{J}_3 r^2   \right]  \left[ \bar{\mathcal{J}}_1 \le 1+\bar{r}^2  \ri^2 -4 \bar{\mathcal{J}}_3 \bar{r}^2   \right]  -(\mathcal{J}_0/2)^2 \le 1+r^2  \ri^2  \le 1+\bar{r}^2  \ri^2
 \, .
\eeq
\end{subequations}
The quantity $\mathcal{T}$ is obtained by substituting  $\mathcal{J}_2 = \bar{\mathcal{J}}_2 =0 $ into eq.~\eqref{eq:definition_T}, while the other definitions $\mathcal{S}, \bar{\mathcal{S}}$ are introduced here for convenience.
The cost function~\eqref{eq:cost_PF2_CFT2} admits two conserved charges associated with the angular coordinates $\theta, \bar{\theta}$ (that appear as additional cyclic variables), defined by
\begin{subequations}
\beq
K_{\theta} = \frac{1}{2} \frac{\partial \mathcal{L}_{\rm PF ,D \bar{D},iso}}{\partial \dot{\theta}} = 
-\frac{\mathcal{J}_3 r^2}{\mathcal{T}} \le \mathcal{J}_3 \bar{\mathcal{S}} \,  \dot{\theta} - 8 \mathcal{J}_0 \bar{\mathcal{J}}_3 \bar{r}^2 \, \dot{\bar{\theta}}  \ri \, ,
\eeq
\beq
K_{\bar{\theta}} = \frac{1}{2} \frac{\partial \mathcal{L}_{\rm PF ,D \bar{D},iso}}{\partial \dot{\bar{\theta}}} = 
-\frac{\bar{\mathcal{J}}_3 \bar{r}^2}{\mathcal{T}} \le \bar{\mathcal{J}}_3 \mathcal{S} \,  \dot{\bar{\theta}} - 8 \mathcal{J}_0 \mathcal{J}_3 r^2 \, \dot{\theta}  \ri \, .
\eeq
\label{eq:charges_PF2_CFT2}
\end{subequations}
The boundary conditions~\eqref{eq:bdy_cond_CFT2} at $t=0$ force $r(0)=\bar{r}(0)=0$, which imply $K_{\theta} = K_{\bar{\theta}}=0$.
Since these charges are conserved along the full motion, we conclude that they remain zero at any value of $t \in [0,1]$.
By direct inspecting eq.~\eqref{eq:charges_PF2_CFT2}, we conclude that this is only possible when $\dot{\theta}= \dot{\bar{\theta}}=0$, or equivalently $\theta, \bar{\theta}$ are constant along the circuit.
Plugging this result inside the cost function~\eqref{eq:cost_PF2_CFT2}, we finally get 
\beq
\mathcal{L}_{\rm PF ,D \bar{D},iso} = 4 \left[ \frac{\mathcal{J}_3 \dot{r}^2}{\le 1-r^2  \ri^2} + \frac{\bar{\mathcal{J}}_3 \dot{\bar{r}}^2}{\le 1-\bar{r}^2  
\ri^2}   \right] \, .
\label{eq:final_cost_PF2_CFT2}
\eeq
First of all, this expression is manifestly positive definite, allowing us to interpret the length of geodesics in this space as appropriate measures of complexity.
We will come back to this point in the next subsection, when discussing more general choices of penalty factors. 
Since the previous line element is the sum of two rescaled FS cost functions for the two one-dimensional CFTs, we immediately find that the complexity is
\beq
\mathcal{C}_{\rm PF ,D \bar{D},iso} = \sqrt{\mathcal{J}_3 \mathcal{C}^2_{\rm FS} + \bar{\mathcal{J}}_3 \bar{\mathcal{C}}^2_{\rm FS} } \, ,
\label{eq:complexity_PF2_CFT2}
\eeq
where $\mathcal{C}_{\rm FS} (\bar{\mathcal{C}}_{\rm FS})$ is the FS cost function complexity of the holomorphic (anti-holomorphic) sector.
As we anticipated, this expression is the two-dimensional generalization of the result obtained in eq.~\eqref{eq:complexityConservedQtty}, and the above manipulations parallel the procedure outlined in section~\ref{ssec:analytic_1d}.

In conclusion, we notice that as long as the cost function in the two copies of CFT$_1$ is isotropic along the coset directions ($\mathcal{J}_2 = \bar{\mathcal{J}}_2 = 0$), the state complexity is independent of the penalties $\mathcal{J}_1$ and $\bar{\mathcal{J}}_1$, while $\mathcal{J}_3, \bar{\mathcal{J}}_3$ only act as multiplicative rescalings of the FS cost function in each sector.
More interestingly, we notice that the cost function~\eqref{eq:final_cost_PF2_CFT2} and the complexity~\eqref{eq:complexity_PF2_CFT2} are also independent of the penalty factor $\mathcal{J}_0$ that couples the holomorphic and anti-holomorphic sectors.
In our numerical results, we will see that a small non-trivial dependence on $\mathcal{J}_0$ occurs for non-vanishing penalty factors $\mathcal{J}_2, \bar{\mathcal{J}}_2$.

\textit{En passant}, we notice that the case of trivial penalty factors $\mathcal{J}_1=\bar{\mathcal{J}}_1=\mathcal{J}_3 = \bar{\mathcal{J}}_3=1$ and $\mathcal{J}_2 = \bar{\mathcal{J}}_2 = \mathcal{J}_0=0$ can be recovered as a special case of the above computation.
As mentioned below eq.~\eqref{eq:naive_CFT2}, in this latter case the geodesics factorize and the complexity is the quadratic sum of the lengths of the separate trajectories.

\subsection{Coupling \texorpdfstring{$(D, \bar D)$}{(D,\bar D)}: perturbative expansion}
\label{ssec:2d_perturbative}

When $\mathcal{J}_2, \bar{\mathcal{J}}_2 \ne 0$, we could not find a general analytic expression for the geodesics on the space of states.
Nevertheless, we were able to make progress by performing a perturbative expansion around the solution with trivial penalty factors associated with the FS cost function.
To this aim, let us take the following ansatz for the penalty factors and the generic trajectory over the space of states:
\begin{subequations}
\beq
\mathcal{J}_i = \mathcal{J}_{i}^{\rm FS} + \varepsilon \beta_i \, , \qquad
\bar{\mathcal{J}}_i = \bar{\mathcal{J}}_{i}^{\rm FS} + \varepsilon \bar{\beta}_i \, , \qquad
\mathcal{J}_0 = \varepsilon \beta_0 \, ,
\label{eq:ansatz_perturbative_penalties_CFT2}
\eeq
\beq
X^\mu(t) =
  X^\mu_0(t) + \varepsilon X^\mu_1(t) + \varepsilon^2 X^\mu_2(t) +
  \mathcal{O}(\varepsilon^3)~, 
  \eeq
  \beq
  \bar{X}^\mu(t) =
  \bar{X}^\mu_0(t) + \varepsilon \bar{X}^\mu_1(t) + \varepsilon^2 \bar{X}^\mu_2(t) +
  \mathcal{O}(\varepsilon^3)~,
\eeq
\label{eq:ansatz_traj_CFT2}
\end{subequations}
where $\mathcal{J}_i^{\rm FS} = \bar{\mathcal{J}}_i^{\rm FS} = \lbrace 1, 0, 1 \rbrace$, and $X_0^{\mu} = (\lambda_0, \lambda_0^*)$ is the geodesic in eq.~\eqref{eq:fsSol} (the same conventions apply to the anti-holomorphic sector).
For comparison, this ansatz generalizes the one-dimensional expression in eq.~\eqref{eq:pert_expansion_CFT1}.

We truncate the series expansion of the trajectories at second order around $\varepsilon=0$, assuming that $\varepsilon$ is small enough.
Consequently, the ansatz~\eqref{eq:ansatz_traj_CFT2} implies that $\mathcal{J}_0$ is small. 
By directly plugging the above expansion inside the inequality~\eqref{eq:constraints2_pen_CFT2}, we find that the constraint to get a positive-definite metric is satisfied as long as we take 
\beq
\beta_0^2 - 4 \le  \beta_1 - \beta_3 \ri  \le \bar{\beta}_1 - \bar{\beta}_3  \ri  < 0 \, .
\label{eq:constraint_pert_pen_CFT2}
\eeq
In other words, when $\mathcal{J}_0$ is taken to be small, the gap between the penalties $\mathcal{J}_1, \mathcal{J}_3$ should be comparable in size (and not smaller). For instance, we cannot select $\beta_1-\beta_3=\bar\beta_1-\bar\beta_3=0$.
Working under this assumption, we perturbatively determined the geodesics on the Hilbert space and we computed their length using Wolfram Mathematica.
The state complexity with boundary conditions~\eqref{eq:bdy_cond_CFT2} is given by
\begin{subequations}
    \beq
    \mathcal{C} = \sqrt{\mathcal{C}_{\rm pert.}^2 + \bar{\mathcal{C}}_{\rm pert.}^2} + \mathcal{O}(\varepsilon^3) \, , 
    \eeq
    \beq
    \begin{aligned}
   \frac{\mathcal{C}_{\rm pert.}}{\mathcal{C}_{\rm FS}} & =  1 + \varepsilon \left[ \beta_3 - 2 \beta_2 \cos(2
  \theta_\tar) \right]  \\
  &  - \varepsilon^2 \left[ \dfrac{\left(\beta_3 - 2 \beta_2 \cos(2
    \theta_\tar)\right)^2}{8} - \dfrac{2 \beta_2^2  \sin(2 \theta_\tar)^2}{r_\tar}\left(
      (1
  + r_\tar^2) \,\arctanh(r_\tar) - r_\tar \right) \right] ~,
  \end{aligned}
\eeq
\end{subequations}
where $\mathcal{C}_{\rm FS}$ is the FS cost function, and $\bar{\mathcal{C}}_{\rm pert.}$ is obtained by exchanging $(r_{\rm T}, \theta_{\rm T})  \leftrightarrow (\bar{r}_{\rm T}, \bar{\theta}_{\rm T})$ and 
$\{\beta_1,\beta_2,\beta_3\} \rightarrow \{\bar{\beta}_1,\bar{\beta}_2,\bar{\beta}_3\}$ 
in the definition of $\mathcal{C}_{\rm pert.}$.
Surprisingly, we found that the complexity is still unaffected by the coupling $\mathcal{J}_0$ between the left and right copies of the CFT, instead it is simply given by the quadratic sum of the perturbative result obtained for one-dimensional CFTs in eq.~\eqref{eq:complexity-perturbation-theory}. 

In the next subsection, we will show that the complexity depends on the penalty factor $\mathcal{J}_0$.
The reason we did not capture a dependence on this penalty is that it enters at higher order in the perturbation theory (likely at third order).

\subsection{Coupling \texorpdfstring{$(D, \bar D)$}{(D,\bar D)}: numerical results}
\label{ssec:2d_numerical}

We perform a numerical analysis of the geodesics associated with the cost function~\eqref{eq:cost_PF_CFT2_general} by solving the associated Euler-Lagrange equations with a shooting method.
We then plug back the solutions inside the cost function, and compare the result to the decoupled case where $\mathcal{J}_0=0$.
In other words, we study the relative complexity
\beq
\mathcal{C}_{\rm rel} = \frac{\mathcal{C} - \mathcal{C}(\J_0 = 0)}{\mathcal{C}(\J_0 = 0)} . 
\label{eq:relative_complexity}
\eeq
Note that we are normalizing the complexity with respect to its value with $\mathcal{J}_0=0$ but where the other penalties are not necesarilly trivial. This is done in order to isolate the effect of the coupling between the two different copies, from the effect of penalties within each copy. 
Since the numerical analysis for two-dimensional CFTs is technically hard and resource-consuming, we focus on regions of the space of states where either $|r|, |\bar{r}| \leq 0.8$ (in case 1), or $|r|, |\bar{r}| \leq 0.6$ (in case 2).
We have checked for several sample target states that the behavior of the complexity in regions closer to the boundary ($|r|=|\bar{r}|=1$) is qualitatively similar to the plots reported below.

\subsubsection{Case 1}

\begin{figure}[ht]
    \centering
    \includegraphics[width=1\linewidth]{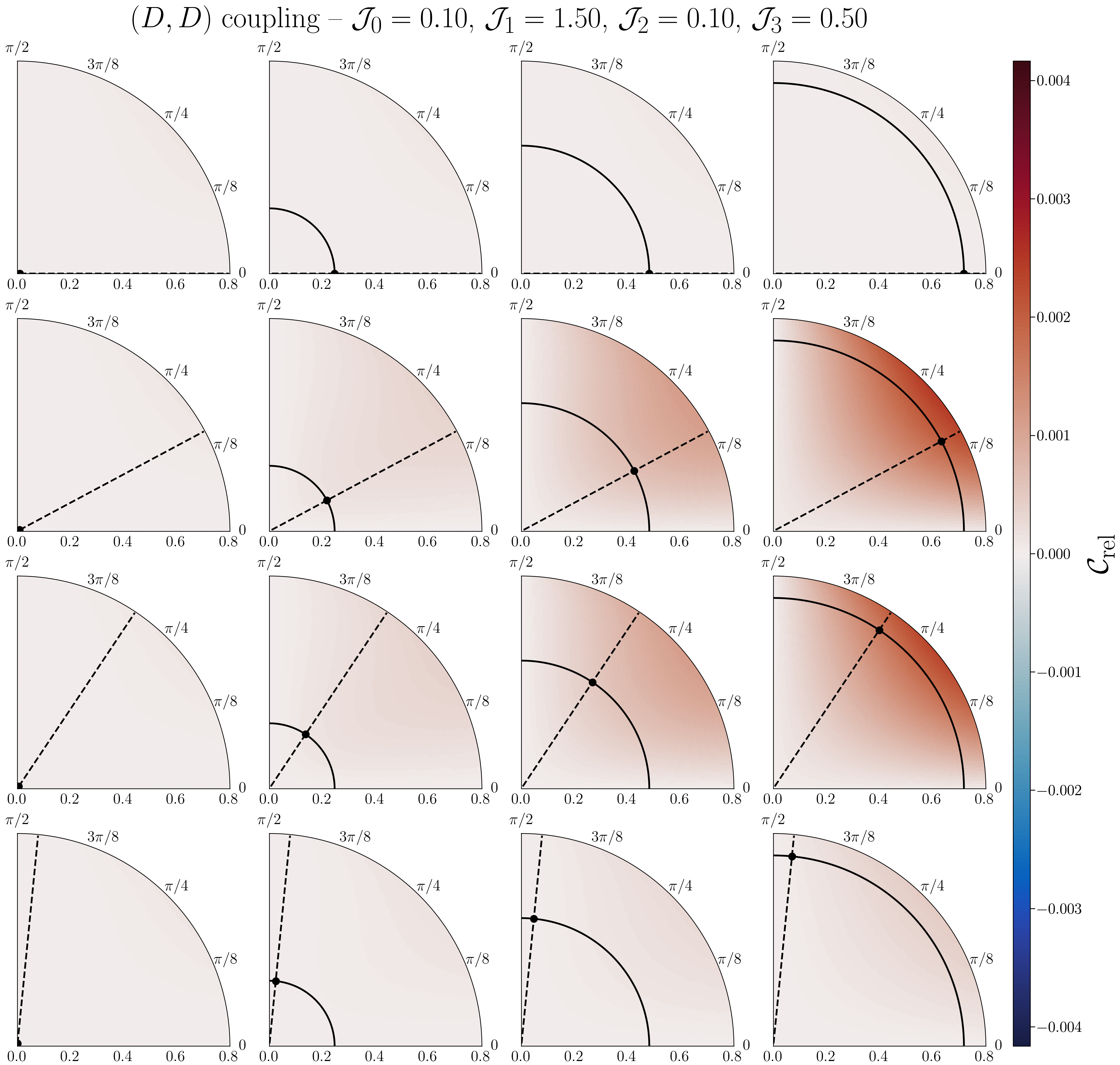}
    \caption{Relative complexity~\eqref{eq:relative_complexity} for a coupling $(D,D)$ between the two copies of one-dimensional CFTs for the penalties in case 1, see eq.~\eqref{eq:case1_penalties}.
    The black dot denotes the position of the right-copy target state, while each point in the quadrant indicates the position of the left-copy target state. The value of complexity is indicated by the color according to the color bar on the left. A small missing area appears around the origin since we have considered $r_{\rm T}\geq 0.01$ to improve numerical stability.}
    \label{fig:plotDD_2dCFT}
\end{figure}

We consider the set of penalty factors denoted with case 1 in eq.~\eqref{eq:case1_penalties}, where the coupling $\mathcal{J}_0$ between the left and right copies is small, and there is a gap between $\mathcal{J}_1, \mathcal{J}_3$ (and the corresponding penalties in the anti-holomorphic copy).
This setting is close to the perturbative regime investigated in subsection~\ref{ssec:2d_perturbative}.
The relative complexity is plotted in fig.~\ref{fig:plotDD_2dCFT}.

First of all, we find a non-trivial dependence on the penalty factor $\mathcal{J}_0$, since the relative complexity is non-vanishing.
The order of magnitude is $\mathcal{C}_{\rm rel} \sim 10^{-3}$, compared to a coupling between the two sectors of order $\mathcal{J}_0 = 0.1$.
This dependence was not captured by our previous perturbative expansion, which only went to order $\varepsilon^2$.

Second, we observe that the relative complexity is always positive, meaning that the existence of a non-vanishing coupling between the left and right copies of the two-dimensional CFT makes it harder to move along the space of states.
Naively, this result seems surprising because we would expect that the inclusion of an additional coupling $\mathcal{J}_0$ does not affect the cost of the optimal trajectory within each copy and that one could always construct a trajectory by moving first in one copy and then in the other in such a way that the coupling of the two copies does not matter.
However, this naive expectation is incorrect since 
the projection from the group manifold to the coset space involves a non-trivial minimization which ties the coefficients corresponding to different generators in the Hamiltonian. 
As a result, we notice that setting $\dot{\bar{\lambda}}= \dot{\bar{\lambda}}^*=0$ in eq.~\eqref{eq:cost_PF_CFT2_general}, there is still a residual dependence of the cost function on $\mathcal{J}_0$ in the first line.
At the numerator, there is a contribution of kind $\mathcal{J}_2 \mathcal{J}_0^2 \dot{\lambda}^2$ (and similarly for the complex conjugate term), which arises when the motion is purely along a single copy.
This observation motivates why the inclusion of the coupling $\mathcal{J}_0$ can make the motion in the space of states easier or harder, depending on the choice of penalty factors. For instance, we find here that the complexity increases, while we will show below that the opposite behavior happens in case 2. Of course, when $\mathcal{J}_2 =0$, the dependence of the final result for complexity on $\mathcal{J}_0$ disappears, as we already observed in subsection~\ref{ssec:isotropic_2dCFT}. When $\mathcal{J}_2 \neq 0$, the dependence on $\mathcal{J}_0$ enters at third order in perturbation theory around the solution with trivial penalties, as anticipated in subsection~\ref{ssec:2d_perturbative}. 

Finally, we notice that the complexity increases when both the left- and right-copy target states are located nearby $\theta = \bar{\theta} = \pi/4$, and for points closer to the boundary.
This behavior suggests that the maximal increase of complexity happens when we symmetrically couple the target states between the two copies, and when they are as far as possible from the reference state.

\subsubsection{Case 2}

\begin{figure}[ht]
    \centering
    \includegraphics[width=1\linewidth]{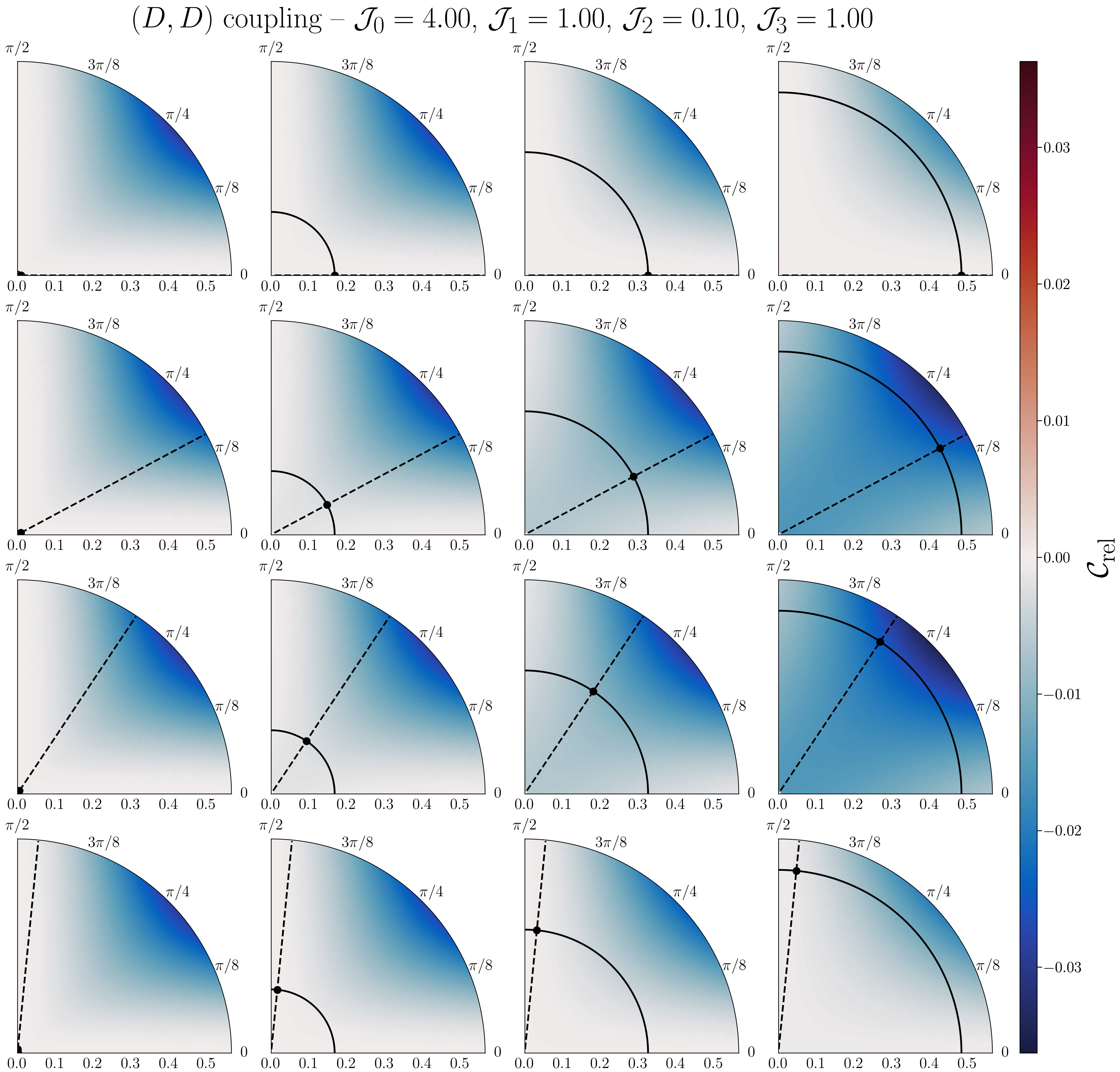}
    \caption{Relative complexity~\eqref{eq:relative_complexity} for a coupling $(D,D)$ between the two copies of one-dimensional CFTs for the penalties in case 2, see eq.~\eqref{eq:case2_penalties}.
    The black dot denotes the position of the right-copy target state, while each point in the quadrant indicates the position of the left-copy target state. The value of complexity is indicated by the color according to the color bar on the left. A small missing area appears around the origin since we have considered $r_{\rm T}\geq 0.01$ to improve numerical stability.}
    \label{fig:plotDD2_2dCFT}
\end{figure}

An alternative set of penalty factors such that the metric over the space of states is positive-definite, denoted with case 2 in eq.~\eqref{eq:case2_penalties}, consists of taking a large coupling $\mathcal{J}_0$ between the two copies, while at the same time $\mathcal{J}_1 = \mathcal{J}_3$ in each sector.
This setting is far from the perturbative regime considered in subsection~\ref{ssec:2d_perturbative}.
We depict the corresponding relative complexity in fig.~\ref{fig:plotDD2_2dCFT}.

First, we observe that the order of magnitude of $\mathcal{C}_{\rm rel} \sim 10^{-2}$ is larger than the case with smaller $\mathcal{J}_0$. 
This supports the idea that the contribution of this penalty factor to complexity arises at higher orders in the perturbative expansion~\eqref{ssec:2d_perturbative}.
Second, we find that the relative complexity in case 2 is always negative, therefore providing a setting where the inclusion of an additional coupling between the holomorphic and anti-holomorphic copies of the two-dimensional CFT allows for shorter trajectories over the geometry.
In analogy with case 1, the relative complexity increases (in absolute value) when the left- and right-copy target states are taken closer to the angular direction $\theta = \bar{\theta} = \pi /4$.

Finally, we find cases where the relative complexity is non-vanishing even when the state of one of the CFT copies is around the origin. This is apparent in the left-most (and right-most) plots in the second and third lines of fig.~\ref{fig:plotDD2_2dCFT}.
To explain this behavior, we observe that the cost function~\eqref{eq:cost_PF_CFT2_general} depends non-trivially on $\mathcal{J}_0$ even when we set $\dot{\bar{\lambda}} = \dot{\bar{\lambda}}^*=\bar{\lambda} = \bar{\lambda}^*=0$. This is again due to the non-intuitive behavior of the Riemannian submersion procedure. Even though the state within the second copy is not changing, we are forced to use its stabilizer direction in order to set to zero the relevant conserved charge.

\subsection{Other couplings between one-dimensional CFTs}
\label{ssec:comments_other_coupling}

In subsections~\ref{ssec:2d_analytic}--\ref{ssec:2d_numerical}, we investigated  Nielsen's state complexity of two-dimensional CFTs associated with a cost function of the form~\eqref{eq:Lagrangian_CFT2}, where $\mathcal{Q}=D, \, \bar{\mathcal{Q}}=\bar{D}$.
In this setting, we could obtain both analytical and numerical results.
There are other choices of the couplings between the left and right copies of one-dimensional CFTs where we can have numerical control over the complexity.
Let us focus on the following two possibilities:
\begin{itemize}
    \item A coupling between $D$ and the linear combination $\bar{P}+\bar{K}$, with Lagrangian
    \beq
\mathcal{L}_{\rm D, \bar{P}+ \bar{K}} =  \mathcal{L}_{\mathrm{CFT}_1} +
\bar{\mathcal{L}}_{\mathrm{CFT}_1}  + \mathcal{J}_0 \langle H, D \rangle  \langle \bar{H}, \bar{P} + \bar{K} \rangle \, .
\label{eq:Lagrangian_DPK}
    \eeq
    \item A coupling between the symmetric combination of $(P, \bar{K})$ and $(K, \bar{P})$, namely
    \beq
\mathcal{L}_{\rm (P, \bar{K})+(K, \bar{P})} =  \mathcal{L}_{\mathrm{CFT}_1} +
\bar{\mathcal{L}}_{\mathrm{CFT}_1}  +  \mathcal{J}_0 \le \langle H, P \rangle  \langle \bar{H}, \bar{K}  \rangle + \langle H, K \rangle \langle \bar{H}, \bar{P} \rangle \ri \, .
\label{eq:Lagrangian_PKKP}
    \eeq
\end{itemize}
Furthermore, we will assume for simplicity that the Lagrangian $\mathcal{L}_{\rm CFT}$ of the left-copy CFT is given by eq.~\eqref{eq:pf-0-metric} with trivial penalty factors $\mathcal{J}_1=\mathcal{J}_3=1, \, \mathcal{J}_2 =0$ (and similarly for the right-copy CFT).

Given the above cost functions, we perform a projection over the coset space as outlined at the beginning of subsection~\ref{ssec:2d_analytic}.
After employing this procedure, the infinitesimal metric induced from any of the above cost functions to the space of states reads
\beq
\dd s^2_{\rm coupled} =  \dd s^2_{\rm coupled} (\mathcal{J}_0=0)
+ \dd s^2_{\rm int.} \, .
\label{eq:general_2d_metric}
\eeq
When the penalty factors in each copy are trivial, the line element $\dd s^2_{\rm coupled} (\mathcal{J}_0=0)$ is the FS expression in eq.~\eqref{eq:fs-metric-1d}, summed over two copies of a one-dimensional CFT.
The interacting contribution $\dd s^2_{\rm int.}$ contains information about the coupling between the two copies.

\subsubsection{Coupling $(D, \bar{P}+\bar{K})$}
\label{ssec:coupling_DPK}

In the case of the cost function~\eqref{eq:Lagrangian_DPK}, the interacting contribution reads
\begin{align}
\begin{split}
  \dd s^2_{\mathrm{int.}} & = \frac{\J_0}{\mathcal R} \biggl[\J_0 \frac{\bigl(
    1 + |\lambda|^2 \bigr)^2}{\bigl( 1 - |\lb|^2 \bigr)^2} \dd \mathcal P^2 + 4
    i
    \dd \mathcal P \dd \mathcal Q + 4 \J_0 \frac{\bigl( \lb - \lbs
  \bigr)^2}{\bigl( 1 - |\lambda|^2 \bigr)} \dd \mathcal Q^2\biggr]~,
  \label{eq:interacting_metric_CFT2_DPK}
  \\
  \dd \mathcal P & \equiv \Bigl[(\lbs)^2 - 1\Bigr] \dd \lb + \Bigl[(\lb)^2 - 1\Bigr]
  \dd \lbs~, \quad \dd \mathcal Q = \lls \dd \lambda - \lambda \dd \lls~,\\
  \mathcal R & \equiv \Bigl( 1 - |\lambda|^2 \Bigr)^2 \Bigl( 1 - |\lb|^2 \Bigr)^2 +
  \J_0^2 \Bigl( 1 + |\lambda|^2 \Bigr)^2 \Bigl( \lb - \lbs \Bigr)^2~.
\end{split}
\end{align}
Let us analyze the signature of this metric over the space of states.
To this aim, it is convenient to perform the change of variables from the two complex coordinates $(\lambda, \bar{\lambda})$ to the four real coordinates $(\lambda_R, \lambda_I, \bar{\lambda}_R, \bar{\lambda}_I)$, parametrized by 
\beq
\lambda = \lambda_R + i \lambda_I \, , \qquad
\lambda^* = \lambda_R - i \lambda_I \, , \qquad
\bar{\lambda} = \bar{\lambda}_R + i \bar{\lambda}_I \, , \qquad
\bar{\lambda}^* = \bar{\lambda}_R - i \bar{\lambda}_I \, .
\eeq
After converting the metric~\eqref{eq:general_2d_metric} including the interacting part~\eqref{eq:interacting_metric_CFT2_DPK} to this new coordinate system, we compute all the principal minors of the metric.
One can show that the metric is always positive-definite at the origin, since the determinant of the four principal minors (evaluated at $\lambda=\bar{\lambda}=0$) are
\beq
256 \le \mathcal{J}_0^2 +1 \ri \, , \qquad
64 \le \mathcal{J}_0^2 +1 \ri \, , \qquad
16 \, , \qquad
4 \, .
\eeq
which are strictly positive for any choice of penalties.
At the same time, we can easily argue that the signature of the metric is not constant through all the space of states. 
To show this, let us compute the ratio between the determinant of the full metric $g_{\mu\nu}$ and its top-left principal minor $g_{\mu\nu}^{(3)}$ of dimension $3 \times 3$, which reads 
\beq
\mathfrak{R} =  \frac{\det \le g_{\mu\nu} \ri }{\det \le g_{\mu\nu}^{(3)} \ri} = 
\frac{4 \le \mathcal{J}_0^2 +1 \ri}{\le  \bar{\lambda}_R^2 + \bar{\lambda}_I^2 -1  \ri^2 + \mathcal{J}_0^2 \left[ \bar{\lambda}_I^4 + \le \bar{\lambda}_R^2 -1 \ri^2 -2 \bar{\lambda}_I^2 \le \bar{\lambda}_R^2 +1 \ri  \right]  } \, .
\eeq
To get a positive-definite metric, this expression needs to be positive
over all the Poincar\'{e} disk, parametrized by $\lambda_R^2 + \lambda_I^2 \leq 1$ (and the same for the anti-holomorphic copy).
However, we can easily show that this expression changes sign across the space of states by studying two limiting cases:
\beq
\mathfrak{R}|_{\bar{\lambda}_R = \bar{\lambda}_I = 0 } = 4 >0  \, , \qquad
\mathfrak{R}|_{\bar{\lambda}_R = \bar{\lambda}_I = \frac{1}{\sqrt{2}} } = - \frac{4 \le \mathcal{J}_0^2 +1 \ri}{\mathcal{J}_0^2} <0 \, .
\eeq
For this reason, we cannot interpret the line element with the interacting part~\eqref{eq:interacting_metric_CFT2_DPK} as a valid notion of complexity geometry over all the space of states.
Nonetheless, following the perspective advocated around eq.~\eqref{eq:pfIpm0condition}, we can still define a meaningful notion of complexity by restricting the geometry to the region around the origin where the metric is positive-definite.
The locus of points where this condition stops being satisfied defines a non-trivial boundary in the Poincar\'{e} disk.
We study the Nielsen complexity corresponding to this scenario in appendix~\ref{app:ssec:coupling_DPK}.

\subsubsection{Coupling \texorpdfstring{$(P,\bar K) + (K,\bar P)$}{(P,\bar K) + (K,\bar P)}}
\label{ssec:coupling_PK}

Then interacting part of the line element corresponding to the coupling in eq.~\eqref{eq:Lagrangian_PKKP} reads
\begin{equation}
  \begin{split}
   \label{eq:interacting_metric_CFT2_PK}
    \dd s^2_{\mathrm{int.}} & = -\mathcal K^{-1} \left[  \dd \mathcal P^2 + \dd
      \mathcal Q^2 + 4 \J_0 \bigl(1-|\lambda|^2\bigr) \bigl(1-|\bar{\lambda}|^2\bigr)
      \right.
      \\
      & \left.   \quad \times \left(\mathcal A(\lambda,\bar{\lambda}) \dd \lambda \dd \lb
        +
        \mathcal A(\lambda,\bar{\lambda})^* \dd \lls \dd \lbs + \mathcal B(\lambda,\bar{\lambda})
        \dd \lls
    \dd \lb + \mathcal B(\lambda,\bar{\lambda})^* \dd \lambda \dd \lbs \right)\right]~,
    \\
    \dd \mathcal P & \equiv 2 \J_0 \bigl(1 - |\lb|^2 \bigr) \left[ \bigl( (\lls)^2
      \lb + \lbs \bigr) \dd \lambda - \bigl( \lb + \lambda^2 \lbs \bigr) \dd
      \lls
    \right]~, \quad \dd \mathcal Q \equiv \dd \mathcal P(\lambda \leftrightarrow
    \lb)~,\\
    \mathcal A(\lambda,\lb) & \equiv 4 \J_0^2 \lls \lbs \bigl( \lls \lb + \lambda
    \lbs \bigr) + \bigl(1 - |\lambda|^2 \bigr) \bigl(1 - |\lb|^2 \bigr) \bigl(
    (\lls)^2 + (\lbs)^2 \bigr)~,
    \\
    \mathcal B(\lambda,\lb) & \equiv 4 \J_0^2 \lambda \lbs \bigl( \lls \lb + \lambda
    \lbs \bigr) - \bigl(1 - |\lambda|^2 \bigr) \bigl(1 - |\lb|^2 \bigr) \bigl(
      1 +
    \lambda^2 (\lbs)^2 \bigr)~,\\
    \mathcal K & \equiv \bigl( 1 - |\lambda|^2 \bigr)^4 \bigl( 1 - |\lb|^2 \bigr)^4
    - 4 \J_0^2 \bigl( 1 - |\lambda|^2 \bigr)^2 \bigl( 1 - |\lb|^2 \bigr)^2
    \bigl(
    \lls \lb + \lambda \lbs \bigr)^2~.
  \end{split}
\end{equation}
The analysis of the positivity of this metric is much harder compared to the case studied in subsection~\ref{ssec:coupling_DPK}.
However, one can similarly show that it is possible to choose $\mathcal{J}_0$ such that the metric is positive-definite at the origin, and that there is no choice of $\mathcal{J}_0$ that ensures the positivity of the metric over all the space of states.
The four principal minors at the origin $\lambda = \bar{\lambda}=0$ have the following determinants
\beq
256 \le \mathcal{J}_0^2 -1 \ri^2 \, , \qquad
-64 \le  \mathcal{J}_0^2 -1 \ri \, , \qquad
16 \, , \qquad
4 \, .
\eeq
Therefore, the metric is positive-definite at the origin when $\mathcal{J}_0 <1$.
Taking the alternative perspective advocated below eq.~\eqref{eq:pfIpm0condition}, we assume that the interpretation of the manifold as a complexity geometry stops holding beyond a non-trivial boundary in the space of states where the metric changes signature. 
We perform a numerical investigation of Nielsen's complexity with this metric in the region where the metric is positive-definite in appendix~\ref{app:ssec:coupling_PK}.

\section{Discussion}
\label{sec:discussion}

In this work, we developed a systematic procedure to obtain a complexity metric over the coset space induced from the cost function of a theory invariant under a generic Lie group.
This includes the case of the conformal group $\mathrm{SO}(d,2)$, relevant for the investigation of complexity in \acp{cft}.
We have shown that any right-invariant norm on the group manifold can be expressed as
\beq
\mathcal{F}_{\mathcal{I}}^2 = \le \mathcal{F}_{\mathcal{I}}^{\mathrm{state}} \ri^2
+ \tilde{\mathcal{I}}_{ab} f_a f_b \, ,
\label{eq:projection_conclusions}
\eeq
where $\mathcal{F}_{\mathcal{I}}$ is the cost function on the Lie group,
$\mathcal{F}_{\mathcal{I}}^{\mathrm{state}}$ the induced metric on the coset space,
$\tilde{\mathcal{I}}_{ab}$ a certain symmetric matrix which depends on
the penalty factors, and $f_a$ a vector which depends (among other things) on the variations in the stabilizer coordinates which do not modify the state. For more details,
see section~\ref{ssec:general_projection}.

We have shown that the map $f_a=0$ defines a (pseudo-)Riemannian submersion from the Lie group to the coset space, generalizing the technology developed in
\cite{Auzzi:2020idm} for the unitary group.
We further demonstrated that our procedure of setting $f_a=0$ is equivalent to an extremization of the cost function over the group manifold, thus generalizing the minimization prescription advocated in \cite{Brown:2019whu} for the unitary case. 
When the stabilizer is abelian, we demonstrated that the quantity $f_a$ can be related to a conserved charge $K_a$ for the cost function $\mathcal{F}_{\mathcal{I}}^2$. The precise identification is given in eq.~\eqref{eq:conserved_charge}, showing that the vanishing of the function $f_a$ is equivalent to the vanishing of the conserved charge.  Finally, we commented on the relation with coadjoint orbits which was explored in \cite{Chagnet:2021uvi} for the case without penalties.

As a first application of our method, we studied the state complexity in the
presence of penalty factors for one- and two-dimensional \acp{cft}.
We followed three approaches. First, for special choices of the penalty factors, we computed the exact geodesics of the metric and their length.
Second, we performed a perturbative expansion around the \acl{fs} case
(which corresponds to the case with trivial penalty factors) to gain some understanding of the general trend of complexity.
Finally, we numerically solved the Euler-Lagrange equations associated with the metric.
The results are collected in Table~\ref{tab:results}.

In one-dimensional CFTs, the state complexity presented only a mild dependence on the penalty factor along the direction of the dilatation operator. This mild dependence was only present for anisotropic penalty factors for the generators  $L_{\pm}$.
More generally, penalizing the $L_+$ ($L_-$) directions (\ie associating them with larger penalties) led to larger costs of moving along the real (imaginary) directions and hence the complexity showed signs of favoring alternative trajectories involving the cheaper generator combined with dilatations. 

For two-dimensional CFTs, we mainly focused on the setting where two copies of one-dimensional CFTs were coupled via their dilatation control functions. The dependence on this dilatation-dilatation coupling  $\mathcal{J}_0$ was again mild, similarly to the dependence on the dilatation penalty in one-dimensional CFTs, and could only be observed for anisotropic penalties along the directions $(L_+,L_-)$ in each copy.  This mild dependence was only apparent in the numerical studies. We numerically investigated two regimes,  where $\mathcal{J}_0$ was taken to be small (large).
In the former case, we found that the  state complexity increased compared to the case with vanishing $\mathcal{J}_0$; in the latter case, the state complexity decreased compared to the case with vanishing $\mathcal{J}_0$.
This shows that the projection over the coset space induces non-trivial and counterintuitive effects.

An important point of our analysis is that the metric over the space of states  is not guaranteed to be positive-definite, which we believe is an essential input to interpret the geodesic's length as a measure of complexity.
We proposed two possible approaches to overcome this issue. 
The first approach, that we employed in the bulk of the paper, is to constrain the set of penalty factors such that the cost function is positive-definite across all the space of states.
The second approach, that we explored in appendix~\ref{app:sec:cost_nontrivial_bdy}, is to determine a boundary in the Hilbert space where the metric ceases to be positive-definite; consequently, we restrict the target states to live in the region delimited by this boundary. It would be interesting to develop a physical intuition for these different choices. We will comment on this further below.

\vskip 2mm

While this work provides a step forward towards defining complexity in \ac{cft}
states, there are still several future directions that are left open. We list those below.

\textbf{Choice of penalties.} We have seen that the procedure presented in this paper, inspired by the ideas of Riemannian submersions, imposes constraints on the penalty factors, see \eg equation \eqref{eq:pfIpm0condition} in order for the metric to be positive definite.
It would be interesting to interpret these constraints from the CFT point of view. We know that unitarity bounds impose constraints on CFT data, including the dimensions of operators, and it would be interesting to see if similar physical considerations can be imposed on the values of the penalty factors in our complexity construction.
    
\textbf{Generalization to higher dimensions.}
    The general procedure resulting in the decomposition
    \eqref{eq:projection_conclusions} is valid for any Lie group, including
    $\mathrm{SO}(d,2)$ for general dimension $d$.
    Therefore, a natural continuation of this work would be to study state
    complexity for higher-dimensional ($d \geq 3$) \acp{cft} in the presence of non-trivial
    penalty
    factors (the case with an isotropic cost function was considered in
    \cite{Chagnet:2021uvi}).
    The main obstacle to this generalization is technical. Since the
    dimensionality of the space increases, solving the
    Euler-Lagrange equations associated with the cost function becomes more difficult.
    A first step in this direction could be to consider special choices of the penalty factors where analytic or perturbative results are accessible, similar to the cases in section~\ref{ssec:interlude:simple_metric}.

\textbf{Relation to holography.}
    We discuss a relation between Nielsen's complexity in CFT and holography in appendix~\ref{app:holographic_interpretation} (for alternative approaches, \eg see Refs.~\cite{Caputa:2018kdj,Flory:2020dja,Flory:2020eot,Erdmenger:2020sup,Chandra:2021kdv,Erdmenger:2021wzc,Erdmenger:2022lov,Chandra:2022pgl,Erdmenger:2024xmj}).
    The central observation is that the phase space described by CFT coherent states coincides with the one of a massive particle in \ac{ads} spacetime. It was already shown in \cite{Chagnet:2021uvi} that the \ac{fs} metric (without penalties) can be mapped to the bulk symplectic form for this particle defined using its position and momentum.
    In the appendix, 
    we show that the condition $f_a=0$ to project the metric from the unitary manifold to the coset space also has a natural interpretation in terms of setting to zero the associated symplectic potential.
    So far, we only performed this matching in the case of an isotropic cost function. We plan to generalize this mechanism to arbitrary penalty factors in the future, by exploiting this geometric relation.

\textbf{Negative curvature.}
    One of the reasons to define a right-invariant metric on the Lie group (as opposed to a bi-invariant one) is that the resulting geometry admits regions with negative curvature.
    References \cite{Brown:2016wib,Brown:2017jil} proposed that this requirement is necessary to describe the time evolution induced by a chaotic Hamiltonian, since negative curvature implies that nearby geodesics deviate from each other.
    Furthermore, the same authors also argued that negative curvature, together with an appropriate scaling of the typical sectional curvatures, was necessary for the Nielsen complexity to manifest a switchback effect.
    Notice that one can always find regions with negative curvature in the complexity geometry if the commutator structure of the algebra satisfies
    $[easy \, , easy ]= \, hard$, where \textit{easy} and \textit{hard} refer to a smaller or larger penalty factor associated with the given generator in the cost function, respectively
    \cite{Brown:2019whu}.
    It would be interesting to investigate whether the curvature of the penalized complexity geometry associated with the conformal group, extended along the lines of \cite{Erdmenger:2024xmj} to allow for motion between different  Verma modules, can be used to diagnose chaotic/integrable properties of CFT Hamiltonians. 
The question of diagnosing quantum chaos using complexity has also been addressed from the perspective of  Krylov/spread complexity recently (see, \eg the reviews \cite{Nandy:2024evd,Baiguera:2025dkc,Rabinovici:2025otw} and references therein).

\textbf{Generalization to Virasoro.}
In this work, we focused on the global conformal group and studied specific examples in dimensions $d=1,2$.
However, the symmetry groups of one and two-dimensional CFTs admit an infinite-dimensional extension provided by the Virasoro group.
The analysis of Nielsen complexity in this case was initiated in \cite{Caputa:2018kdj}, and then further pursued in \cite{Erdmenger:2020sup,Flory:2020eot,Flory:2020dja,Erdmenger:2021wzc,Erdmenger:2022lov,deBoer:2023lrd,Erdmenger:2024xmj}.
In the above-mentioned works, the authors mainly focused on studying the FS metric, which assigns vanishing cost to the generators of the stabilizer of a state. 
It would be interesting to apply the techniques developed in this paper to study Nielsen complexity for a general choice of penalty factors (this problem was initially mentioned in the appendix of \cite{Caputa:2018kdj}).
In this context, the insights provided by Ref.~\cite{Erdmenger:2024xmj} will also allow to move between different conformal families, thus allowing for a complete exploration of the conformal group.

\textbf{Relation to binding complexity.}
Binding complexity \cite{Balasubramanian:2018hsu} is a notion of complexity tailored to systems composed of multiple subsystems. It assigns a significantly higher cost to non-local operations acting between different subsystems, in contrast to local operations within each subsystem.
This framework can be used to model \textit{distributed quantum computation}~\cite{10.1145/1324177.1324179,Beals_2013,Caleffi:2022wxp}, where a computational task is shared between small quantum computers or nodes.
Operations within each node are considered more coherent, while those between different nodes are given lower priority.
A schematic (albeit hand-wavy) analogy can be drawn to our setting of complexity in CFT$_2$. 
We may consider a regime in which gates acting within each CFT copy are much cheaper than those connecting the two copies. It would be interesting to explore this limit in our results and investigate whether it admits a meaningful physical interpretation.
Exact techniques developed in~\cite{Baiguera:2023bhm} might be adaptable to this setting. In particular, that work relates binding complexity to entanglement entropy, potentially offering a new route to connect complexity and entanglement in two-dimensional CFTs.

\textbf{Golden gates.} 
The mathematical community investigated the optimal way to cover a group manifold, \ie the specific choice of gates that reaches all its points within a certain tolerance with the least number of applications, see \eg~\cite{PARZANCHEVSKI2018869,sarnak2637letter}.
It would be interesting to see if these  ideas, referred to as the \textit{golden-gate problem}, can be adapted to our context, thus providing optimal sets of gates allowing to move within an entire Verma module of the conformal group.


\section*{Acknowledgments}
We are happy to thank Asaf Arzi, Adam Chapman and Andrea Legramandi for valuable comments. We are particularly grateful to Asaf Arzi for helping us understand the mathematics of Pseudo-Riemannian submersions.
The work of SB is supported by the INFN grant \textit{Gauge Theories and Strings (GAST)} via a research grant on \textit{Holographic dualities, quantum information and gravity}.
The work of SC and OS is supported by the Israel Science Foundation (grant No. 1417/21), by the German Research Foundation through a German-Israeli Project Cooperation (DIP) grant “Holography and the Swampland”, by Carole and Marcus Weinstein through the BGU Presidential Faculty Recruitment Fund, by the ISF Center of Excellence for theoretical high energy physics and by the ERC starting Grant dSHologQI (project number 101117338). NC is supported in part by the FOM program 167 (Strange Metals), by the Dutch Research Council (NWO) project 680-91-116 (Planckian Dissipation and Quantum Thermalisation: From Black Hole Answers to Strange Metal Questions.), and by the Dutch Research Council/Ministry of Education.

\appendix

\section{(Pseudo-)Riemannian submersions}
\label{app:submersions}

In this appendix, we briefly review the definition of (pseudo-)Riemannian
submersion and state several results that play a key role in projecting a metric from a Lie group to the coset space, as discussed in
section~\ref{ssec:projection_coset}.
We refer to the following textbooks for more details: section~3.5 of
\cite{petersen2006riemannian}, section 9 of \cite{Besse:1987pua} and
sections~7, 11 in \cite{oneill1983semiriemannian}.

\begin{deff}[Submersion]\label{Def1}
  Let $M, B$ be smooth manifolds, and $\pi : M \rightarrow B$ be a smooth map.
  We denote its differential as $d\pi: TM \rightarrow TB$, which for any $y \in M$ induces a linear map between the vector spaces $[d\pi]_y : T_{y} M \rightarrow T_{x}B$ with $x= \pi(y)$.
  The map $\pi$ is called a \emph{submersion} if $[d\pi]_y$ is surjective for all $y \in M$.
\end{deff}

\begin{deff}[Pseudo-Riemannian submersion]\label{Def2}
  Let $\pi : M \rightarrow B$ be a surjective submersion between
  pseudo-Riemannian manifolds with respective dimensions $m>b$.
  We define the \emph{vertical space} at a point $y \in M$ as $\mathcal{V}_y
  \equiv \ker(d \pi_y)$, and the \emph{horizontal space} $\mathcal{H}_y$ as its
  orthogonal complement inside $T_y M$.
  Then $\pi$ is called a \emph{pseudo-Riemannian submersion} if $[d\pi]_y$ maps
  $\mathcal{H}_y$ isometrically onto $T_x B$.
  In other words, we have
  \beq
  \langle X, Y \rangle = \langle [d\pi]_y X, [d\pi]_y Y  \rangle \, ,
  \qquad
  \forall \,\, X,Y \in \mathcal{H}_y \, ,
  \eeq
  where $\langle \cdot, \cdot \rangle$ denotes an inner product over $T_y M$ on
  the left-hand side, and the induced inner product over $T_x B$ on the
  right-hand side.
\end{deff}

\begin{figure}[ht]
  \centering
  \includegraphics[scale=1]{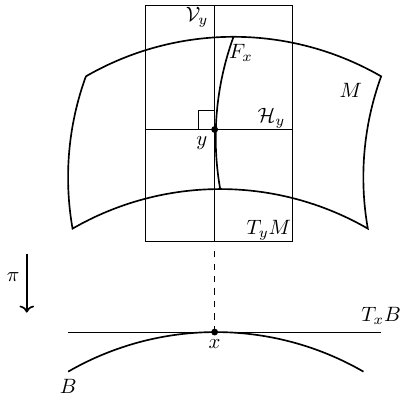}
  \caption{Pictorial representation of a submersion from the manifold $M$ to the subspace $B$. Picture inspired from \cite{Besse:1987pua,Auzzi:2020idm}.}
  \label{fig:app:submersion}
\end{figure}

Let us discuss the previous definitions. We refer the reader to
figure~\ref{fig:app:submersion} for  clarification.
Definition \ref{Def1} requires the map $d\pi$ to be surjective in order to define a submersion.
Due to the rank-nullity theorem, the map $d\pi_y$ has maximal rank, therefore its kernel $\ker(d\pi_y)$ at any point has dimension $f = m-b$.
In definition \ref{Def2}, the previous statements imply that the tangent space
decomposes in terms of vertical and horizontal spaces as
\beq
T_y M = \mathcal{V}_y \oplus \mathcal{H}_y \, .
\eeq
For the purposes of this work, we will identify $M$ as the group manifold on
which
a Lie group $G$ acts.
After introducing a Hilbert space and fixing a reference state, we will then
identify $B$ as the coset space obtained after quotienting $M$ with the
stabilizer group of the reference state.
In section \ref{ssec:projection_coset}, we use the following theorem (see, \eg theorem 9.12 in \cite{Besse:1987pua} for the Riemannian case, plus the definition 23 and lemma 24 of \cite{oneill1983semiriemannian} for the pseudo-Riemannian one):

\begin{theo}[Projection over coset space]
  \label{thm:induced_metric_submersion2}
  Let $M$ be a pseudo-Riemannian manifold with metric $g$ and $H$ a closed
  subgroup of the isometry group $G$ of $M$.
  Assume that the projection $\pi : M \rightarrow M/H$ is a smooth surjective
  submersion.
  Then there exists one and only one pseudo-Riemannian metric $\hat{g}$ on the
  coset space $M/H$ such that $\pi$ is a pseudo-Riemannian submersion.
\end{theo}

In other words, quotients of manifolds by the action of an
isometry group define uniquely a pseudo-Riemannian submersion and an induced
metric over the coset space.
Since we give a recipe in section~\ref{ssec:projection_coset} to determine these pseudo-Riemannian submersion, this theorem is crucial to demonstrate that the metric obtained by the procedure outlined there is unique.

\section{Fundamental representation of the conformal group}
\label{app:fund_CFT}

The global part of the $d$--dimensional conformal group is described by the
non-compact group $\mathrm{SO}(d,2)$.
In this work we build unitary circuits \eqref{eq:circuits-ansatz} in Lorentzian signature by using instead the generators of the Euclidean conformal algebra $\mathfrak{so}(d+1,1)$, and then imposing appropriate conjugation properties, \eg given by eq.~\eqref{eq:conjugation_so12gen} in the one-dimensional case.
This procedure is based on the map between the Euclidean and Lorentzian
conformal generators reviewed in appendix~A of reference
\cite{Chagnet:2021uvi} (see also \cite{Minwalla:1997ka,Luscher:1974ez} for more details); here we summarize the main ingredients required for this work.
The commutation relations satisfied by the generators of the Euclidean
conformal group read
\beq
\begin{aligned}
  & [D, P_{\mu}] = P_{\mu} \, , \qquad
  [L_{\mu\nu}, P_{\rho}] = \delta_{\nu\rho} P_{\mu} -
  \delta_{\mu\rho} P_{\nu} \, ,  \\
  & [D, K_{\mu}] = - K_{\mu} \, , \qquad
  [L_{\mu\nu}, K_{\rho}] =  \delta_{\nu\rho} K_{\mu} -
  \delta_{\mu\rho} K_{\nu} \, ,  \\
  & [K_{\mu}, P_{\nu}] = 2 \le \delta_{\mu\nu} D - L_{\mu\nu} \ri \, , \\
  & [L_{\mu\nu}, L_{\rho\sigma}] = - \delta_{\nu\sigma} L_{\mu\rho} +
  \delta_{\nu\rho} L_{\mu\sigma} - (\mu \leftrightarrow \nu) \, .
\end{aligned}
\eeq

\subsection{Inner product on a Lie group}
\label{app:inner_product_Lie}

Let us define an inner product on a generic Lie group, with the idea to then
specialize to the conformal case.
In general, a Lie algebra $\mathfrak{g}$ admits a non-degenerate bilinear
symmetric \textit{Killing} form $B$ defined as (see, \eg eq.~(4.8.24) in the lecture notes \cite{Tomasiello})
\beq
B(x,y) = \frac{1}{2} \Tr[\mathrm{ad}_x \circ \mathrm{ad}_y]~,\quad x,y \in
\mathfrak g~,
\eeq
where $\mathrm{ad}_x(z)=[x,z]$ denotes the adjoint operation of  the algebra, $\circ$ denotes the composition operation, and the factor of $1/2$ is chosen for convenience.
It is important to stress that in this formula, the trace does not necessarily refer to a finite-dimensional matrix representation, but is generally performed on the linear operator.
By decomposing two arbitrary elements on the Lie algebra in terms of the orthogonal Hermitian generators $\lbrace \omega_I \rbrace \in \mathfrak{g}$ as $X = X^I \omega_I, Y= Y^J \omega_J$, we obtain an explicit expression for the bilinear form
\beq
B(X,Y)  = \frac{1}{2} X^I Y^J \sum_{AB} f_{JA}^B f_{IB}^A~,
\label{eq:bilinear_form_Lie_group}
\eeq
where $f$ are the structure constants defined by $[\omega_A, \omega_B] =
f_{AB}^C \omega_C$.

In this paper, we will focus on the fundamental representation of the conformal
algebra $\mathfrak{so}(d,2)$ in terms of matrices in $\mathcal
M_{d+2,d+2}(\mathds C)$ spanned by 
\beq
(M_{AB})^C_{\,\,D} \equiv \delta_A^{\,\, C} \eta_{BD} - \delta_B^{\,\, C}
\eta_{AD} \, , \qquad
\eta = \mathrm{diag} \, \le -1, -1, 1, \dots, 1 \ri \, ,
\eeq
where capital Latin indices run over the range $\{-1,0,1,\ldots d\}$.
In this representation, the generators of the conformal group read (see also appendix~D of reference
\cite{Chagnet:2021uvi}):
\begin{equation}
\begin{split}
    R(D)&\equiv -i M_{-1,0}, \qquad
    R(L_{\mu\nu}) \equiv M_{\mu\nu}, \qquad\\
    R(P_\mu) &\equiv M_{-1,\mu}-iM_{0,\mu}, \qquad
    R(K_\mu) \equiv -(M_{-1,\mu}+iM_{0,\mu}),
\end{split}
\end{equation}
where Greek letters run over the range $\{1,\ldots d\}$ and $R$ denotes the fundamental representation of a given generator.
The fundamental representation of the Hermitian conjugate of a given generator is given by
\begin{equation}\label{eq:HermConj}    
R(X^\dagger) = \eta^{-1} R(X)^\dagger \eta,
\end{equation}
where $R(X)^\dagger$ denotes the matrix transpose plus the complex conjugation, not
to be confused with the $R(X^\dagger)$ operation defined above. 
Indeed this satisfies the expected relations:
\begin{equation}
    R(P_\mu^\dagger)=R(K_\mu),\qquad
    R(D^\dagger)=R(D),\qquad
    R(L_{\mu\nu}^\dagger)=-R(L_{\mu\nu}).
\end{equation}
In the case of the fundamental representation, the bilinear form can be mapped to the simple trace over matrices. That is, if we define the inner product\footnote{Here the $\Tr$ is performed over the matrix indices of the fundamental (finite-dimensional) representation.}
\beq
\inner{X}{Y} \equiv \frac{1}{2} \Tr[R(X) \,\cdot R(Y)] \, , 
\label{eq:inner_product_so12}
\eeq
we can use bilinearity $ \langle X, Y \rangle = X^I Y^J \langle \omega_I, \omega_J \rangle$, the assumption that the generators are normalized, and apply standard identities on the structure constants to show that the bilinear form \eqref{eq:bilinear_form_Lie_group} and the inner product \eqref{eq:inner_product_so12} are related as
\beq
\langle X, Y \rangle = \mathcal{N} \, B(X,Y) \, .
\label{eq:identity_inner_bilinear}
\eeq
Here, the relative normalization $\mathcal{N}$ depends on the quadratic Casimir of the adjoint representation of the specific Lie group, see \eg eqs.~(15.78) and (15.93) in \cite{Peskin:1995ev}.
As an aside, let us mention that it is sometimes convenient to work with non-Hermitian generators in terms of a complexified inner product
\beq
\inner{X}{Y} = \frac{1}{2} \Tr[R(X^\dagger) \,\cdot R(Y)] \, ,
\label{eq:inner_product_so12_complexified}
\eeq
in which case we can maintian the normalization condition $\inner{\omega_I}{\omega_J}=\delta_{IJ}$ for non-Hermitian generators, but we will avoid using this notation in this manuscript to avoid confusion. For simplicity, in the main text, we will often omit the $R$ indication for the fundamental representation, but it will be clear from the context.

Before proceeding, let us mention the following useful property. Both the inner product and the bilinear form can be shown to satisfy the following property
\beq
\langle X, \mathrm{Ad}_{p}(Y) \rangle = \langle \mathrm{Ad}_{p^{-1}}(X), Y
\rangle \, .
\label{eq:self_adjoint_app}
\eeq
where $\mathrm{Ad}_p(X) = p X p^{-1}$ denotes the adjoint action of $G$ on $\mathfrak{g}$.
For the case of the inner product, this follows from the ciclicity of the trace, while for the bilinear form this can be easily checked by using the orthogonality property of the Killing form
\beq
B(\mathrm{Ad}_p(X), \mathrm{Ad}_p(Z)) = B(X, Z) \, ,
\eeq
and then choosing  $Z = \mathrm{Ad}_{p^{-1}}(Y)$. In this way, we get
$B(\mathrm{Ad}_p(X), Y) = B(X, \mathrm{Ad}_{p^{-1}}(Y))$.

\subsection{Explicit example -- SO(1,2)}
\label{app:ssec:algebra_so12}

Let us now focus on $d=1$, where the global part of the conformal algebra
reduces to $\mathfrak{so}(1,2)$.
The algebra can be spanned by generators $\omega_I = \lbrace D, P, K \rbrace$
satisfying the conjugation rules \eqref{eq:conjugation_so12gen} and the
commutation relations
\beq
 [K, P] = 2 D~,  \qquad [P, D] = - P~, \qquad [K, D] = K~.
\label{eq:map_twobases_so12}
\eeq
The $\mathfrak{so}(1,2)$ algebra is locally isomorphic to $\mathfrak{sl}(2,\mathbb{R})$. We can consider the  Hermitian generators in eq.~\eqref{eq:hermitian_SL2R},
\beq
L_0 = D~, \qquad  L_+ = \frac{1}{2} \le P+K \ri~, \qquad L_- = \frac{i}{2} (P - K)~,
\eeq
whose Lie brackets read 
\beq
[L_0, L_+] = -i L_- \, , \qquad
 [L_0, L_-] = i L_+ \, , \qquad 
 [L_+, L_-] = i L_0  \, .
 \label{eq:algebraSL2R}
 \eeq
The finite-dimensional fundamental representation maps the previous generators to matrices in $\mathcal M_{3,3}(\mathds C)$ chosen as
\begin{align}
  \begin{gathered}
    R(P) =
    \begin{bmatrix}
      0 & 0 & 1\\
      0 & 0 & -i\\
      1 & -i & 0
    \end{bmatrix}~, \quad
    R(K) =
    \begin{bmatrix}
      0 & 0 & -1\\
      0 & 0 & -i\\
      -1 & -i & 0
    \end{bmatrix}~, \quad
    R(D) = L_0 =
    \begin{bmatrix}
      0 & i & 0\\
      -i & 0 & 0\\
      0 & 0 & 0
    \end{bmatrix}~,\\
    R(L_+)  =
    \begin{bmatrix}
      0 & 0 & 0\\
      0 & 0 & -i\\
      0 & -i & 0
    \end{bmatrix}~, \quad R(L_-)  =
    \begin{bmatrix}
      0 & 0 & i\\
      0 & 0 & 0\\
      i & 0 & 0
    \end{bmatrix}~.
  \end{gathered}
  \label{eq:generators_so12}
\end{align}
In this representation, the Hermitian conjugation
is defined using eq.~\eqref{eq:HermConj}, where $\eta = \mathrm{diag}(-1,-1,1)$ is the flat metric in this
three-dimensional space, and the inner product is defined by eq.~\eqref{eq:inner_product_so12}.
One can explicitly check that the application of the conjugation rule
\eqref{eq:HermConj} on the explicit generators
\eqref{eq:generators_so12} is consistent with the identities
\eqref{eq:conjugation_so12gen}.

Since the $\mathrm{SO}(1,2)$ group is non-compact, the quadratic bilinear form is indefinite.
Indeed, it turns out that the non-vanishing inner products between the
generators read
\beq
\inner{P}{K} = \inner{K}{P} = -2 \, , \qquad
\inner{D}{D} = \inner{L_0}{L_0}= 1 \, , \qquad
\inner{L_\pm}{L_\pm} = -1 \, .
\label{eq:normalization_generators_so12}
\eeq
Finally, we explicitly show that the inner product
\eqref{eq:inner_product_so12} coincides with the bilinear form
\eqref{eq:bilinear_form_Lie_group} in this simple case.
The non-vanishing structure constants in the normalized Hermitian basis are given by $f_{+-}^0 = i$, $f_{0+}^- = -i$, $f_{0-}^+ = i$, as can be read directly from
eq.~\eqref{eq:algebraSL2R}.
It is then sufficient to use these results to compute 
\begin{align}
  B(L_0,L_0) & = \frac{1}{2} f_{0A}^B f_{0B}^A = \frac{1}{2} \le f_{0+}^- f_{0-}^+ + f_{0-}^+ f_{0+}^- \ri = 1~,\\
  B(L_{\pm}, L_{\pm}) & = \frac{1}{2} f_{\pm A}^B f_{\pm B}^A = \frac{1}{2} \le f_{\pm 0}^{\mp} f_{\pm \mp}^0 + f_{\pm \mp}^0 f_{\pm 0}^{\mp}  \ri = -1~,
\end{align}
which coincides with eq.~\eqref{eq:normalization_generators_so12}.
By using the bilinearity of the inner product, one can then show that the identity \eqref{eq:identity_inner_bilinear} holds for any two elements of the Lie algebra, where in this case $\mathcal{N}=1$.
\section{Cost functions with non-trivial boundary over the space of states}
\label{app:sec:cost_nontrivial_bdy}

In the main text, we imposed restrictions on the penalty factors to impose that the cost functions were positive-definite through all the space of states.
Here, we take the alternative perspective discussed around eq.~\eqref{eq:pfIpm0condition}: we keep the penalty factors arbitrary, and identify a boundary in the space of states within which the metric is positive definite.
We interpret the metric over the space of states as an appropriate complexity geometry only in the region delimited by this boundary.
The existence of the boundary is determined by the locus where the metric is singular and changes signature.
If we believe that the distances over the coset space quantify the cost of constructing states, the existence of a boundary acts as an obstruction to reach certain states in the Hilbert space.

We quantitatively explore the existence of a non-trivial boundary in the complexity geometry in the case of one-dimensional CFTs in appendix~\ref{app:ssec:bdy_1dCFT}.
We then move to the two-dimensional case in appendices~\ref{app:ssec:coupling_DPK} and \ref{app:ssec:coupling_PK}, where we numerically compute the state complexity for two different cases of couplings between the left and right copies.

\subsection{Boundary in one-dimensional CFTs}
\label{app:ssec:bdy_1dCFT}

Let us consider the metric~\eqref{eq:metricpmbasisfinal} describing the complexity geometry on the coset space of a one-dimensional CFT.
By computing the determinant of the two leading principal minors (the full metric and its top-left entry), we find that the cost function is positive-definite if the following conditions hold:
\beq
 \alpha (\theta) > 0 \, , \qquad
 \mathcal{I}_0 (r^2 +1)^2 - 2 r^2 \alpha (\theta) > 0 \, ,
    \label{eq:constraints_positive_1dCFT}
\eeq
where we have already assumed $\mathcal{I}_0>0$. 
In the main text, we used the above requirements to identify the constraints~\eqref{eq:pfIpm0condition} on the penalty factors such that the metric is positive-definite through all the Poincar\'{e} disk.
Here, we take the opposite perspective: we keep the penalties arbitrary, and use eqs.~\eqref{eq:constraints_positive_1dCFT} to find the locus of points where the metric changes signature.

Assuming that we only consider positive penalty factors $\mathcal{I}_{\pm} \geq 0$, the boundary $\mathfrak{B}$ is identified by the curve 
\beq
\mathfrak{B} \equiv \lbrace  (r, \theta) \in \mathcal{D} :  
\mathcal{I}_0 (r^2+1)^2 -2 r^2 \left[  \mathcal{I}_+ + \mathcal{I}_- + \le \mathcal{I}_- - \mathcal{I}_+ \ri \cos (2 \theta)  \right] = 0 
\rbrace \, ,
\eeq
where $\mathcal{D}$ was defined in eq.~\eqref{eq:statedisk}.
Therefore, the computation of the complexity geometry makes sense only for target states located inside the set $\mathfrak{B}$.
We expect that if the region inside the boundary forms a convex set, then any geodesic connecting the origin with an allowed target point should be fully contained in the region where the metric is positive-definite.

\subsection{Coupling \texorpdfstring{$(D, \bar P+\bar K)$}{(D,\bar P+\bar K)}}
\label{app:ssec:coupling_DPK}

\begin{figure}[tb]
  \centering
  \includegraphics[width=\textwidth]{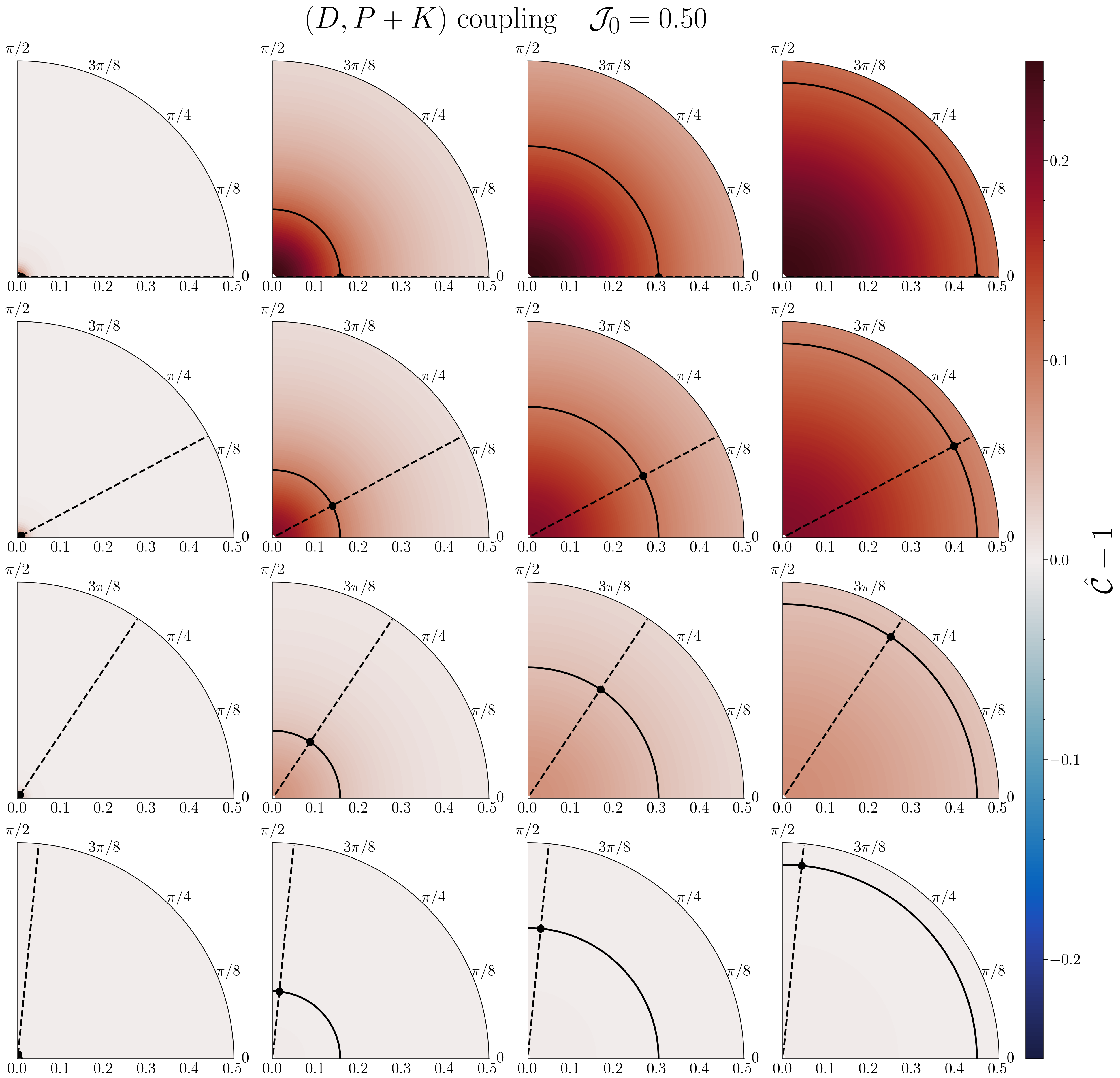}
  \caption{Relative complexity \eqref{eq:relative_complexity} in the $(D,P+K)$
    coupling case for a CFT$_2$ with $\J_0 = 1/2$. The black dot denotes the
    position of the right-copy target state, while each point in the quadrant
  indicates the position of the left-copy target state.}
  \label{fig:dp-coupling-cft2}
\end{figure}

Let us consider the line element~\eqref{eq:general_2d_metric} with interacting term~\eqref{eq:interacting_metric_CFT2_DPK}.
We compute the relative complexity~\eqref{eq:relative_complexity} in a region, centered around the origin $\lambda = \bar{\lambda}=0$ of the space of states, where the metric is positive-definite.
The results are plotted in fig.~\ref{fig:dp-coupling-cft2}.
We observe that the relative complexity \eqref{eq:relative_complexity} monotonically increases along the radial direction for both the left- and right-copies.
The cost to build the optimal circuit seems to be independent of the angular coordinate of the target state in the left copy of the CFT$_1$, while complexity monotonically increases as the angle of the right-copy target state approaches zero.

Some of the previous conclusions can be supported by a perturbative analysis at $\J_0
\ll 1$.
First of all, we notice from our numerical data that even in the
non-perturbative case at $\J_0 = 1/2$, the profile of the right-hand
trajectory remains extremely close to its free shape \eqref{eq:fsSol} (within
roughly $2\%$).
For this reason, a good approximation can be obtained by expanding $\dd
s^2_{\mathrm{int.}}$ up to $\mathcal O(\J_0^2)$, and replacing the right-hand
copy by its free trajectory $\bar{\lambda} = e^{i \bar{\theta}_\tar} \tanh(t \,
\bar{\rho}_\tar)$, where we defined $\bar{\rho}_\tar \equiv \arctanh \bar{r}_\tar$.
If we then parametrize the left-hand trajectory as $\lambda = \tanh(t \rho(t))
e^{i \theta(t)}$, we get
\beq
\begin{aligned}
  \label{eq:ds2-dp-int-series}
  \dd s^2_{\mathrm{int.}} & = 4 \J_0 \theta^\prime(t) \Bigl[\sinh \bigl( 2 t
  \rho(t) \bigr) \Bigr]^2 \cos \bar{\theta}_\tar \bar{\rho}_\tar\\
  & + \J_0^2 \biggl[\Bigl( 2 \bar \rho_\tar \cos \bar \theta_\tar  \cosh \bigl( 2 t
    \rho(t) \bigr) \Bigr)^2 + \Bigl( \sinh(2 t \bar \rho_\tar) \sinh\bigl(2 t \bar
  \rho(t) \bigr) \theta^\prime(t) \sin \bar \theta_\tar \Bigr)^2 \biggr]~.
\end{aligned}
\eeq
First, we notice that the leading-order term in the series around $\J_0=0$
vanishes when $\bar{\theta}_\tar = \frac{\pi}{2}$, while it achieves its maximum
value when $\bar{\theta}_\tar = 0$.
Second, we observe that both terms in the line element are increasing functions
of the radial distance $\bar{\rho}_\tar$.
Finally, while we do not report it here, a more careful perturbative analysis can be used to explain why at fixed $\bar{\theta}_\tar, \bar{\rho}_\tar$, the relative complexity decreases with increasing $r_\tar$, where $r_\tar$ refers to the radial coordinate system in eq.~\eqref{eq:parametrization_CFT2_trajectories}.

\subsection{Coupling \texorpdfstring{$(P,\bar K) + (K,\bar P)$}{(P,\bar K) + (K,\bar P)}}
\label{app:ssec:coupling_PK}

\begin{figure}[tb]
  \centering
  \includegraphics[width=\textwidth]{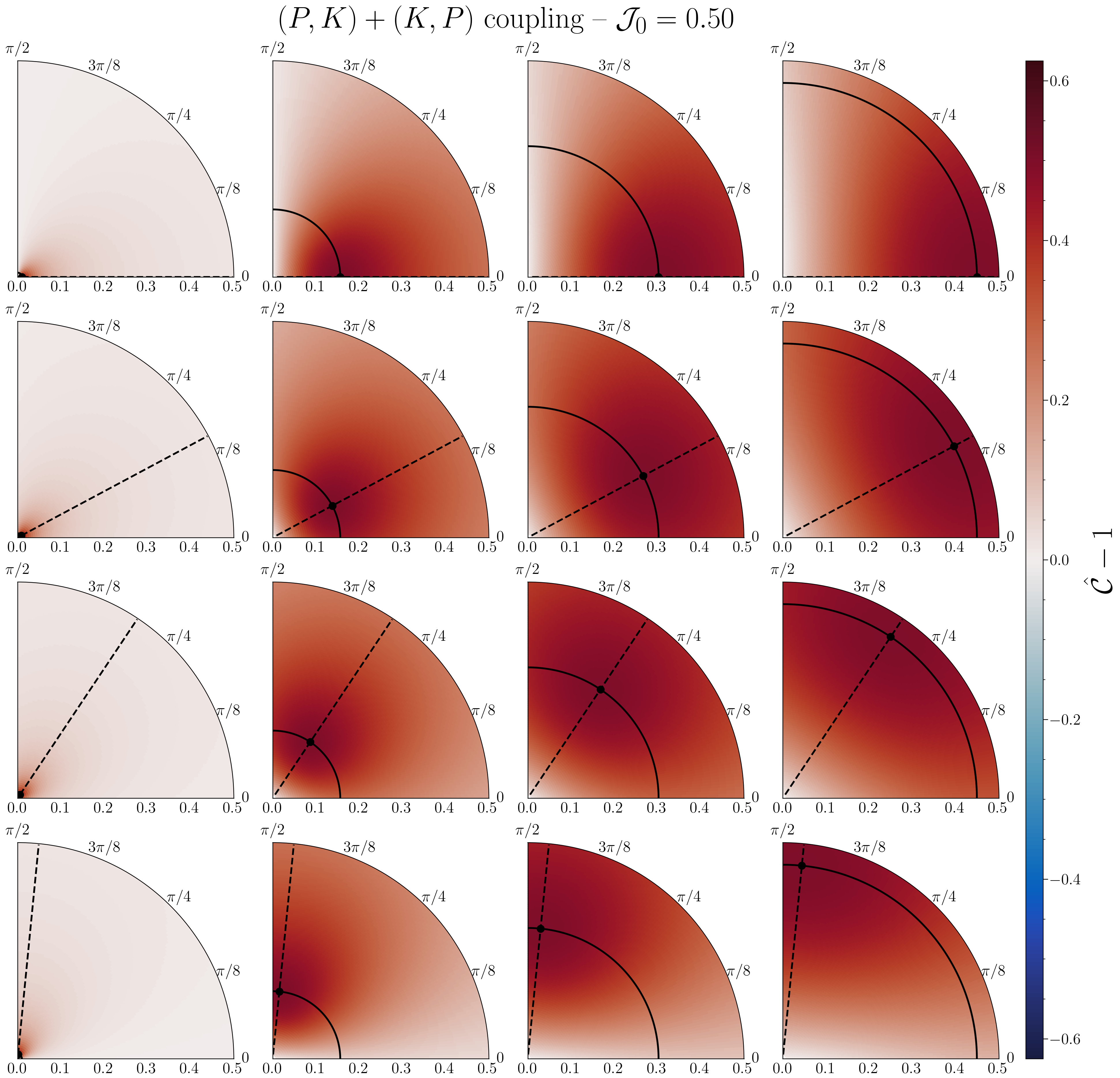}
  \caption{Relative complexity \eqref{eq:relative_complexity} in the $(P,K) +
    (K,P)$ coupling case for a CFT$_2$ with $\J_0 = 1/2$. The black dot denotes
    the position of the right-copy target state while each point in the quadrant
  indicates the position of the left-copy target state.}
  \label{fig:pk-coupling-cft2}
\end{figure}

Let us now consider a coupling of the form~\eqref{eq:interacting_metric_CFT2_PK}.
As we did for the previous case, we study in fig.~\ref{fig:pk-coupling-cft2} the relative complexity \eqref{eq:relative_complexity} compared to the \ac{fs} metric.
The numerical analysis reveals that the relative complexity shows a maximum
centered around the right-copy target point.
Furthermore, we observe that increasing the radial coordinate of the right-hand
target state leads to an overall increase of the cost, while there is no clear
dependence on the angular orientation of the right target state.

To understand these properties, we first notice that while the complexity shows a lot of structure depending on the endpoints of each side, the trajectories themselves remain very similar to the \ac{fs} geodesics.
Indeed, this is confirmed by the following facts. 
First, one can plug the ansatz $\lambda(t) = e^{i \theta_\tar} \tanh \bigl( \rho(t) \bigr)$ (and similarly for $\lb$) inside the line element with interacting part~\eqref{eq:interacting_metric_CFT2_PK}, and analytically check that whenever we impose $\theta_\tar = \bar{\theta}_\tar$, then the EOM imply $\rho(t)= \rho_1 t$. In other words, the FS trajectories are extremal solutions of the metric whenever $\theta_\tar = \bar{\theta}_\tar$.
Secondly, we can numerically estimate the difference in cost between the geodesics of the metric \eqref{eq:interacting_metric_CFT2_PK} and the optimal paths in the \ac{fs} background.
If we fix $\J_0 = 1/2$, $\theta_\tar = \bar \theta_\tar = \pi/4$, $\bar r_\tar = 1/10$ and we vary $r_\tar \in [10^{-2}, 1/2]$, we find that
\beq
\int_0^1 \dd t \sqrt{\bigl( \operatorname{Re}\lambda(t) -
  \operatorname{Re}\lambda_{\fs}(t) \bigr)^2+\bigl( \operatorname{Im}\lambda(t)
- \operatorname{Im}\lambda_{\fs}(t) \bigr)^2} \sim 10^{-8} \, ,
\eeq
suggesting that the \ac{fs} trajectories remain the extremal solutions, and it is only the
complexity on these trajectories which is affected by the penalties.
With this information at hand, we can then plug the \ac{fs} trajectories
inside the complexity functional.
In this way, we find that the resulting quantity (which we do not report here explicitly) only depends on $\theta_\tar - \bar \theta_\tar$ thus explaining the
angular dependency of figure~\ref{fig:pk-coupling-cft2}.
An easier explicit expression can be found by setting $\theta_\tar = \bar
\theta_\tar$, \ie
\beq
\hat{\mathcal C} = 2 \J_0 \frac{\rho_\tar \bar \rho_\tar}{\rho_\tar^2 + \bar \rho_\tar^2}
\, ,
\eeq
which admits a maximum at $\rho_\tar = \bar \rho_\tar$.

\section{Holographic interpretation}
\label{app:holographic_interpretation}

In this appendix, we discuss a holographic interpretation for the projection of the metric over the coset space determined in section~\ref{ssec:projection_coset} for $d$--dimensional \acp{cft}.
The following analysis refers to the case with trivial penalty factors, while the possible extension to the general case is discussed in section~\ref{sec:discussion}.

\subsection{Geodesics of massive particles in embedding space}

First, let us summarize the relation between coherent states in CFT$_d$ and
timelike geodesics of massive particles in AdS${}_{d+1}$, whose study was
initiated in \cite{Chagnet:2021uvi} (see \cite{Dorn:2005jt} for the details on
the following manipulations in \ac{ads} space).
The geodesics are described by a vector $X^A$ in embedding space such that $X^2
= - R^2$ (where $A = 0,0^\prime,1,\ldots,d$), and by
the conjugate momentum $P^A \equiv m \frac{\dot X^A}{\sqrt{- \dot X^2}}$ with,
by definition, $P^2 = -m^2$ --- thus defining a particle with mass $m$.
The action of a massive particle reads \cite{Dorn:2005jt}
\beq
S= - \int \dd \tau \left[  - \frac{\dot{X}^2(\tau)}{2 e(\tau)} + \frac{e(\tau)
m^2}{2} - \frac{\mu(\tau)}{2} \le X(\tau)^2 + R^2 \ri  \right] \, ,
\label{eq:particle_action}
\eeq
where $e(\tau)$ is an einbein and $\mu(\tau)$ acts as a Lagrange multiplier
that imposes the above-mentioned constraint $X(\tau)^2 = -R^2$ along all the
trajectory.
The Euler-Lagrange equations are given by
\beq
e(\tau)^2 = - \frac{\dot{X}(\tau)^2}{m^2} \, , \qquad
\mu(\tau) = - \frac{m}{R^2} \sqrt{-\dot{X}(\tau)^2} \, .
\label{eq:EOM_particle}
\eeq
The $\mathrm{SO}(d,2)$ symmetry of the action \eqref{eq:particle_action} in
embedding space leads to multiple conserved charges of the form
\beq
J_{AB} = P_A X_B - P_B X_A \, ,
\eeq
where $J_{00^\prime} = E$ is the energy of the particle and $J_{mn}$ (with $m,n
= 1, \dots, d$) are the angular momenta associated with the spatial rotations.
For convenience, we define a pair of complex vectors
\beq
z_n = J_{0^\prime n} - i J_{0 n} \, , \qquad
z^*_n = J_{0^\prime n} + i J_{0 n} \, ,
\eeq
which uniquely describe the reduced phase space of timelike geodesics.
To confirm this, we notice that the following identities
\begin{align}
  \label{eq:holo-constraints}
  C = \frac{1}{2} J_{AB}J^{AB} = m^2 R^2~, \qquad J_{mn} = \frac{z_m^* z_n -
  z_n^* z_m}{2 i E}~,
\end{align}
successfully constrain all the conserved charges of the system in terms of
$z,z^*$.

Next, one can show that the solutions to geodesic equations in AdS$_{d+1}$ are
successfully parametrized by
\begin{align}
  \begin{gathered}
    X_0 = r(t) \cos(t/R)~, \quad X_{0^\prime} = r(t) \sin(t/R)~,\\
    X_n = \frac{J_{0 n} X_{0^\prime} - J_{0^\prime n} X_0}{E}~,
  \end{gathered}
\end{align}
showing that the spatial components $X_n$ are uniquely determined by the
conserved charges.
Finally, there exists a change of coordinates which maps the complex vectors $(z,z^*)$ to the parameters $\lambda$ defining the unitary circuits
\eqref{eq:circuits-ansatz} as \cite{Chagnet:2021uvi}
\begin{align}
  z_n = 2 m R \, \frac{\lambda^*_n - (\lambda^*)^2 \lambda_n}{1 - 2 |\lambda|^2
  + \lambda^2 (\lambda^*)^{2}}~, \qquad
  E = m R \,\frac{1 - (\lambda^*)^2\lambda^2}{1 - 2 |\lambda|^2 + \lambda^2
  (\lambda^*)^{2}}~.
\end{align}
A traditional result of classical mechanics is that the symplectic form on the
phase space can be obtained as
\beq
\omega_{\mathrm{bulk}} = \delta P^A \wedge \delta X_A = \omega_{\fs} \, ,
\label{eq:bulk_sympl}
\eeq
where $\omega_{\fs}$ is the symplectic form associated with the \ac{fs} geometry. In particular, the latter equality was shown to hold in reference~\cite{Dorn:2005jt}. 
The following symplectic potential can then be determined
\beq
Q(t) = P^A(t) \delta X_A(t) \, , \qquad
\dd Q = \omega_{\mathrm{bulk}} \, .
\eeq
The symplectic potential $Q$ is formally $t$-dependent,
but it can be matched to the field theory \ac{mc} expectation value
$\ev{U^\dagger \dd U}{\Delta}$ and the bulk pre-symplectic potential $\Omega$
derived in ref.~\cite{Dorn:2005jt} at the special time $t_* = \frac{R}{2i}
\log \bigl( \frac{1 + (\lambda^*)^2}{1 + \lambda^2}\bigr)$, 
where $U$ is a unitary circuit \eqref{eq:circuits-ansatz} and $\ket{\Delta}$ a
primary state with conformal dimension $\Delta = m R$. In our notations, 
\begin{align}
  \Omega = Q(t_*) = \ev{U^\dagger \dd U}{\Delta}\vert_{\gamma_R = 0} = i \Delta
  \frac{\lambda^* \dd \lambda - \lambda \dd \lambda^*}{1 - |\lambda|^2}~.
\end{align}
In reference \cite{Chagnet:2021uvi}, it was shown that the following identity holds $\Delta \delta s^2_\fs =
\delta s^2_1$, in terms of the line element
\begin{align}
  \delta s_1^2 = \frac{m^2}{2} \left( \delta \hat X^2(t_-) + \delta \hat
  X^2(t_+) \right)
  \label{eq:relation_ds1_X}
\end{align}
where $\delta \hat X^2$ denotes the quantity $\delta X^2$ subject to the
constraint $\dot X^A \delta X_A = 0$, and $t_\pm$ denote the times defined by the condition $\partial_t \delta \hat X^2(t_\pm) = 0$, equal to $t_\pm =\frac{R}{2i} \log\Bigl( \frac{(\lambda^*)^2 \dd \lambda \pm \dd \lambda^*}{\dd
\lambda \pm \lambda^2 \dd \lambda^*}\Bigl)$.
In the next subsection, we give a novel interpretation to
eq.~\eqref{eq:relation_ds1_X}, and connect it to the \ac{fs} metric on the coset
space of a $d$--dimensional \ac{cft}.

\subsection{Novel interpretation}

In order to give a novel interpretation to the holographic
formula~\eqref{eq:relation_ds1_X}, we first notice that the condition $\dot
X^A \delta X_A = 0$ is equivalent to setting $Q = 0$, which is reminiscent of
the prescription to define a pseudo-Riemannian submersion that maps the metric
from the Lie group to the coset space, as discussed in
section~\ref{ssec:general_projection}.
In the particular case of the \ac{fs} line element \eqref{eq:susskindBeforeMin},
this condition reads $K_D=0$.

In order to make further progress in interpreting the holographic system, we
define yet another line element
\begin{align}
  \label{eq:holo-fs-all-time}
  \delta s_2^2 = \frac{1}{2} \left(m^2 \delta X^2 + R^2 \delta P^2\right) - P^A
  \delta X_A X^B \delta P_B~.
\end{align}
Surprisingly, not only is $\delta s_2^2$ time-independent, but one can also
show that it satisfies $\delta s_2^2 = \Delta \delta s^2_\fs$.

Next, we show that there exists a relation between this line element and
eq.~\eqref{eq:relation_ds1_X}.
Firstly, we notice that $P^A \delta X_A X^B \delta P_B = - Q^2$.
If we express $\delta s_2^2$ in terms of the constrained variation which
satisfies $Q=0$, we then get $\delta s_2^2 = \frac{1}{2} \left(m^2 \delta \hat
X^2 + R^2 \delta \hat P^2\right)$.
Secondly, we use that $\delta \hat X^2(t_-) = \delta \hat P^2(t_+) = 0$ such that
in reality, $\Delta \delta s^2_\fs = \delta s_1^2 = \frac{m^2}{2} \delta \hat
X^2(t_+) = \delta s_2^2(t_+)$.
This establishes the relation between the two line elements.

So far, we have shown the following relations between the different quantities
\begin{align}
  \begin{array}{c@{}c@{}c}
    \omega_{\text{bulk}} & ~\Longleftrightarrow~ & \delta s^2_\fs \\
    \Updownarrow\, ? & & \Downarrow \\
    \delta s_2^2 & \Longleftrightarrow & \delta s_1^2
  \end{array}
\end{align}
The main issue in this diagram is that it is difficult to see how $\delta
s_1^2$ is related to $\delta s_\fs^2$.
The final link is to show that there is a natural connection between the line
element \eqref{eq:holo-fs-all-time} and the bulk symplectic form
\eqref{eq:bulk_sympl}. This will confirm the status of $\delta s^2_2$ as an
equivalent, \emph{time-independent} bulk measure of the \acl{fs} metric
using more intuitive quantities. The crucial observation
is that in a K\"ahler manifold, the relation between symplectic form and the associated metric is governed by a complex map $\bm J$ as $g(X,Y) = \omega(X, \bm
J Y)$. In the case of the bulk symplectic form, this means that we can write
\begin{align}
  \omega_{\text{bulk}} & = \frac{1}{2} \delta P^A \otimes \delta X_A -
  \frac{1}{2} \delta X^A \otimes \delta P_A~,\\
  \Delta \delta s_{\fs}^2 & = \frac{\Delta}{2} \delta P^A \otimes ( \bm J
  \delta X)^A - \frac{\Delta}{2} \delta X^A \otimes  ( \bm J \delta P)^A~.
  \label{eq:fs-from-symplectic}
\end{align}
If we focus on specific times $t = t_*$ for which $Q(t_*) = \Omega$, then we
can use the field theory results for the complex map $\bm J \dd \lambda = -i
\dd \lambda$ and $\bm J \dd \lambda^* = i \dd \lambda^*$. Doing so, we find
that
\begin{align}
  \bm J \delta X(t_*) & = \frac{R^2}{\Delta} \delta P(t_*) +
  \frac{1}{\Delta}\Bigl( X(t_*) \cdot \delta P(t_*) \Bigr) X(t_*) +
  \frac{R^2}{\Delta} P(t_*) \delta \log \bigl(1 - |\lambda|^2\bigr)~,\\
  \bm J \delta P(t_*) & = -\frac{m^2}{\Delta} \delta X(t_*) -\frac{1}{\Delta}
  \Bigl( P(t_*) \cdot \delta X(t_*) \Bigr) P(t_*) - \frac{m^2}{\Delta} X(t_*)
  \delta \log \bigl(1 - |\lambda|^2\bigr)~.
\end{align}
Plugging these expressions inside the metric \eqref{eq:fs-from-symplectic} and
remembering that $P \cdot \delta P = X \cdot \delta X = 0$, we then obtain $\delta s_2^2$.
Since the line element $\delta s_2^2$ is time-independent, we finally conclude that the relation between the metrics holds at all times.

\bibliographystyle{JHEP}
\bibliography{refs}

\end{document}